\documentclass{aa}

\usepackage{graphicx}
\usepackage{amsmath,amssymb}
\usepackage[utf8]{inputenc}
\usepackage{txfonts}
\usepackage{hyperref}

\newcommand{\xe}{x_{\rm e}}
\newcommand{\xeres}{x_{\rm e}^{\rm res}}
\newcommand{\lG}{\lambda_G}
\newcommand{\aG}{\alpha_G}
\newcommand{\GN}{G_{\rm N}}
\newcommand{\dd}{\mathrm{d}}
\newcommand{\rdamp}{r_{\rm damp}}
\newcommand{\ldamp}{\ell_{\rm damp}}
\newcommand{\lmfp}{\lambda_{\rm mfp}}
\newcommand{\CLASS}{\texttt{CLASS}}
\newcommand{\Neff}{N_{\rm eff}}

\usepackage{xcolor}

\begin{document}

\title{The gravitational constant and the cosmic microwave background}

\author{B. Lamine\inst{1}
   \and S. Beauchamp\inst{1}
   \and A. Blanchard\inst{1}
   \and J. Delhomelle\inst{1,3}
   \and R. Kou\inst{1}
   \and J. Rich\inst{2}}

\institute{Univ Toulouse, CNES, CNRS, IRAP, Toulouse, France\\
\email{brahim.lamine@utoulouse.fr}
\and
IRFU, CEA, Universit\'e Paris-Saclay, F-91191 Gif-sur-Yvette, France
\and
Department of Physics, University of Helsinki, P.O. Box 64, FI-00014 Helsinki, Finland}

\date{Received ...; accepted ...}

\abstract
{The cosmic microwave background (CMB) constrains the value of the
gravitational constant $G$ at the recombination epoch to within a few percent
of its laboratory value, and a number of studies have measured this
sensitivity from the full temperature and polarization spectra. In most of
them the effect of $G$ on the detailed physics of recombination is not made
explicit, being absorbed into a global fit.}
{We aim to provide a physical interpretation of the CMB sensitivity to $G$ by
returning to the semi-analytic description of the acoustic spectrum, and to
determine how the three characteristic angular scales of the CMB, the acoustic
scale $\ell_A$, the equality scale $\ell_{\rm eq}$, and the diffusion damping
scale $\ell_{\rm damp}$, respond to a variation of $G$ during
recombination.
We also ask where, and how strongly, $G$ enters the physics of the residual
ionization left in the dark ages.}
{We use a modified version of the Boltzmann code \CLASS{} in which $G$ is
varied through the dimensionless parameter $\lG \equiv \sqrt{G/\GN}$, scanning
$\lG \in [0.25,2]$ at fixed density parameters and fixed CMB temperature, with
the primordial helium fraction recomputed for each model. Analytic rate
arguments are confronted with the numerical ionization histories, the
visibility function, the Hu \& White damping integral, and direct fits to the
damping envelope of the simulated temperature spectra (TT).}
{While $\ell_{\rm eq}$ and $\ell_A$ are nearly independent of $G$, we
  show that non-equilibrium recombination leads to
  $\ldamp \propto \lG^{-0.304}$. The dependence exists only above a
  critical value of $G$, determined by fundamental and cosmological
  parameters, which turns out to be a factor of $49$ below $\GN$.
  We also examine the effect of $G$ on
  the residual ionization left in the dark ages and find it weak, the dark-ages
  optical depth varying by only $16\%$ when $G$ is multiplied by $64$.}
{Because $G$ enters only through the expansion rate, the whole imprint comes
  from recombination proceeding out of equilibrium. The competition between
  the atomic rates and that expansion is governed by the single
  dimensionless combination $\alpha_G \equiv G m_e \mu/\hbar c$, with $\mu$
  the mean mass of matter per photon.
  Each angular scale is the ratio
  of the distance to last scattering to a comoving length imprinted before it,
  so a modification of gravity confined to low redshift rescales all three
  alike. The ratio $\ldamp/\ell_A$ therefore isolates the value of $G$ at
  recombination and, up to the lensing of the damping tail and to any screening
  mechanism, compares it directly with the laboratory value measured today.}

\keywords{cosmic microwave background --
cosmological parameters --
early Universe --
cosmology: theory}

\maketitle

\section{Introduction}

Whether the gravitational constant takes the same value at all epochs
is one of the oldest questions of cosmology
\citep{Dirac1937,Uzan2011}. In scalar-tensor theories the effective
$G$ is set by the local value of a scalar field and evolves with the
expansion, so that a mismatch between the value inferred from
early-Universe observables and the laboratory value $\GN$ is a generic
prediction rather than an exotic one. Big Bang Nucleosynthesis (BBN)
has long provided the classical constraint, through the sensitivity of
the frozen neutron-to-proton ratio to the expansion rate
\citep{Steigman1977,Alvey2020}, and the CMB provides a second,
independent epoch, which turns out to be sensitive to the value of
$G$.

The response of the CMB to a modified $G$ has by now been studied in
considerable detail. Following the early analyses of \citet{Zahn2003}
and \citet{Galli2009}, constraints have been obtained from successive
Planck releases, both within the phenomenological varying-$G$
parametrisation \citep{Wang2020,Sakr2022} and within specific
scalar-tensor and harmonic-attractor models
\citep{Ooba2016,Ballardini2022}, and the statistical precision has
steadily improved to the point where the value of $G$ relevant to
recombination is now pinned to within a few percent of the laboratory
measured value $\GN$ \citep{CODATA2018}. The most recent such
determination, using Planck PR4 together with DESI baryon acoustic
oscillations and the BBN helium fraction, reaches $1.8\%$ and is
compatible with the laboratory value at one standard deviation
\citep{Lamine2025}. These works establish that the CMB is genuinely
sensitive to $G$, and that the sensitivity is not degenerate with the
standard cosmological parameters, because the physics of recombination
involves non-gravitational processes, in particular Thomson
scattering, in addition to the expansion rate.

Two of these works do discuss the mechanism, in words.
\citet{Zahn2003} observe that a faster expansion leaves hydrogen less
able to recombine, that the visibility function broadens as a result,
and that the equality redshift does not move. \citet{Galli2009} add
that the extra small-scale damping can be offset by raising the scalar
spectral index. Neither, however, treats the sound horizon, the
damping scale or the equality scale analytically, and neither gives
the $G$-dependence of any of them in closed form.

What none of these analyses provides, being global fits to the full
spectra, is an explicit and detailed account of \emph{how} $G$ enters
the physics of the last-scattering epoch, and therefore which features
of the temperature spectrum carry the signal. This is the gap the
present paper addresses. Rather than fitting the spectra, we return to
the semi-analytic description of the acoustic oscillations
\citep{HuSugiyama1995,HuWhite1997} and follow the response of its
ingredients to a variation of $G$, so as to obtain the $G$-dependence
of the three characteristic angular scales of the CMB: the acoustic
scale $\ell_A$, the equality scale $\ell_{\rm eq}$, and the diffusion
damping scale $\ell_{\rm damp}$. The interest of this approach is that
it isolates the physical origin of the effect, which turns out to be
almost entirely kinetic.

The reason the effect is subtle is that the gravitational constant
does not appear in the atomic physics of recombination. It enters the
problem only through the Friedmann equation, and therefore only
through the rate at which the Universe expands while recombination
proceeds. This makes the CMB sensitivity to $G$ qualitatively
different from its sensitivity to, say, the fine-structure constant or
the electron mass, which act directly on energy levels and
cross-sections. A useful reference point is the exploratory
three-scale treatment of \citet{Rich2015}, in which decoupling is
defined by equating the Thomson scattering rate to the expansion
rate. That definition forces the photon mean free path and the Hubble
radius to coincide at decoupling, so that the damping scale, being
their geometric mean, becomes a Hubble radius itself. All three scales
of the spectrum are then comoving Hubble radii carrying the same
factor of $\sqrt{G}$, which cancels in their ratios and leaves only a
logarithmic residual. We show that this identification is precisely
the step that suppresses the signal, quantify by how much the two
lengths actually differ, and compute what replaces the cancellation
once the finite thickness of last scattering and the
out-of-equilibrium kinetics of recombination are retained.

The paper is organised as follows. A reader who wants the results
rather than the derivations will find each of them in a single
equation, listed in Sec.~\ref{sec:conclusion}.
Section~\ref{sec:framework} defines the family of models and the
numerical setup. Section~\ref{sec:geometric} shows that the equality
scale $\ell_{\rm eq}$ of Eq.~\eqref{eq:leq} is invariant under a
rescaling of $G$, and that the acoustic scale $\ell_A$ drifts by only
a few percent, so that the sensitivity is carried almost entirely by
the damping scale. Section~\ref{sec:recombination} establishes the
structural point that the Saha equilibrium is $G$-free,
Eq.~\eqref{eq:saha}, and then isolates the single dimensionless
combination through which $G$ does enter the kinetics,
$\aG = Gm_e\mu/\hbar c$ in Eq.~\eqref{eq:alphaG}, together with the
criterion $\mathcal{Q}$ of Eq.~\eqref{eq:Qcrit} that decides how far
recombination departs from equilibrium.
Section~\ref{sec:decoupling} studies decoupling and the visibility
function, and the characteristic time and length scales that enter the
damping. Section~\ref{sec:damping} propagates the kinetic ionization
history through the diffusion damping integral, writes the damping
length and the damping multipole in fundamental constants,
Eqs.~\eqref{eq:rdamp} and \eqref{eq:ldamp_micro}, reaches the central
result in the decomposition of Eq.~\eqref{eq:decomposition}, locates
the critical value of $G$ below which the whole effect disappears,
Eq.~\eqref{eq:aGcrit_explicit}, and analyses the damping tail of the
\CLASS{} spectra. Section~\ref{sec:residual} turns to the residual
ionization left in the dark ages and to its scaling with $G$.
Section~\ref{sec:finaldiscussion} discusses what the result does and
does not establish, in particular the distance-free ratios of
Eq.~\eqref{eq:threeratios}, and Sec.~\ref{sec:conclusion} states the
conclusions in brief.

\section{Framework and numerical setup}
\label{sec:framework}

\subsection{The family of models}

We parametrise a modified gravitational constant by the dimensionless,
redshift-independent quantity
\begin{equation}
\lG \;\equiv\; \sqrt{G/\GN}\,,
\end{equation}
so that the Friedmann equation for a spatially flat universe reads
\begin{equation}
H(z) \;=\; \sqrt{\frac{8\pi G \rho(z)}{3}} \;=\; \lG\, \tilde H(z)\,,
\label{eq:friedmann}
\end{equation}
where $\tilde H$ denotes the Hubble rate evaluated at $G = \GN$. To
isolate the effect of $G$ alone, we consider models in which the
physical energy densities $\rho_{i,0}$ at the present epoch are held
fixed, and we further lock the CMB temperature to its measured value
$T_0 = 2.7255$~K \citep{Fixsen2009}. The densities themselves evolve
with redshift, but they are identical across the family at a
given $z$. With $\rho_{i,0}$ fixed,
$\omega_i \equiv \Omega_i h^2 \propto \lG^2$. The Friedmann equation
\eqref{eq:friedmann} is then satisfied provided
$h = H_0/100 \propto \lG$, which leaves the density parameters
$\Omega_i$ unchanged as well as the critical density
$\rho_{c,0} \equiv 3H_0^2/8\pi G$. More precisely, the matter density
parameter $\Omega_m \equiv \rho_{m,0}/\rho_{{\rm crit},0}$ is one of
the quantities held fixed. The present matter density can be written
as $\rho_{m,0} = \mu\,n_{\gamma,0}$, where
\begin{equation}
\label{eq:mu}
\mu \;\equiv\; \frac{\rho_m}{n_\gamma} \;=\; m_p\,\eta_b\,\frac{\rho_m}{\rho_b}
\end{equation}
is a constant of the model since both quantities scale as
$(1+z)^3$, and $\eta_b = n_b/n_\gamma$ is the ratio between the
baryon and photon number densities. Here $\rho_m$ is the total
non-relativistic density, cold dark matter and baryons, so that
$\mu = 6.52\times10^{-36}\,$kg for our reference model. The Friedmann
equation taken at $z = 0$ then gives
\begin{equation}
H_0 = \sqrt{\frac{8\pi G\,\mu\,n_{\gamma,0}}{3\,\Omega_m}}\,,
\label{eq:H0G}
\end{equation}
where $\mu$, $n_{\gamma,0}$ and $\Omega_m$ are all held fixed across
the family of models, so that the entire $G$-dependence of $H_0$ is
the explicit $G^{1/2}\propto \lG$. The combination that actually
enters the pre-recombination physics is not $H_0$ itself but
$8\pi G\mu n_{\gamma,0}/3$, equal to $\Omega_m H_0^2$, in which no
density parameter appears, and it is the one we use from
Sec.~\ref{sec:rate} onwards.

Since the physical densities are unchanged, so is their ratio. The
matter-radiation equality redshift is therefore an invariant of the
family, and we have verified numerically that
\begin{equation}
  z_{\rm eq} = 3405.8 \quad \text{for all models}, \qquad \lG \in [0.25,2]\,.
\end{equation}
This is worth emphasising because it separates a rescaling of $G$ from
the addition of dark radiation since an increase of $\Neff$ at fixed
$\omega_m$ displaces $z_{\rm eq}$, whereas a rescaling of $G$ does
not.

\subsection{Numerical implementation}

We use a modified version of the Boltzmann code \CLASS{}
\citep{Blas2011} in which $\lG$ is a free parameter. The patch that
produces this version, the reference input files for the ten models and
the documentation of the modified BBN table are publicly available at
\url{https://github.com/blamine/varying-constants-class}, archived at
\url{https://doi.org/10.5281/zenodo.23399410}. Recombination is
computed with \texttt{HyRec} \citep{AliHaimoud2011}. We have checked
that switching to \texttt{RECFAST} \citep{Seager1999} changes none of
the scalings reported below at the percent level. It is worth
stressing, since the analytic discussion of
Secs.~\ref{sec:recombination} to \ref{sec:residual} is cast at the
level of the Peebles equation, that every number quoted in this paper
comes from \texttt{HyRec}'s full multilevel treatment, for every
model and therefore including the helium-rich end of the scan. The
Peebles-level algebra is used to interpret those numbers and never to
produce them. We scan ten models,
$\lG \in \{0.25, 0.5, 0.75, 0.9, 1, 1.1, 1.25, 1.5, 1.75, 2\}$. The wide
spacing spans a range far beyond any observationally allowed interval, so that
power-law behaviour can be identified unambiguously, while the close triplet
$0.9, 1, 1.1$ brackets the standard value tightly enough for the logarithmic
derivatives quoted throughout to be obtained without any fit. The helium mass
fraction is the self-consistent BBN value in every case, obtained as described
below, and we recover the value
\CLASS{} actually used from the fully ionized plateau of $\xe$ at high
redshift, which reproduces the input to $10^{-13}$.

The primordial helium mass fraction is not an external constant in
this context, since $Y_{\rm He}$ responds to the expansion rate at
BBN. A faster expansion freezes the neutron-to-proton ratio earlier
and raises the helium yield. The sensitivity of $Y_{\rm He}$ to $G$ is
known analytically at the linear level \citep{DSW2007}, but because
our scan spans a wide range of $\lG$, well outside the regime in which
a linear sensitivity coefficient would be adequate, we do not rely on
a fitting formula and instead recompute $Y_{\rm He}$ self-consistently
for each model with a modified version of \texttt{PArthENoPE}
\citep{Gariazzo2022} in which the expansion rate is rescaled by $\lG$,
and pass the result to the recombination module.

The reionisation optical depth requires a similar treatment. We do not
want $\tau_{\rm reio}$ to drift from model to model as a mere
by-product of the rescaled clock, since that would confuse a change of
the damping tail with a change of the overall amplitude. With
\begin{equation}
\tau_{\rm reio} = \sigma_T c \int_0^{z_{\rm reio}}
\frac{\xe(z)\,(1-Y_{\rm He})\,n_{b,0}\,(1+z)^2}{H(z)}\,\dd z\,,
\label{eq:taureio}
\end{equation}
and holding the reionisation history $\xe(z)$ fixed, the only
$G$-dependent factors left are the expansion rate and the electron
budget, so that
\begin{equation}
\frac{\tau_{{\rm reio};G}}{\tau_{{\rm reio};\GN}} =
\lG^{-1}\,\frac{1-Y_{{\rm He};G}}{1-Y_{{\rm He};\GN}}\,,
\label{eq:taureio_rescale}
\end{equation}
and we rescale the input $\tau_{\rm reio}$ of each model accordingly,
using the $Y_{\rm He}$ obtained above. This keeps the late-time
optical depth physically equivalent across the family. The residual
ionization left by recombination is not treated this way, since
$\xe(z)$ in the dark ages is itself an output of the kinetics and
carries its own $G$-dependence, which is the subject of
Sec.~\ref{sec:residual}.

\section{The three angular scales of the CMB}
\label{sec:geometric}

The temperature power spectrum is characterised by three angular
scales: the equality scale $\ell_{\rm eq}$, which marks the transition
to the radiation-driven regime; the acoustic scale $\ell_A$, which
sets the spacing of the peaks; and the diffusion damping scale
$\ell_{\rm damp}$, which governs the fall-off of power at high
multipole. Before treating recombination and decoupling in detail, we
establish how these three scales respond to a variation of $G$. The
result is simple and organises the rest of the paper, and we present
the scales in order of increasing sensitivity to $G$. The equality
scale is invariant to better than a tenth of a percent. The acoustic
scale is almost invariant, drifting by only a few percent through the
displacement of the last-scattering surface. And essentially the
entire sensitivity of the CMB to $G$ is carried by $\ell_{\rm damp}$,
which is not a purely geometric ratio and moves substantially.

\subsection{The equality scale}
\label{sec:equality}

The equality scale is the angular projection of the comoving
wavenumber $k_{\rm eq}$ that entered the horizon at matter-radiation
equality \citep{HuSugiyama1995},
\begin{equation}
\ell_{\rm eq} \;=\; k_{\rm eq}\,D_c(z_{\rm dec})\,,
\qquad
k_{\rm eq} = \sqrt{2}\,a_{\rm eq} H(z_{\rm eq})\,,
\label{eq:leq}
\end{equation}
where $D_c(z_{\rm dec}) = \int_0^{z_{\rm dec}} c\,\dd z/H(z)$ is the
comoving distance to last scattering, which coincides with the
comoving angular-diameter distance entering the projection since our
models are spatially flat. We evaluate it at the visibility-peak
redshift $z_{\rm dec}$ of Sec.~\ref{sec:decoupling}. The standard
alternative, the redshift $z_\ast$ defined by $\tau(z_\ast)=1$,
differs from it by only a few units and leaves the discussion of this
section unchanged. It does \emph{not} leave the damping scale
unchanged, however, and the sensitivity of $\rdamp$ to that choice is
quantified in Sec.~\ref{sec:silkintegral}. The behaviour of
$\ell_{\rm eq}$ is controlled by $z_{\rm eq}$, which, as established
in Sec.~\ref{sec:framework}, is strictly invariant under a rescaling
of $G$ because it depends only on the ratio of the matter and
radiation physical densities, both left unchanged in our
$\lambda_G$-models. We find $z_{\rm eq} = 3405.8$ for all ten
models. At fixed $z_{\rm eq}$ the comoving Hubble radius at equality
scales as $a_{\rm eq} H(z_{\rm eq}) \propto \lG$, so that
$k_{\rm eq} \propto \lG$, while the physical densities being fixed
makes every background length an integral of $c/H \propto \lG^{-1}$,
so that $D_c(z_{\rm dec}) \propto \lG^{-1}$. The two factors cancel,
and
\begin{equation}
\ell_{\rm eq} = 203.9 \quad \text{for all } \lG \in [0.25,2]\,.
\end{equation}
This cancellation is very nearly exact because $k_{\rm eq}$ is
anchored to $z_{\rm eq}$, which is \emph{strictly} invariant. The only
residual $G$-dependence is the tiny one carried by $D_c$ through the
drift of the integration bound $z_{\rm dec}$, and it amounts to just
$0.07\%$ across the whole range $\lG \in [0.25,2]$. For all practical
purposes $\ell_{\rm eq}$ is invariant, and it carries no information
on $G$ (as long as one considers that $\lambda_G$ does not vary with
the redshift).

\subsection{The acoustic scale}
\label{sec:acoustic}

\begin{figure}[h]
\centering
\includegraphics[width=\columnwidth]{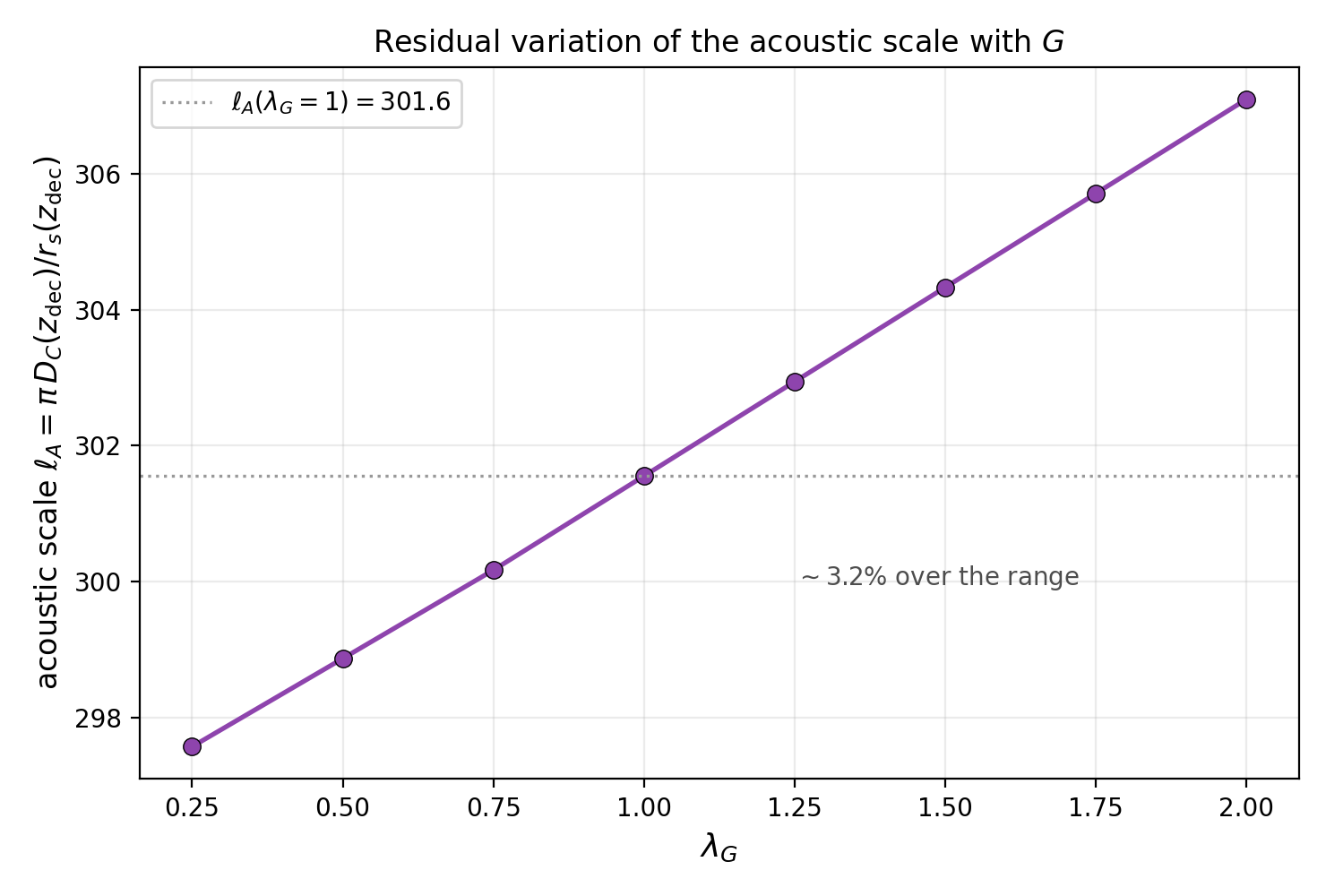}
\caption{Acoustic scale
  $\ell_A = \pi\,D_c(z_{\rm dec})/r_s(z_{\rm dec})$ of
  Eq.~\eqref{eq:lA} for the eight models, evaluated at the visibility
  peak $z_{\rm dec}(\lG)$. The geometric cancellation
  $D_c \propto r_s \propto \lG^{-1}$ would make $\ell_A$ exactly
  invariant at fixed $z_{\rm dec}$. The residual $\sim 3\%$ drift is
  due entirely to the displacement of $z_{\rm dec}$ with $G$, and
  matches the widening of the acoustic-peak spacing in the \CLASS{}
  spectra.  The dotted line marks $\ell_A(\lG=1) = 299.6$, in
  agreement with Table~\ref{tab:sensitivities}.}
\label{fig:ellA}
\end{figure}

The acoustic scale is the angle subtended by the comoving sound
horizon at last scattering \citep{HuSugiyama1995,HuWhite1997},
\begin{equation}
\ell_A \;=\; \pi\,\frac{D_c(z_{\rm dec})}{r_s(z_{\rm dec})}\,,
\qquad
r_s(z_{\rm dec}) = \int_{z_{\rm dec}}^{\infty} \frac{c_s(z)}{H(z)}\,\dd z\,,
\label{eq:lA}
\end{equation}
with $r_s$ the comoving sound horizon, $c_s$ the photon-baryon sound
speed (which depends only on the fixed baryon-to-photon ratio), and
$D_c(z_{\rm dec})$ the same comoving distance as in
Eq.~\eqref{eq:leq}. Both integrals are again evaluated at
$z_{\rm dec}$. At \emph{fixed} $z_{\rm dec}$ the sole $G$-dependence
is once more the overall factor $1/H \propto \lG^{-1}$, common to
sound horizon and distance,
\begin{equation}
r_s(z_{\rm dec}) \propto \lG^{-1}\,, \qquad D_c(z_{\rm dec}) \propto \lG^{-1}\,,
\end{equation}
so that their ratio, and hence $\ell_A$, would be exactly independent
of $G$, with $\ell_A \simeq 299.6$ at $\lG=1$. Here, however, the
cancellation is less perfect than for $\ell_{\rm eq}$. The reason is
that $r_s$ has its lower integration limit \emph{at} $z_{\rm dec}$,
which itself drifts with $G$. As the last-scattering surface moves
(Sec.~\ref{sec:decoupling}), $z_{\rm dec}$ rises from $\simeq 1059$ at
$\lG = 0.25$ to $\simeq 1104$ at $\lG = 2$, shifting the lower limit
of $r_s$ and the upper limit of $D_c$ by unequal amounts. These two
shifts are far from symmetric. For
$D_c = \int_0^{z_{\rm dec}} c\,\dd z/H$ the moving endpoint sits at
high redshift, where the integrand $c/H$ is small, so the displaced
sliver is negligible against an integral dominated by low
redshifts. Once the overall $\lG^{-1}$ scaling is removed,
$D_c(z_{\rm dec})$ varies by only $\sim 0.07\%$ across the range. For
$r_s = \int_{z_{\rm dec}}^{\infty} c_s\,\dd z/H$, by contrast,
$z_{\rm dec}$ is the \emph{lower} limit, right where the integrand is
largest, so the same displacement removes or adds a sliver from the
dominant part of the integral. Its residual variation is $\sim 3\%$,
some forty times larger.  The sensitivity of $\ell_A$ to $G$ therefore
comes almost entirely from $r_s$, not from $D_c$. Evaluating
Eq.~\eqref{eq:lA} at the actual $z_{\rm dec}(\lG)$ leaves a residual
variation (Fig.~\ref{fig:ellA}),
\begin{equation}
\ell_A \simeq 296.1 \to 304.4 \quad \text{as}\ \lG:\,0.25\to2\,,
\end{equation}
a $\sim 3\%$ drift. This is also why $\ell_A$ is less stable than
$\ell_{\rm eq}$. The equality scale $k_{\rm eq}$ is anchored to the
invariant $z_{\rm eq}$ rather than to the moving $z_{\rm dec}$, so it
has no equivalent of the sensitive lower limit of $r_s$. The drift is
confirmed directly by the spacing of the acoustic peaks in the
\CLASS{} spectra, which widens from $\simeq 299$ to $\simeq 307$ over
the same range. The acoustic scale is thus \emph{almost} untouched by
a rescaling of the gravitational constant (the geometric cancellation
of $c/H$ integrals is exact at fixed $z_{\rm dec}$) with a small
residual set entirely by the displacement of the last-scattering
surface, the same effect that governs the visibility and the damping
scale in Secs.~\ref{sec:decoupling} and \ref{sec:damping}.

Before turning to the damping scale, it is worth saying why $\ell_A$
and $\ell_{\rm eq}$ are as close to one another as they are, $299.6$
against $204$. Both are $D_c$ divided by a comoving length, and the
two lengths, the sound horizon at decoupling and the comoving Hubble
radius at equality to which $\ell_{\rm eq}$ is tied through
$k_{\rm eq} = \sqrt{2}\,a_{\rm eq}H(z_{\rm eq})$, are themselves
comparable, $\ell_A/\ell_{\rm eq} =
(\pi/\sqrt{2})\,(a_{\rm eq}H_{\rm eq}r_s)^{-1} = 1.48$ against the
$1.47$ read directly off the two multipoles. They are comparable
because the two epochs themselves are, $1+z_{\rm eq} = 3407$ against
$1+z_{\rm dec} = 1079$. Each of these redshifts is an energy
divided by $k_BT_0$, so the present photon temperature cancels in
their ratio, and the two energies are the mean mass per photon $\mu$
of Eq.~\eqref{eq:mu} for equality and the hydrogen binding energy for
recombination (Sec.~\ref{sec:recombination}). The two epochs fall
close together, therefore, because $\mu c^2 = 3.7\,$eV is within a
factor of four of $B_1 = 13.6\,$eV. The matter content enters through
$\mu$ alone, to which the cold component contributes most of the
weight. Without it, equality would follow recombination rather than
precede it and the two multipoles would lie hundreds of multipoles
apart.

\subsection{The damping scale}
\label{sec:whydamping}

The two results above share a common origin. Each geometric scale is
the ratio of a length at last scattering to the distance to last
scattering, and both scale as $\lG^{-1}$ through $c/H$, so the
cancellation in the projected angle leaves only the few-percent drift
of $\ell_A$ discussed above.

The damping scale is different precisely because it is not a purely
geometric ratio. As the following sections show, $\rdamp$ is not built
from $c/H$ alone but from the Thomson opacity, and therefore from the
whole ionization history $\xe(z)$, which carries a $G$-dependence of
its own, as will be shown in Sec.~\ref{sec:recombination}. The
geometric cancellation that renders $\ell_{\rm eq}$ and $\ell_A$
nearly invariant breaks down significantly for $\ell_{\rm damp}$,
and that residual is exactly the effect that is the subject of this
paper.

More precisely, the main result of this paper is to show that the
damping scale responds to a small change of $G$ as
$\ell_{\rm damp} \propto \lG^{-0.30}$, the exponent being a logarithmic
derivative at $\lG = 1$, which is the regime any observational constraint
occupies. Its dependence on the choice of last-scattering cut-off is
quantified in Sec.~\ref{sec:silkintegral}, and the behaviour at large
departures from $\lG = 1$ in Sec.~\ref{sec:wide}. We establish this result with explicit
analytical expressions of the damping tail and check it against the
\CLASS{} outputs of our modified code. Two approximations must be
dropped to reach it. Identifying the photon mean free path with the
Hubble radius at decoupling, as \citet{Rich2015} does, leaves only a
logarithmic residual dependence of $\ell_{\rm damp}$ on
$\lambda_G$. Relaxing that identification but keeping the ionization
history in Saha equilibrium gives instead $\propto \lambda_G^{-1/2}$,
which overshoots. The two approximations therefore act in opposite
directions, and recovering the correct scaling requires dropping both,
that is accounting for the finite thickness of last scattering and for
the out-of-equilibrium kinetics of recombination. The remainder of the
paper is devoted to computing this response:
Sec.~\ref{sec:recombination} treats recombination as the extended
process it is, Sec.~\ref{sec:decoupling} the width of last scattering,
and Sec.~\ref{sec:damping} the damping scale itself.

\section{Recombination}
\label{sec:recombination}

\subsection{The Saha equilibrium contains no $G$}
\label{sec:saha}

The starting point is a structural observation. Written per hydrogen
nucleus, the Saha equation reads
\begin{equation}
\frac{\xe^2}{1-\xe} \;=\; \frac{1}{n_{\rm H}(z)}
\left(\frac{m_e k_B T}{2\pi\hbar^2}\right)^{3/2} e^{-B_1/k_B T}\,,
\label{eq:saha}
\end{equation}
with $B_1 = 13.6$~eV the ionization energy and $n_{\rm H}$ the
hydrogen number density. Every quantity on the right-hand side is
either atomic or thermodynamic, there is no $H$, and therefore no
$G$. An equilibrium ionization history is a universal function of
temperature and baryon density, identical in all the models of our
family.

The consequence is central to this paper: describing the ionization
history as an equilibrium function of temperature removes the
\emph{explicit} $G$-dependence of $\xe$. On its own, however, this
does not make the CMB insensitive to $G$, as we show in
Sec.~\ref{sec:damping}. The finite thickness of the shell plays a
different part, examined in Sec.~\ref{sec:decoupling}, since it is
what places decoupling away from the rate crossing
$\Gamma_{\rm T} = H$. For atomic constants such as $\alpha$ or $m_e$,
which appear explicitly in Eq.~\eqref{eq:saha}, the equilibrium
description retains the leading effect and the kinetics supplies a
correction. For $G$ the situation is inverted since the equilibrium
description retains nothing explicit, and the kinetic, finite-width
physics beyond equilibrium supplies the entire effect.

\subsection{Kinetic channels}
\label{sec:peeblesfactor}

Recombination is not in local equilibrium. Direct recombination to the
ground state is ineffective, because the emitted photon immediately
ionises a neighbouring atom, so the net formation of neutral hydrogen
must proceed through the $n=2$ level by one of two slow channels: the
forbidden two-photon decay $2s \rightarrow 1s$, with
$\Lambda_{2s \rightarrow 1s} \simeq 8.22\ {\rm s^{-1}}$, and the
$2p \rightarrow 1s$ Lyman-$\alpha$ transition, whose resonant photon
is reabsorbed almost immediately and escapes only by redshifting out
of the line. The effective three-level evolution
\citep{Peebles1968,ZKS1968}, refined into the modern multi-level
treatment by \citet{SeagerSasselovScott2000}, is
\begin{equation}
\frac{\dd \xe}{\dd t} \;=\; -\,C_{\rm P}
\left[\, \alpha_B(T)\, n_{\rm H}\, \xe^2
\;-\; \beta_B(T)\,(1-\xe)\, e^{-(B_1-B_2)/k_B T} \,\right],
\label{eq:peebles}
\end{equation}
where $B_2 = B_1/4 = 3.4$~eV is the binding energy of the $n=2$ level
and $B_1 - B_2 = 10.2$~eV the Lyman-$\alpha$ energy, so that the
exponential is the Boltzmann population of $n=2$ relative to the
ground state. Here $\alpha_B$ is the case-B recombination coefficient, a rate per
unit density with the dimensions of a volume rate,
${\rm m^3\,s^{-1}}$, and $\beta_B$ the corresponding photoionization
rate, a pure frequency, ${\rm s^{-1}}$. The
latter is not an independent quantity, detailed balance fixing it from
$\alpha_B$,
\begin{equation}
\beta_B(T) = \alpha_B(T)\left(\frac{m_ek_BT}{2\pi\hbar^2}\right)^{3/2}
e^{-B_2/k_BT}\,,
\label{eq:betaB}
\end{equation}
and with this definition the bracket of Eq.~\eqref{eq:peebles}
vanishes identically on the Saha solution of
Eq.~\eqref{eq:saha}. Finally $C_{\rm P}$ is the
Peebles factor, the fraction of atoms reaching $n=2$ that go on to the
ground state rather than being photoionized back to the continuum,
\begin{equation}
C_{\rm P} = \frac{\Lambda_{2s \rightarrow 1s} + \Lambda_\alpha}
{\Lambda_{2s \rightarrow 1s} + \Lambda_\alpha + \beta_B}\,,
\qquad
\Lambda_\alpha = \frac{8\pi H}{\lambda_\alpha^3\, n_{1s}}\,,
\label{eq:peeblesC}
\end{equation}
where $n_{1s}$ is the number density of hydrogen atoms in the $1s$
ground state and $\lambda_\alpha$ the Lyman-$\alpha$ wavelength. What
is worth noting here, and is less often stressed, is that the Sobolev
escape rate $\Lambda_\alpha$ is itself proportional to $H$. A faster
expansion drags Lyman-$\alpha$ photons out of resonance more
efficiently, so that the probability of escaping the bottleneck, and
hence $C_{\rm P}$ itself, depends on $G$. This dependence is not a
minor one. At $\lG = 1$ the Sobolev term supplies a fraction
$\Lambda_\alpha/(\Lambda_{2s\rightarrow1s}+\Lambda_\alpha) \simeq
0.36$ of the escape at $z_{\rm rec}$, and at that fixed redshift the
local sensitivity is $C_{\rm P} \propto \lG^{+0.45}$. This is a channel of its
own, distinct from the competition between the recombination rate and
the expansion rate that Eq.~\eqref{eq:efolds} makes explicit below,
and which Eq.~\eqref{eq:peeblesC} does not by itself display.

The departure of recombination from equilibrium is not a small effect
and must be taken into account. It is already apparent in the ionization
histories of Fig.~\ref{fig:xe}, where even the standard-$G$ curve lags
visibly behind the Saha prediction throughout the transition. The two
channels above are the reason, and the rest of this section quantifies
how their competition with the Hubble rate depends on $G$.

\subsection{Numerical ionization histories}
\label{sec:numxe}

\begin{figure}[t]
\centering
\includegraphics[width=\columnwidth]{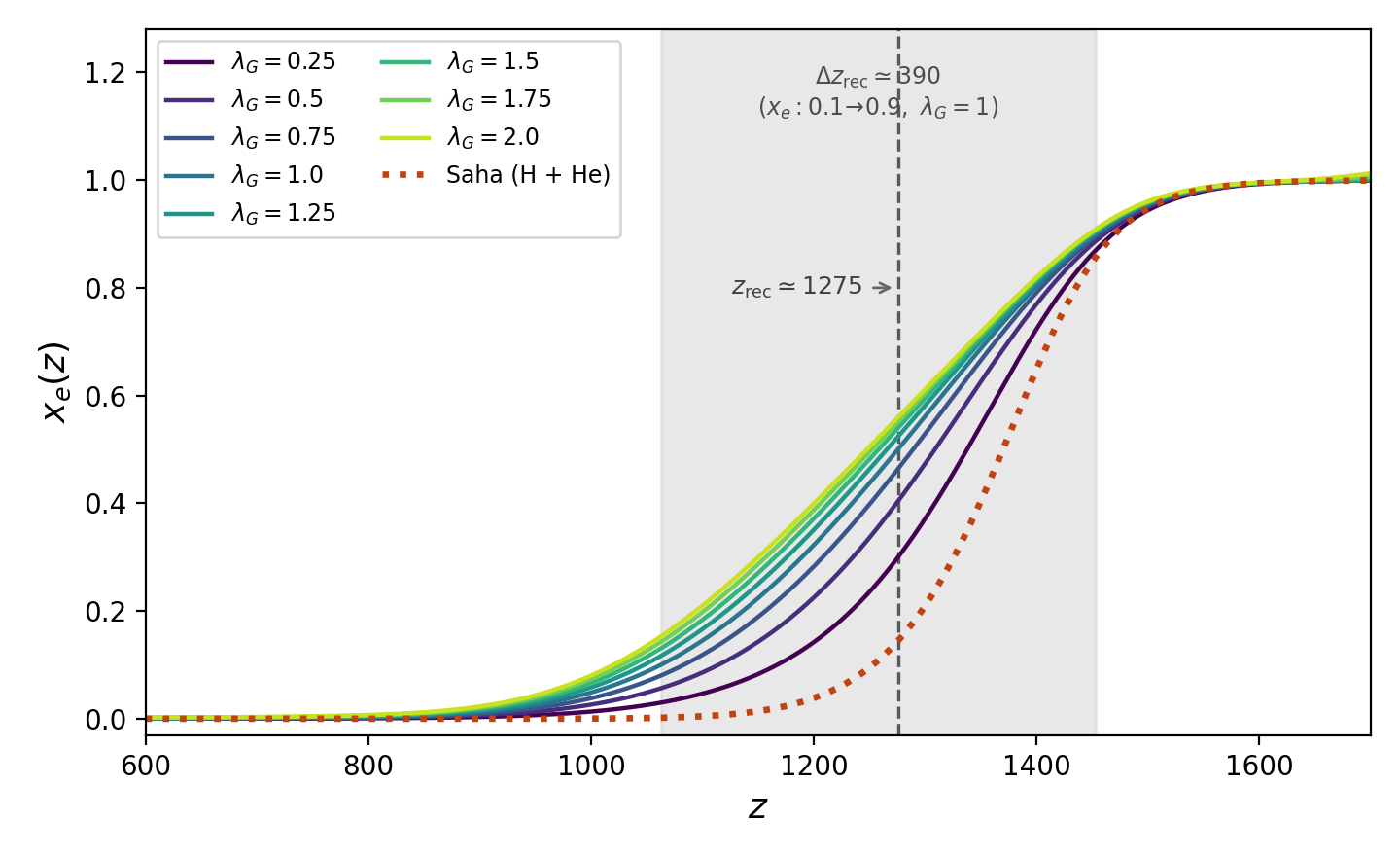}
\caption{Ionization histories $\xe(z)$ for $\lG \in [0.25,2]$ (solid,
  dark to light) with the Saha equilibrium superposed (dotted), the
  latter computed with the standard-$G$ cosmological parameters. The
  equilibrium curve is common to all models because
  Eq.~\eqref{eq:saha} contains no $G$. The shaded band marks the
  recombination width for $\lG = 1$,
  $\Delta z_{\rm rec} = z(\xe=0.9) - z(\xe=0.1) \simeq 390$, and the
  dashed line the midpoint $z_{\rm rec}$ where $\xe = 1/2$. The curves
  move towards the Saha curve as $\lambda_G$ decreases.}
\label{fig:xe}
\end{figure}

Figure~\ref{fig:xe} shows the ionization histories obtained by
integrating the Peebles equation~(\ref{eq:peebles}) for the eight
models, together with the Saha equilibrium. The equilibrium curve does
not depend on $G$ according to Eq.~\eqref{eq:saha}, so the entire
spread between the solid curves and the Saha curve is kinetic in
origin. The lag grows with $\lG$, and the transition becomes
progressively less abrupt. In the formal limit $\lG \rightarrow 0$ the
histories converge back onto the Saha curve, and the reason is
simple. Since $G$ enters only through the expansion rate, at fixed
redshift neither the temperature nor the baryon density depends on it,
so all atomic rates are identical from one model to the next and only
the clock changes. Saha equilibrium is precisely the state reached
when the reactions are infinitely fast compared with the expansion,
that is when detailed balance has time to settle at each
temperature. The relevant measure is the ratio $\Gamma_{\rm rec}/H$,
the number of recombination attempts per $e$-fold of expansion, which
scales as $\lG^{-1}$ at fixed composition and diverges as $G \to
0$. One might object that the Lyman-$\alpha$ escape rate is itself
proportional to $H$ and shuts off in the same limit. What keeps
recombination going is the two-photon channel $2s\rightarrow1s$, whose
rate is a fixed atomic constant. It bounds the Peebles factor of
Eq.~\eqref{eq:peeblesC} from below,
$C_{\rm P} \geq
\Lambda_{2s\rightarrow1s}/(\Lambda_{2s\rightarrow1s}+\beta_B)$, so
that the effective recombination rate stays finite while $H \to
0$. The equilibrium description is then recovered exactly and the
$G$-dependence disappears. In that limit $z_{\rm rec}$ becomes
analytic, being fixed by the Saha condition $\xe = 1/2$ alone, which
involves only $B_1$ and the baryon density and yields
$B_1/k_BT \simeq 42$. The present temperature $T_0$ then turns that
recombination temperature into a redshift, $z_{\rm rec} \rightarrow 1370$. It is
the nonzero expansion rate that drives recombination out of
equilibrium, and a larger $G$ does so more strongly. We make this rate
argument more quantitative in Sec.~\ref{sec:rate}.

It is convenient to summarise each history by two numbers. We define
the recombination redshift $z_{\rm rec}$ as the midpoint of the
transition, where the free-electron fraction has fallen to half its
plateau value, $\xe(z_{\rm rec}) = 1/2$, and the recombination width
$\Delta z_{\rm rec} = z(\xe = 0.9) - z(\xe = 0.1)$ as the redshift
interval over which the bulk of the transition takes place. Both are
marked on Fig.~\ref{fig:xe} for $\lG = 1$, and their dependence on $G$
is shown later in Fig.~\ref{fig:redshifts} ($z_{\rm rec}$) and
Fig.~\ref{fig:widths} ($\Delta z_{\rm rec}$): a larger $G$ moves
$z_{\rm rec}$ to lower redshift and widens $\Delta z_{\rm rec}$.

The steepness of the transition is a quantitative diagnostic that will
be used repeatedly below. Keeping the dominant Boltzmann factor of
Eq.~\eqref{eq:saha}, with $B_1/k_BT \simeq 40$ at recombination,
\begin{equation}
\frac{\dd \ln \xe^{\rm Saha}}{\dd \ln (1+z)}
\;\simeq\; +\frac{1}{2}\frac{B_1}{k_B T} \;\simeq\; +20\,,
\label{eq:saha_slope}
\end{equation}
whereas the computed histories using the full Peebles
model~(\ref{eq:peebles}) give $\simeq +6.4$ near $z_{\rm rec}$. The
out-of-equilibrium transition is markedly shallower, and it is
followed by a long tail in which $\xe$ remains far above the
equilibrium value (see Fig.~\ref{fig:xe}).

\subsection{Discussion based on recombination rate}
\label{sec:rate}

The physical origin of the two trends, later and smoother
recombination for larger $G$, follows from a rate argument that we now
make explicit. Much of the physics of the CMB can be understood in
terms of ratios of characteristic rates, in the spirit of
\citet{Zahn2003} and \citet{GreeneCyrRacine2023}. What determines the
outcome of a competing set of processes is how each rate compares with
the expansion rate that drives the system away from equilibrium. For
recombination the relevant comparison is between the recombination
rate $\Gamma_{\rm rec} \equiv C_{\rm P}\alpha_Bn_{\rm H}\xe$ and the
Hubble rate $H$. Using $\dd N = H \dd t$ with $N = \ln a$ the number
of $e$-folds, and dividing Eq.~\eqref{eq:peebles} by $H$, one obtains
\begin{equation}
\frac{\dd \xe}{\dd N} \;=\; -\,\frac{\Gamma_{\rm rec}}{H}\, \xe
\left[\, 1 - \frac{\beta_B (1-\xe) e^{-B_2/k_BT}}
{\alpha_B n_{\rm H} \xe^2} \right]\,.
\label{eq:efolds}
\end{equation}
The bracket measures the departure from equilibrium, it vanishes
identically when $\xe$ sits on the Saha curve (and meanwhile
$\Gamma_{\rm rec}/H\rightarrow\infty$), while it is of order unity
once the system lags behind. Hence the prefactor $\Gamma_{\rm rec}/H$
controls the number of recombination events per electron per $e$-fold
of expansion, that is, the number of attempts the chemistry gets to
correct a given departure before the expansion has moved the target
again. The $G$-dependence of this ratio is not a simple power of
$\lG$, even to a first approximation, as the spread of the histories in
Fig.~\ref{fig:xe} makes clear.  Three effects combine. The denominator
carries $H \propto \lG$, which would alone give
$\Gamma_{\rm rec}/H \propto \lG^{-1}$ at fixed composition. But the
numerator $\Gamma_{\rm rec}$ also responds to $G$, in three ways. The
hydrogen density carries $n_{\rm H} \propto (1-Y_{\rm He})$, and since
$Y_{\rm He} \propto \lG^{0.72}$ this contributes a further
$\lG^{-0.23}$, so that at fixed $\xe$ the ratio scales as
$\lG^{-1.23}$ rather than $\lG^{-1}$, a value we recover as
$\lG^{-1.20}$ from the computed histories. The free-electron fraction
$\xe$ itself then lags differently in each model, and the Sobolev
escape factor contained in $C_{\rm P}$ carries an explicitly
$G$-dependent term $\Lambda_\alpha \propto H$. The net behaviour is
therefore better read off the computed curves of Fig.~\ref{fig:xe}
than summarised by a single exponent. A larger $G$ lowers
$\Gamma_{\rm rec}/H$ at fixed redshift, gives recombination fewer
attempts per $e$-fold, and drives the ionization history further from
equilibrium. This is the root of every $G$-dependence that follows.

It is instructive to rewrite this ratio in terms of fundamental
constants, since doing so identifies the dimensionless combination
through which $G$ enters. Let us write
$n_{\rm H} = (1-Y_{\rm He})\,\eta_b n_\gamma$, with
$\eta_b \equiv n_b/n_\gamma$ the baryon-to-photon number ratio and
$n_\gamma = (2\zeta(3)/\pi^2)(k_BT/\hbar c)^3$ the blackbody photon
density.  The expansion rate follows from the Friedmann equation with
matter and radiation,
$H^2 = (8\pi G/3)\,\mu n_\gamma\,[1+(1+z)/(1+z_{\rm eq})]$, where
$\mu$ is defined by Eq.~\eqref{eq:mu} and can equivalently be written
$\mu = m_{\rm cdm}\eta_{\rm cdm} + m_p\eta_b$. Here $m_{\rm cdm}$ is the mass of the
cold dark matter particle and $\eta_{\rm cdm} \equiv n_{\rm cdm}/n_\gamma$ is
the ratio of the number density of these particles to that of photons,
the dark matter analogue of $\eta_b$, so that
$\rho_m = m_{\rm cdm} n_{\rm cdm} + m_p n_b$ is a mass density. For the case-B
coefficient two standard expressions are in use, the hydrogenic result
of \citet{Seaton1959}, obtained from the Kramers cross-sections, and
the numerical fit of \citet{Pequignot1991} adopted by
\texttt{RECFAST}. We work with the first throughout,
$\alpha_B = D\,\Lambda^{1/2}\phi_B(\Lambda)$ with
$D = (2^6/3)(\pi/3)^{1/2}\,\alpha^2\hbar^2/(m_e^2c)$,
$\Lambda = B_1/k_BT$ with $B_1$ the ionization energy of hydrogen
introduced in Eq.~\eqref{eq:saha} and
$\phi_B(\Lambda) =
\sum_{n\geq2}(\Lambda/n^3)\,e^{\Lambda/n^2}E_1(\Lambda/n^2)$.  The
function $\phi_B$ varies only logarithmically and is close to $1.8$ at
recombination. Two things recommend it. It is the only one of the two
whose dependence on temperature can be written down in closed form,
which is what will make Eqs.~\eqref{eq:rdamp} and \eqref{eq:Qcrit}
below possible at all, and it is the closer of the two to
\texttt{HyRec} \citep{AliHaimoud2011}, the multilevel calculation that
actually produces our histories. Deep in the ionization tail the
effective coefficient can be extracted from the computed $\xe(z)$ with
no fit at all, as Sec.~\ref{sec:discussion} sets out, and doing so
puts \texttt{HyRec} $6\%$ above the Seaton expression and $12\%$ above
the \citet{Pequignot1991} fit, with little drift from
$z = 600$ down to $z = 100$. The two case-B expressions themselves
differ by a near-constant factor of $1.051$ over that whole range, the
Gaunt factor that corrects the semiclassical Kramers cross-sections
for quantum effects \citep{KarzasLatter1961,RybickiLightman1979}, and
their local logarithmic slopes differ by less than $0.002$, so no
exponent in this paper depends on the choice. Measuring the
temperature in atomic units, $\theta \equiv k_BT/(\alpha^2 m_e c^2)$,
so that $\Lambda = 1/(2\theta)$, one finds
\begin{align}
\frac{\Gamma_{\rm rec}}{H} = \;&\frac{16\sqrt{2\zeta(3)}}{3\pi}\,C_{\rm P}\,\xe\,
(1-Y_{\rm He})\,\eta_b \nonumber\\
&\times\alpha^5\,\theta\,\phi_B(\theta)
\left(1+\frac{1+z}{1+z_{\rm eq}}\right)^{-1/2}\alpha_G^{-1/2}\,,
\label{eq:alphaG}
\end{align}
with
\begin{equation}
\alpha_G \equiv \frac{G m_e \mu}{\hbar c}\,.
\end{equation}
At recombination $\theta \simeq 0.011$. The electron mass in
$\alpha_G$ is not a property of the expansion, which involves $G$ only
through the product $G\mu$. It appears because $\alpha_G$ measures that
product in atomic units, $G\mu = \aG\,c^2\hbar/m_ec$, so that the
electron mass enters from the atomic side of the ratio and not from the
Hubble rate. The gravitational constant
enters Eq.~\eqref{eq:alphaG} only through $\alpha_G$, explicitly
through the factor $\aG^{-1/2}$ and through the $\aG$ dependence of
$C_{\rm P}$. For $\lG = 1$ one has $\alpha_G \simeq 1.3\times10^{-50}$, and
Eq.~\eqref{eq:alphaG} gives
$\Gamma_{\rm rec}/H \simeq 3.8\times10^3\,C_{\rm P}\,\xe$ at
$z = 1300$. Dividing the numerator
and the denominator of Eq.~\eqref{eq:peeblesC} by $\beta_B$ makes
$C_{\rm P}$ a function of the two ratios
$\Lambda_{2s\rightarrow1s}/\beta_B$ and $\Lambda_\alpha/\beta_B$
alone. The first is purely atomic, and the second carries the whole
$G$-dependence, since
$\Lambda_\alpha/\beta_B \propto H/(\lambda_\alpha^3 n_{1s}\beta_B)$
and $H/n_{1s}$ reduces to $\alpha_G^{+1/2}$ times atomic quantities by
the same algebra that put $\alpha_G^{-1/2}$ in
Eq.~\eqref{eq:alphaG}. The two occurrences differ only in the sign of
the exponent. This contrasts with the equilibrium treatment of
\citet{Rich2015}, where a combination of the same nature,
$\alpha^7/(Gm_{\rm cdm} m_e)$, appears only inside a logarithm, because in
equilibrium the decoupling temperature depends on the expansion rate
only logarithmically.

It is worth writing these atomic rates out in fundamental constants, since it
is their independence of $G$ that isolates $\alpha_G$ in
Eq.~\eqref{eq:alphaG}. The two-photon rate is a hydrogenic quantity varying as
$Z^6$, and each emission vertex carries one power of $\alpha$ beyond the
atomic structure, so the two together give
\begin{equation}
  \Lambda_{2s\rightarrow1s} = C_{2\gamma}\,\alpha^6\,
  \frac{\alpha^2m_ec^2}{\hbar}\,,
  \qquad C_{2\gamma} = 1.317\times10^{-3}\,.
  \label{eq:lam2g}
\end{equation}
The exponents follow from that counting, and $C_{2\gamma}$ is a
dimensionless numerical factor that has no closed form. Indeed, the
second-order matrix element involves a sum over
the complete intermediate spectrum, bound states and continuum alike,
so the two-photon rate is known only numerically. The value above is
the one that reproduces
$\Lambda_{2s\rightarrow1s} = 8.22\ {\rm s^{-1}}$
\citep{Goldman1989,Labzowsky2005,ChlubaSunyaev2008}. This is the
one coefficient in this section that is not analytic. For the other
two rates, the \citet{Seaton1959} form and detailed balance give
everything explicitly. Writing $\lambdabar_e = \hbar/m_ec$ for the
reduced Compton wavelength of the electron, and recalling that
$\theta = k_BT/(\alpha^2m_ec^2)$ is the temperature in atomic units,
\begin{align}
  \alpha_B &= \frac{2^5}{3}\sqrt{\frac{2\pi}{3}}\;
              \alpha^2\,\lambdabar_e^2\,c\;
              \theta^{-1/2}\,\phi_B(\theta)\,,
  \label{eq:alphaB_exact}\\[4pt]
  \beta_B  &= \frac{2^4}{3\sqrt{3}\,\pi}\;
              \alpha^5\,\theta\,\phi_B(\theta)\;
              \frac{c}{\lambdabar_e}\;e^{-1/8\theta}\,,
  \label{eq:betaB_exact}
\end{align}
the numerical coefficients being $15.437$ and $0.980$. The only
dimensioned quantities left are $\lambdabar_e^2c$, a volume rate as a
recombination coefficient must be, and $c/\lambdabar_e$, a pure
frequency as a photoionization rate must be, the electron mass
entering nowhere else than through $\lambdabar_e$ and, in $\beta_B$,
through the argument $B_2/k_BT = 1/8\theta$ of the
exponential. Note that
$\beta_B$ carries the same atomic combination $\alpha^5\theta\phi_B$
that appears in Eq.~\eqref{eq:alphaG}, which is simply because both are
$\alpha_B$ multiplied by a density.

The Sobolev rate is the one that carries the expansion, and it can be
made fully explicit as well. The atomic constants cancel between
$\lambda_\alpha^3$ and $n_\gamma$, since
$\lambda_\alpha = (16\pi/3)\lambdabar_e/\alpha^2$ and
$n_\gamma = (2\zeta(3)/\pi^2)(\alpha^2\theta/\lambdabar_e)^3$ leave
$\lambda_\alpha^3n_\gamma = (2^{13}\pi\zeta(3)/27)\,\theta^3$, so that
with $n_{1s} = (1-\xe)(1-Y_{\rm He})\eta_bn_\gamma$,
\begin{align}
  \Lambda_\alpha &= \frac{27}{2^8\sqrt{3\pi\zeta(3)}}\;
  \frac{\alpha^3\,\theta^{-3/2}\,\aG^{1/2}}
       {(1-\xe)\,(1-Y_{\rm He})\,\eta_b}\;
  \frac{c}{\lambdabar_e}\nonumber\\
  &\quad\times\left(1+\frac{1+z}{1+z_{\rm eq}}\right)^{1/2}\,,
  \label{eq:lamalpha_exact}
\end{align}
with coefficient $0.0313$, the factor $1-\xe$ being the fraction of
hydrogen left in the ground state. This is the explicit form of the
statement made above, that the only $G$ in the Peebles factor is an
$\aG^{1/2}$. At $z_{\rm rec}$ it gives
$\Lambda_\alpha = 4.6\ {\rm s^{-1}}$, hence the escape fraction
$0.36$ quoted in Sec.~\ref{sec:peeblesfactor}. Equation~\eqref{eq:alphaG} is
then no more than the statement that $\alpha_B$, combined with
$n_\gamma \propto (k_BT)^3$ and $H \propto (G\mu n_\gamma)^{1/2}$,
leaves $\alpha^5\theta\,\phi_B\,\aG^{-1/2}$, with no atomic rate
contributing a gravitational parameter of its own.

Equations \eqref{eq:lam2g} to \eqref{eq:lamalpha_exact} are in fact
enough to write the Peebles factor itself in terms of dimensionless
quantities alone, fundamental constants, cosmological parameters and
the temperature $\theta$.
Dividing Eq.~\eqref{eq:peeblesC} through by $\beta_B$ as above gives
\begin{equation}
  C_{\rm P} = \frac{\mathcal{X}}{1+\mathcal{X}}
  \qquad\text{with}\qquad
  \mathcal{X} = \frac{\Lambda_{2s\rightarrow1s}+\Lambda_\alpha}{\beta_B}\,,
  \label{eq:CP_exact}
\end{equation}
and the two ratios follow in closed form,
\begin{align}
  \frac{\Lambda_{2s\rightarrow1s}}{\beta_B} &=
     \frac{3\sqrt{3}\,\pi}{2^4}\,C_{2\gamma}\;
     \frac{\alpha^3}{\theta\,\phi_B}\;e^{1/8\theta}\,,
  \label{eq:CP_2g}\\[4pt]
  \frac{\Lambda_\alpha}{\beta_B} &=
     \frac{81}{2^{12}}\sqrt{\frac{\pi}{\zeta(3)}}\;
     \frac{\aG^{1/2}}{\alpha^{2}\,\theta^{5/2}\,\phi_B}\;
     \frac{e^{1/8\theta}}{(1-\xe)(1-Y_{\rm He})\,\eta_b}\nonumber\\
  &\quad\times\left(1+\frac{1+z}{1+z_{\rm eq}}\right)^{1/2}\,,
  \label{eq:CP_alpha}
\end{align}
the numerical coefficients being $1.34\times10^{-3}$ and $0.0320$. The
two channels are of a different nature, and the contrast is the whole
point. The two-photon term is purely atomic, carrying $\alpha^3$ and no
$\aG$ at all, while the Sobolev term carries $\aG^{1/2}$ together with
the matter content through $\eta_b$, $Y_{\rm He}$ and $\xe$. Both are
lifted by the same large Boltzmann factor $e^{1/8\theta}$, worth
$8.5\times10^4$ at $z_{\rm rec}$, which is what rescues $C_{\rm P}$
from the smallness of $\alpha^3$. There the two terms are
$2.23\times10^{-3}$ and $1.26\times10^{-3}$, and the second over their
sum is the escape fraction $0.36$ of Sec.~\ref{sec:peeblesfactor}.

\begin{figure}[t]
\centering
\includegraphics[width=\columnwidth]{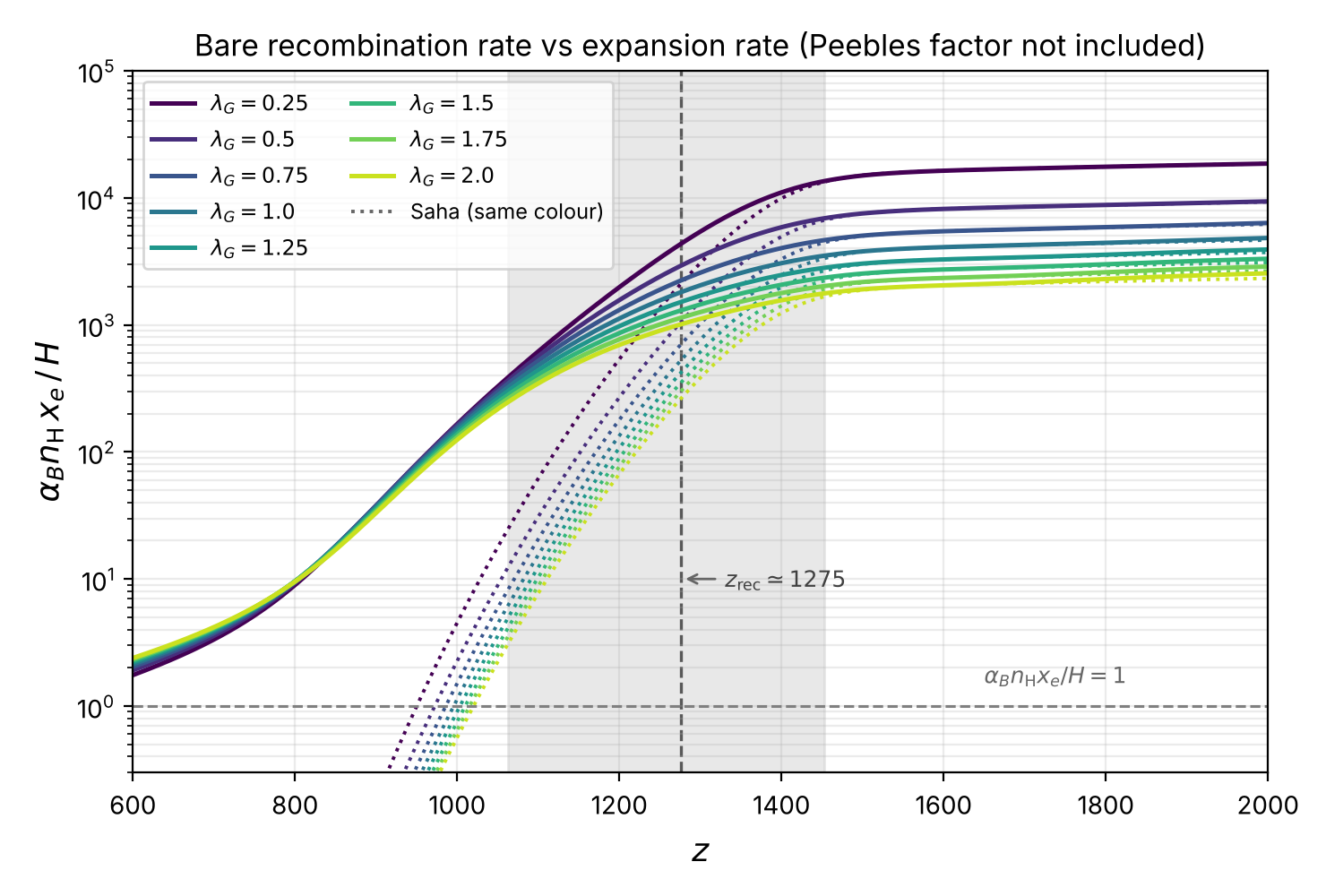}
\caption{Ratio of the \emph{bare} recombination rate to the expansion
  rate, $\alpha_B n_{\rm H}\xe/H$, for the eight models (solid, dark
  to light) and for the associated Saha ionization history
  (dotted). The Peebles factor is deliberately left out here, so that
  this ratio is larger than the $\Gamma_{\rm rec}/H$ of
  Eq.~\eqref{eq:alphaG} by $1/C_{\rm P}$, that is by some two and a
  half orders of magnitude at recombination. Since $\xe$ in Saha
  equilibrium is universal, the dotted curves separate only through
  $H \propto \lG$, so that their spread measures the purely geometric
  effect, whereas the much wider spread of the solid curves measures
  what the kinetics adds on top of it. One sees that the kinetic
  effects delay the recombination epoch considerably. Interestingly,
  the ratio is far above unity throughout recombination and
  crosses one only deep in the residual tail near $z \simeq 405$ (for
  $\lambda_G=1$), so that it does not by itself mark the recombination
  epoch. Nevertheless, its evolution with $\lambda_G$ has a physical
  interpretation: a larger $G$ lowers the ratio at fixed redshift
  during recombination, as $\lG^{-1}$ at fixed composition, which is
  what delays and broadens the transition.}
\label{fig:gammarec}
\end{figure}

We can now understand why recombination is slower than one might
expect. The bare atomic rate
exceeds the expansion rate by a wide margin,
$\alpha_B n_{\rm H}\xe/H \simeq 1.9\times10^3$ at $z_{\rm rec}$
(Fig.~\ref{fig:gammarec}, which shows this bare ratio and not
$\Gamma_{\rm rec}/H$). The Peebles factor, however, is small,
$C_{\rm P} \simeq 3.5\times10^{-3}$ there, since the $n=2$
photoionization rate $\beta_B$ still dwarfs the two escape
channels\footnote{The first two of these numbers depend on the case-B
  expression adopted, the \citet{Pequignot1991} fit lowering the first
  to $1.8\times10^3$ and raising the second to $3.7\times10^{-3}$,
  while their product is unchanged, for the reason given in
  Sec.~\ref{sec:wide}.}. The net rate therefore exceeds the expansion
rate by less than an order of magnitude,
$\Gamma_{\rm rec}/H \simeq 6.6$. It is
this bottleneck, and not the atomic physics itself, that keeps the
plasma from tracking equilibrium.

Although the ratio $\Gamma_{\rm rec}/H$ is what drives recombination,
as the discussion above makes clear, the epoch of recombination is not
obtained from the naive condition $\Gamma_{\rm rec}/H = 1$. The ratio
only falls to unity far down the residual tail, at $z \simeq 405$ for
$\lG = 1$, where the gas is already neutral to better than $99.9\%$,
its residual ionization fraction being $\xe \simeq 6\times10^{-4}$ (we
return to this in Sec.~\ref{sec:residual}). This illustrates a point
worth stating once, since it recurs several times in this paper. Rate
ratios such as $\Gamma_{\rm rec}/H$ govern which way the physics
moves, but their absolute value does not by itself locate the
observable events, which are set by the full history rather than by
the crossing of any single threshold. The recombination epoch is
instead defined observationally, following standard practice, either
by the redshift at which the ionization fraction reaches a reference
value, $z_{\rm rec}$ where $\xe = 1/2$, or equivalently by the point
of steepest fall of $\xe(z)$, and its width by the interval
$\Delta z_{\rm rec} = z(\xe = 0.9) - z(\xe = 0.1)$
\citep{Kaplinghat2000}. For $\lG = 1$ these give
$z_{\rm rec} \simeq 1275$ and $\Delta z_{\rm rec} \simeq 390$, the
shaded band in Fig.~\ref{fig:xe}. This is the characterisation used
throughout. The closely related width of the visibility function,
which defines the last-scattering surface proper, is taken up in the
next section.

\section{Decoupling and the visibility function}
\label{sec:decoupling}

Decoupling, like recombination, is an extended process. Following the
same logic as in the previous section, where we started from the full
ionization history $\xe(z)$ and only then interpreted it through the
rate ratio $\Gamma_{\rm rec}/H$, we begin this section with the
integral quantity that defines the last-scattering surface physically,
namely the visibility function.  We then turn to an interpretation
based on the local rate ratio $\Gamma_{\rm T}/H$, where $\Gamma_T$ is
the Thomson scattering rate. This local rate ratio, as we shall see,
is not straightforward to interpret, yet it is what exposes the evolution of
decoupling with $G$, which is the objective of this paper. We finally
close the section by collecting the characteristic redshifts of the
whole recombination--decoupling sequence and their dependence on $G$.

\subsection{The visibility function}
\label{section:visibility}

A description of the extended shell of recombination that does not
rely on any single rate crossing is captured by the visibility
function $g(z)$, which physically gives the probability that a photon
last scatters at a given redshift $z$. It is built from the optical
depth $\tau(z)$, the mean number of Thomson scatterings a photon
undergoes in travelling from redshift $z$ to the observer,
\begin{equation}
\tau(z) = \int_0^z \frac{c\sigma_T n_e}{(1+z')H}\,\dd z'\,,
\label{eq:tau}
\end{equation}
where we used
$\dd\tau = c\sigma_T n_e\,\dd t = -\,c\sigma_T n_e\,\dd z/[(1+z)H]$.

\begin{figure*}[ht]
\centering
\includegraphics[width=\textwidth]{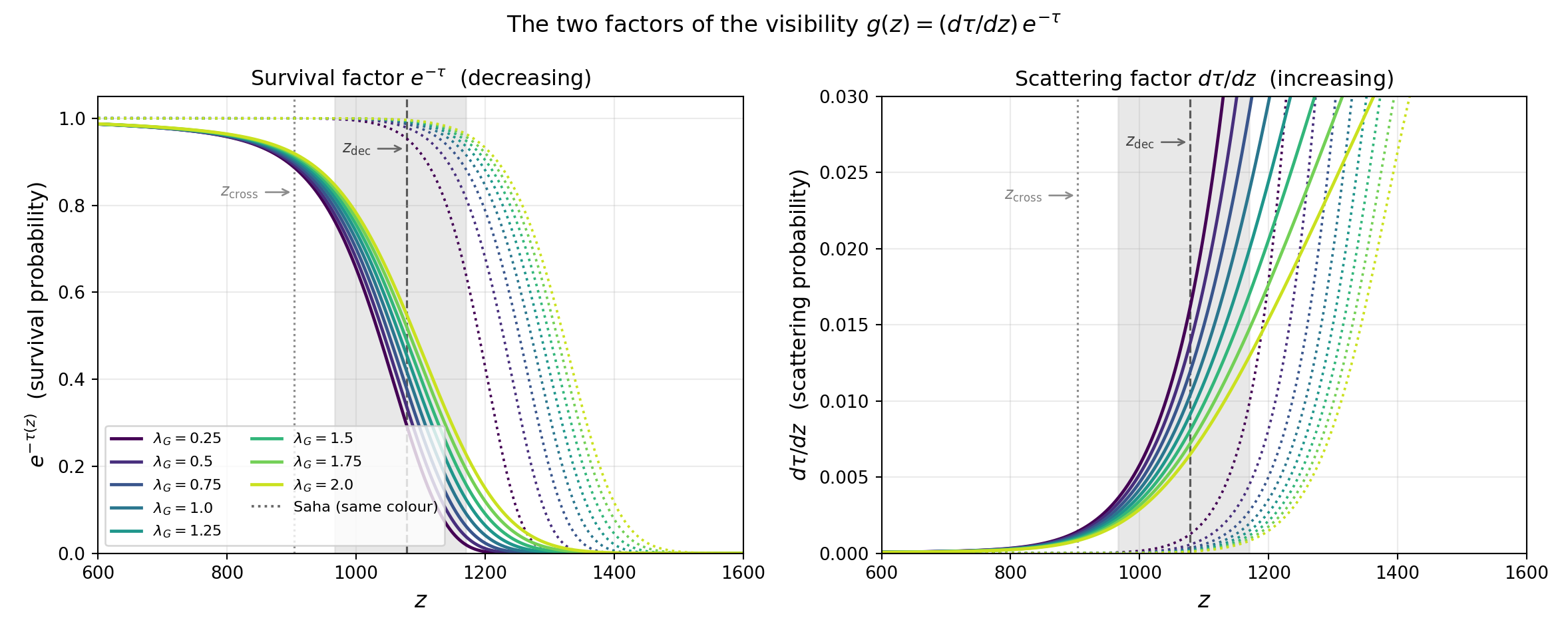}
\caption{The two competing factors of the visibility
  $g(z) = (\dd\tau/\dd z)\,e^{-\tau}$, for the eight models (dotted:
  the corresponding Saha histories). \emph{Left}: the survival factor
  $e^{-\tau}$ decreases with $z$. \emph{Right}: the scattering factor
  $\dd\tau/\dd z \propto \Gamma_{\rm T}/H$ increases with $z$. Their
  product peaks at $z_{\rm dec}$ (dashed line, shaded band), well
  above $z_{\rm cross}$ where $\Gamma_{\rm T} = H$ (dotted vertical
  line). Note that $\tau(z_{\rm dec}) \simeq 0.93$, of order unity,
  which is the coincidence discussed in the text.}
\label{fig:gzfactors}
\end{figure*}

\begin{figure}[h]
\centering
\includegraphics[width=\columnwidth]{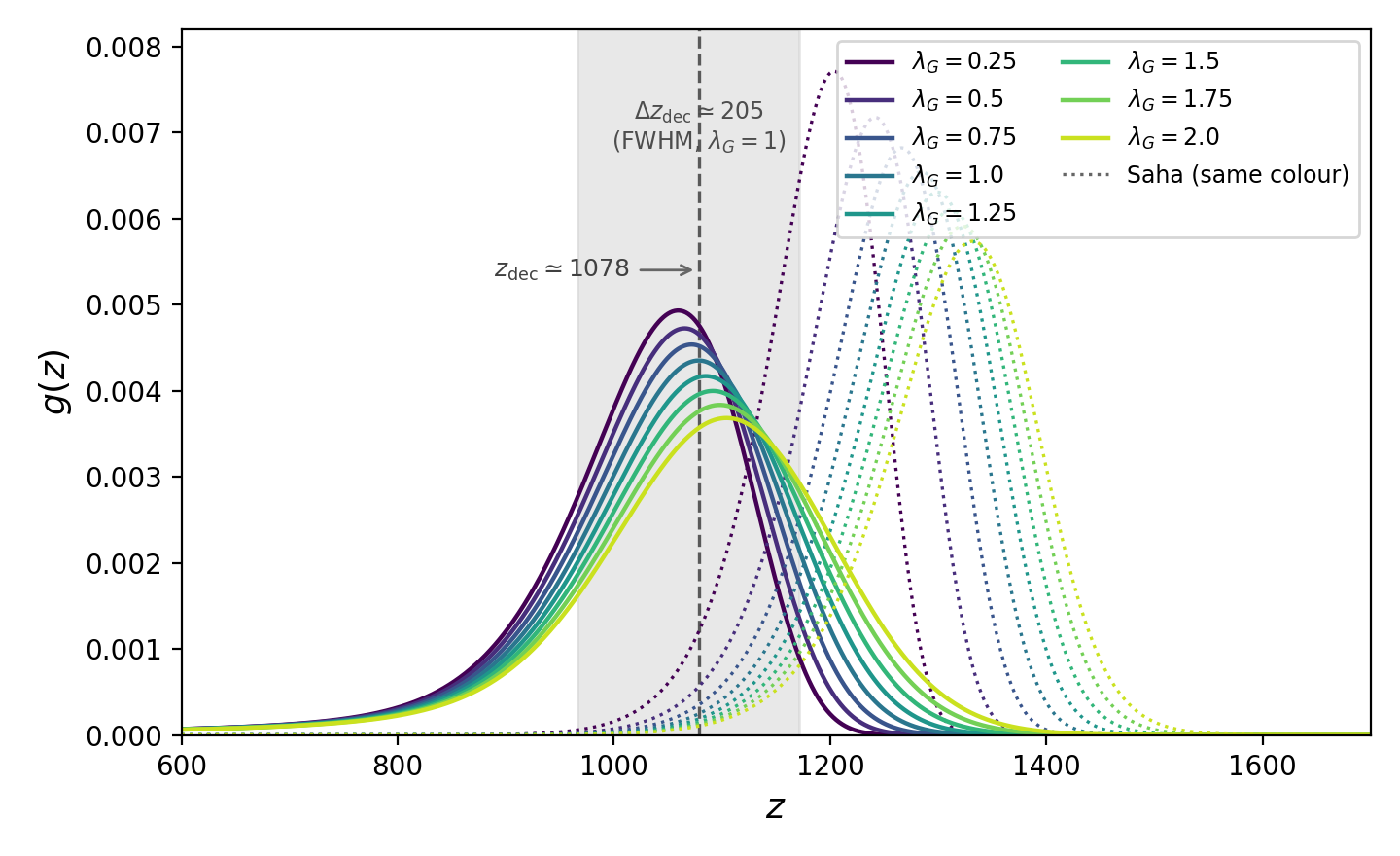}
\caption{Visibility function $g(z)$, normalised so that
  $\int g(z)\,\dd z = 1$. Solid curves are the kinetic (real)
  ionization histories; dotted curves of the same colour are the
  corresponding Saha visibilities, obtained by replacing $\xe(z)$ with
  its equilibrium value at the same $H$. Increasing $\lG$ moves last
  scattering to higher redshift and broadens it, and the shaded band
  marks the FWHM for $\lG = 1$, $\Delta z_{\rm dec} \simeq 204$. For
  every model the Saha visibility is markedly narrower, taller, and
  shifted to higher redshift than its kinetic counterpart (the peak
  separation grows from $\simeq 145$ at $\lG = 0.25$ to $\simeq 230$
  at $\lG = 2$), showing that equilibrium sharpens the last-scattering
  surface but does not reduce its thickness to zero. We stress that
  these Saha curves are illustrative references and \emph{not} the
  $G\to0$ limit of the family: each combines the Saha scattering rate
  with a finite Hubble rate, whereas the true $G\to0$ limit also sends
  $H\to0$, collapsing the shell to zero thickness and $g$ to a Dirac
  spike.}
\label{fig:g}
\end{figure}

The visibility function is then defined as
\citep{Sunyaev1970,Peebles1968}
\begin{equation}
g(z) \;=\; \frac{\dd\tau}{\dd z}\, e^{-\tau(z)}\,,
\label{eq:visibility}
\end{equation}
and gives the probability that a photon last scattered at redshift
$z$. Its peak defines the last scattering \emph{surface}, which is
given by $z_{\rm dec}\simeq1078$ for $\lambda_G=1$. Its FWHM width
defines the thickness of the last scattering shell, and is given by
$\Delta z_{\rm dec}\simeq204$ for $\lambda_G=1$ (shaded band in
Fig.~\ref{fig:g}). It is this width, rather than the recombination
width $\Delta z_{\rm rec}$ of Fig.~\ref{fig:xe}, that enters the
diffusion damping scale of Sec.~\ref{sec:damping}.
The form of Eq.~\eqref{eq:visibility} has a simple probabilistic
meaning. The factor $e^{-\tau(z)}$ is the probability that a photon
travels from redshift $z$ to the observer without scattering again,
while the factor $\dd\tau = (\dd\tau/\dd z)\,\dd z$ is the probability
that it \emph{does} scatter once within $[z, z+\dd z]$. Both factors
are discussed later and represented in Fig.~\ref{fig:gzfactors}. Their
product is therefore the probability that a given scattering at $z$ is
the photon's \emph{last} one, and $g(z)$ so defined is correctly
normalised,
$\int_0^\infty g\,\dd z = \int_0^\infty e^{-\tau}\,\dd\tau = 1$, since
every photon scatters for the last time exactly once.

Figure~\ref{fig:g} shows the visibility $g(z)$ for the eight values of
$\lambda_G$. Increasing $G$ moves the peak to higher redshift, from
$z \simeq 1059$ at $\lG = 0.25$ to $1104$ at $\lG = 2$, and broadens
it, the full width at half maximum growing from $\Delta z \simeq 180$
to $242$. Both trends have the same origin as the recombination trends
of Sec.~\ref{sec:recombination}. A larger $G$ makes the ionization
history both later and shallower, which moves the last-scattering
shell to higher redshift and broadens it. We make this explicit
through the local rate ratio in Sec.~\ref{sec:localrates}.

A first point to note in Fig.~\ref{fig:g} is that the kinetic and Saha
visibilities are clearly \emph{distinct}, each dotted curve being well
separated from the solid curve of the same colour. This is a direct
signature of the out-of-equilibrium nature of recombination. The
recombination bottleneck, that is to say the fact that direct captures
to the ground state produce ionizing photons and that the net
formation of neutral hydrogen is throttled by the slow two-photon
decay and Lyman-$\alpha$ escape, keeps the free-electron fraction well
above its Saha value throughout the tail. This extra scattering caused
by the kinetic freeze-out does shift the peak and thicken the surface,
as can be seen in Fig.~\ref{fig:g}. Far from being a numerical
artefact, this separation is the very quantity the paper exploits. It
is the residual, kinetically-frozen ionization that carries the
dependence on $G$ and shifts the decoupling redshift $z_{\rm dec}$.

That said, the separation of the real visibility from the Saha curves
does shrink as $\lG$ decreases (the peak separation goes from
$\simeq 230$ at $\lG = 2$ to $\simeq 145$ at $\lG = 0.25$), and this
trend has a well-defined endpoint: in the limit $G\to0$ the two
coincide. Both are built from the same Hubble rate and differ only
through $\xe$, and as $\lG\to0$ the expansion freezes relative to the
chemistry, driving the kinetic history to equilibrium,
$\xe \to \xe^{\rm Saha}$, so the two visibilities collapse onto the
\emph{same} Dirac spike. The $G\to0$ last-scattering surface is
therefore unique, independent of whether one starts from the kinetic
or the Saha ionization history. On the range $\lG \in [0.25, 2]$ shown
here, however, one sees only the onset of this convergence, the
asymptotic regime lying well below $\lG = 0.25$. The implication runs
one way only, however. The limit $G\to0$ does collapse the shell to a
Dirac spike, but imposing a Dirac visibility by hand does not place
one in that limit, and it does not by itself remove the
$G$-dependence. As we shall see in Sec.~\ref{sec:silkintegral}, the
damping length is evaluated at the visibility peak, which amounts
precisely to treating $g$ as a Dirac spike, and the resulting $\ldamp$
still carries the full scaling with $G$. What the finite width of the
shell does control is the shape of the damping envelope, and we return
to it in Sec.~\ref{sec:tt}.

\subsection{Local rates and the length interpretation}
\label{sec:localrates}

The rate at which a photon scatters off free electrons is given
  by the Thomson rate $\Gamma_{\rm T} =
c\,n_e\sigma_T$. The ratio $\Gamma_{\rm T}/H$ can also be read as
a ratio of characteristic length scales, since
\begin{equation}
\frac{\Gamma_{\rm T}}{H} \;=\; \frac{n_e \sigma_T c}{H}
\;=\; \frac{c/(aH)}{1/(a\,n_e\sigma_T)} \;=\; \frac{R_H}{\lmfp}\,,
\label{eq:identity}
\end{equation}
is therefore also the ratio of the comoving Hubble radius
$R_H \equiv c/(aH)$ to the comoving photon mean free path
$\lmfp = 1/(a\,n_e\sigma_T)$.

\begin{figure}[t]
\centering
\includegraphics[width=\columnwidth]{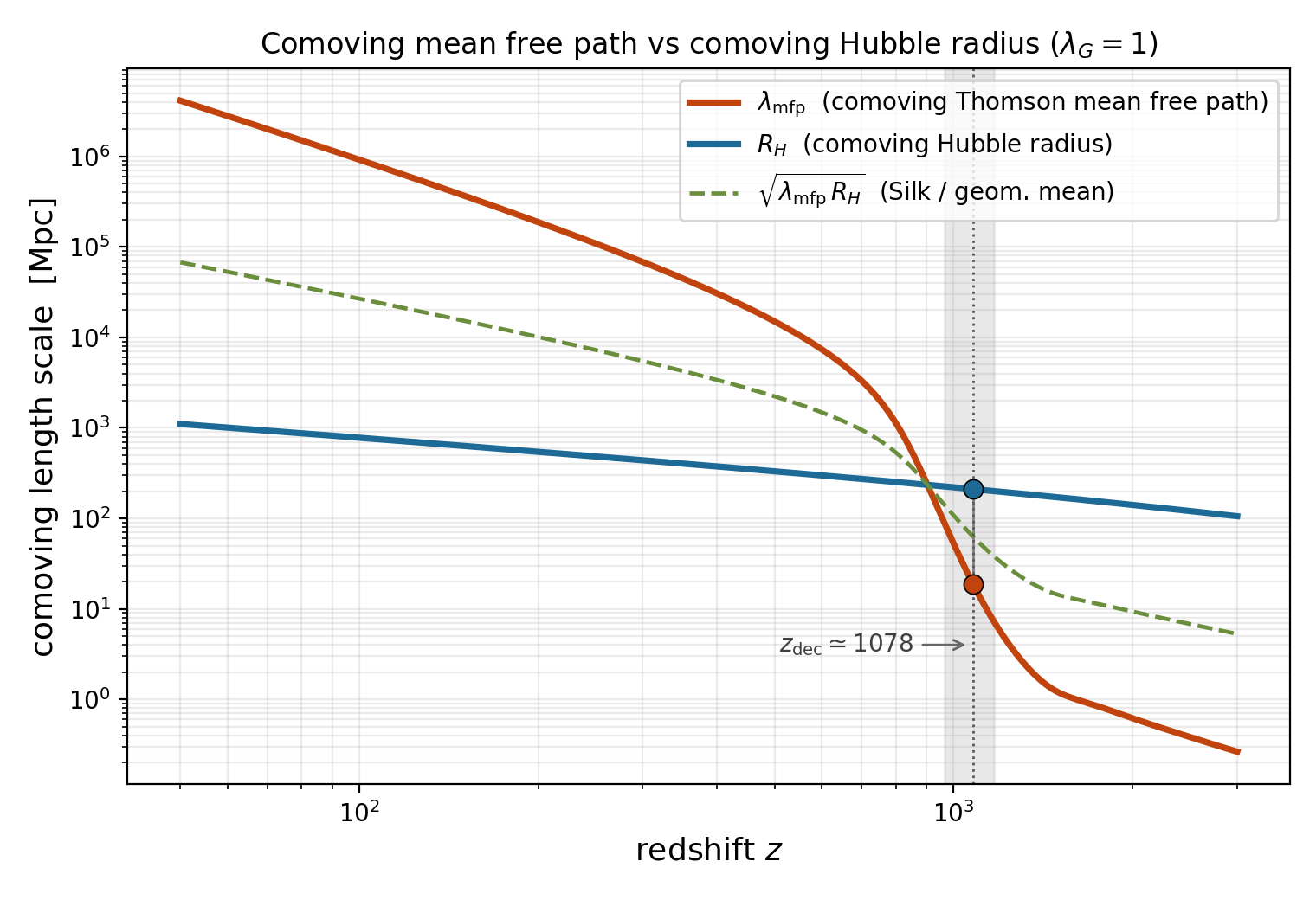}
\caption{Comoving Thomson mean free path $\lmfp$ and comoving Hubble
  radius $R_H$ for $\lG = 1$. The two are equal at
  $z_{\rm cross} \simeq 904$, but cross at very different
  slopes. Their logarithmic derivatives there are $-14.7$ and $-0.6$
  respectively, so the equality holds at that redshift alone. The
  shaded band marks the decoupling width $\Delta z_{\rm dec}$. At the
  visibility peak $z_{\rm dec} \simeq 1078$ the two lengths already
  differ by an order of magnitude. }
\label{fig:mfp}
\end{figure}

The difficulty with this ratio is that it is a poor description of the
\emph{extended} process of decoupling. Indeed, one might expect
  decoupling to occur simply where this rate falls below the expansion
  rate, i.e.\ $\Gamma_{\rm T} = H$. But this is not the case because
the ratio varies enormously across the last-scattering shell. Its logarithmic
slope is
\begin{equation}
\frac{\dd \ln (\Gamma_{\rm T}/H)}{\dd \ln(1+z)}
\;=\; \frac{\dd \ln \xe}{\dd\ln(1+z)} \;+\; \frac{3}{2}\,,
\label{eq:ratio_slope}
\end{equation}
in matter domination. The term $3/2$ is the universal contribution
stemming from $n_{\rm H} \propto (1+z)^3$ so that
$H \propto (1+z)^{3/2}$, while the large first term is the slope of
the ionization history, roughly equal to $+10.8$ near decoupling. This
is steeper than the $+6.4$ quoted near $z_{\rm rec}$ in
Sec.~\ref{sec:numxe}, simply because $\xe(z)$ keeps steepening between
$z_{\rm rec}$ and $z_{\rm dec}$. It is positive because $\xe$ grows
with redshift, and it dominates the universal $3/2$ by nearly an order
of magnitude. The ratio therefore climbs steeply through unity: it is
$\simeq 0.08$ at $z = 700$ and $\simeq 110$ at $z = 1400$, crossing
unity steeply at the redshift $z_{\rm cross}\simeq904$ for
$\lambda_G=1$ (Fig.~\ref{fig:mfp}). Equivalently, $\lmfp$ and $R_H$
are equal at $z_{\rm cross}$ but their derivatives there are very
different, the comoving logarithmic slopes there being $-14.7$ for
$\lmfp$ against $-0.6$ for $R_H$. The equality of the two lengths is
therefore not a robust feature of the shell but an accident of one
redshift (and, moreover, outside the last-scattering shell defined
in Sec.~\ref{section:visibility}). On either side of
$z_{\rm cross}$ one length is already orders of magnitude larger than
the other. It is worth recalling here the distinction drawn in
Sec.~\ref{sec:geometric}. Replacing $\xe$ by its Saha value steepens
its slope from $+6.4$ to $\simeq +20$ (Eq.~\ref{eq:saha_slope}) and
sharpens the crossing further, but it does not collapse the shell to
zero thickness. Hence the transition retains a finite width even in
equilibrium.

\begin{figure}[t]
\centering
\includegraphics[width=\columnwidth]{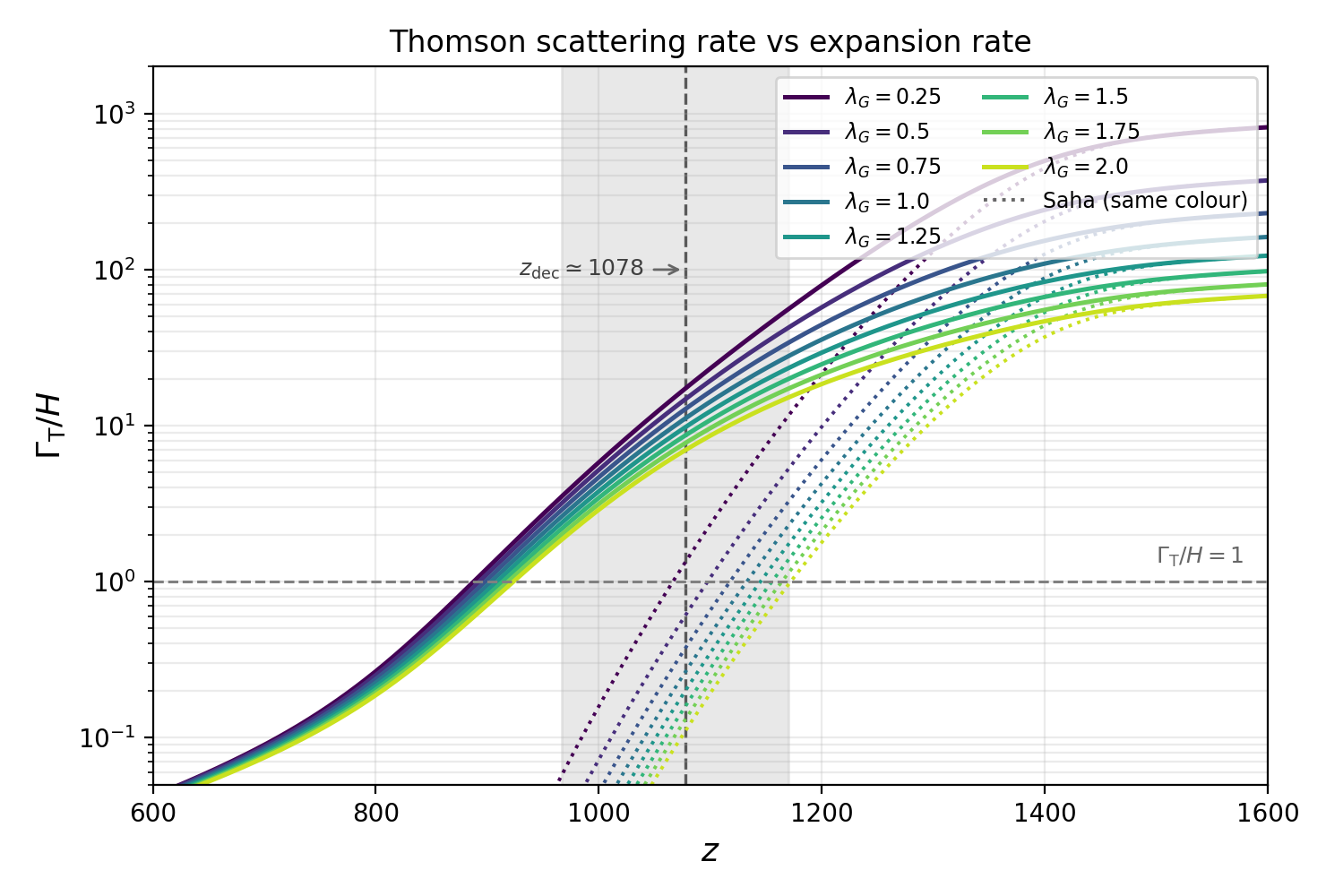}
\caption{Thomson scattering rate relative to the expansion rate,
  $\Gamma_{\rm T}/H$, for the eight models (solid, dark to light) and
  for their Saha counterparts (dotted, same colour). The horizontal
  line marks $\Gamma_{\rm T}/H = 1$; for $\lG = 1$ the kinetic ratio
  crosses it at $z_{\rm cross} \simeq 904$, to the left of the
  visibility peak $z_{\rm dec} \simeq 1078$ (dashed line) and its FWHM
  (shaded band). A larger $G$ lowers the ratio at fixed redshift, as
  $\lG^{-1}$ at fixed composition and somewhat faster once the helium
  response is included, so the crossing and the whole decoupling
  sequence shift with $G$. The Saha histories fall below their kinetic
  counterparts, the gap being the residual ionization that keeps
  photons scattering past the naive $\Gamma_{\rm T}=H$ threshold.}
\label{fig:gammaT}
\end{figure}

As can be seen from Fig.~\ref{fig:gammaT}, the ratio
$\Gamma_{\rm T}/H$ climbs through unity at $z_{\rm cross} \simeq 904$
for $\lG = 1$, well below the visibility peak
$z_{\rm dec} \simeq 1078$, confirming that the naive rate crossing
does \emph{not} coincide with last scattering. The whole family
$\lG \in [0.25,2]$ shifts vertically with $G$. Since
$\Gamma_{\rm T}/H$ falls as $\lG^{-1}$ at fixed redshift and fixed
composition, and as $\lG^{-1.2}$ once the $(1-Y_{\rm He})$ weighting
of the electron density is included, a larger $G$ lowers the ratio and
moves $z_{\rm cross}$ to slightly higher redshift (see
Fig.~\ref{fig:redshifts}), carrying the entire decoupling sequence
with it.

It is worth dwelling on why the last-scattering surface sits at
$z_{\rm dec} \simeq 1078$, where the Thomson rate still exceeds the
expansion rate by an order of magnitude
($\Gamma_{\rm T}/H \simeq 11$), rather than at $z_{\rm cross}$ where
$\Gamma_{\rm T} = H$. The resolution is that the position of the peak
is not set by the local rate alone, but by the integral quantity
$\tau(z)$, and this is exactly the local-versus-integral
distinction. Indeed, since $\dd\tau/\dd z = c\sigma_T n_e/[(1+z)H]$
and the Thomson rate is $\Gamma_{\rm T} = c\sigma_T n_e$, the optical
depth is the running integral of the very rate ratio discussed above,
\begin{equation}
\tau(z) = \int_0^z \frac{\Gamma_{\rm T}}{H}\,\frac{\dd z'}{1+z'}\,,
\label{eq:tau_gammaT}
\end{equation}
so that a large \emph{local} ratio $\Gamma_{\rm T}/H$ need not place
last scattering where $\Gamma_{\rm T} = H$, what matters is the
accumulated integral.

Writing $g = (\dd\tau/\dd z)\,e^{-\tau}$, the two factors pull in
opposite directions: $\dd\tau/\dd z \propto \Gamma_{\rm T}/H$ is local
and \emph{increases} with $z$ (more free electrons, more scattering),
while $e^{-\tau}$ is integral and \emph{decreases} with $z$ (a photon
coming from higher $z$ is more likely to have rescattered on the way
out), as shown in Fig.~\ref{fig:gzfactors}. Their product peaks where
the running optical depth is of order unity,
$\tau(z_{\rm dec}) \simeq 0.93$, and it so happens that
$\tau \simeq 1$ is reached while the local rate is still fast. A large
$\Gamma_{\rm T}/H$ means a photon scatters many times per Hubble time
\emph{locally}, but whether a given scattering is the \emph{last} one
depends on the probability of no further scattering all the way out,
i.e.\ on $e^{-\tau}$, not on the instantaneous rate.

\subsection{Discussion}
\label{sec:discussion}

\begin{figure}[t]
\centering
\includegraphics[width=\columnwidth]{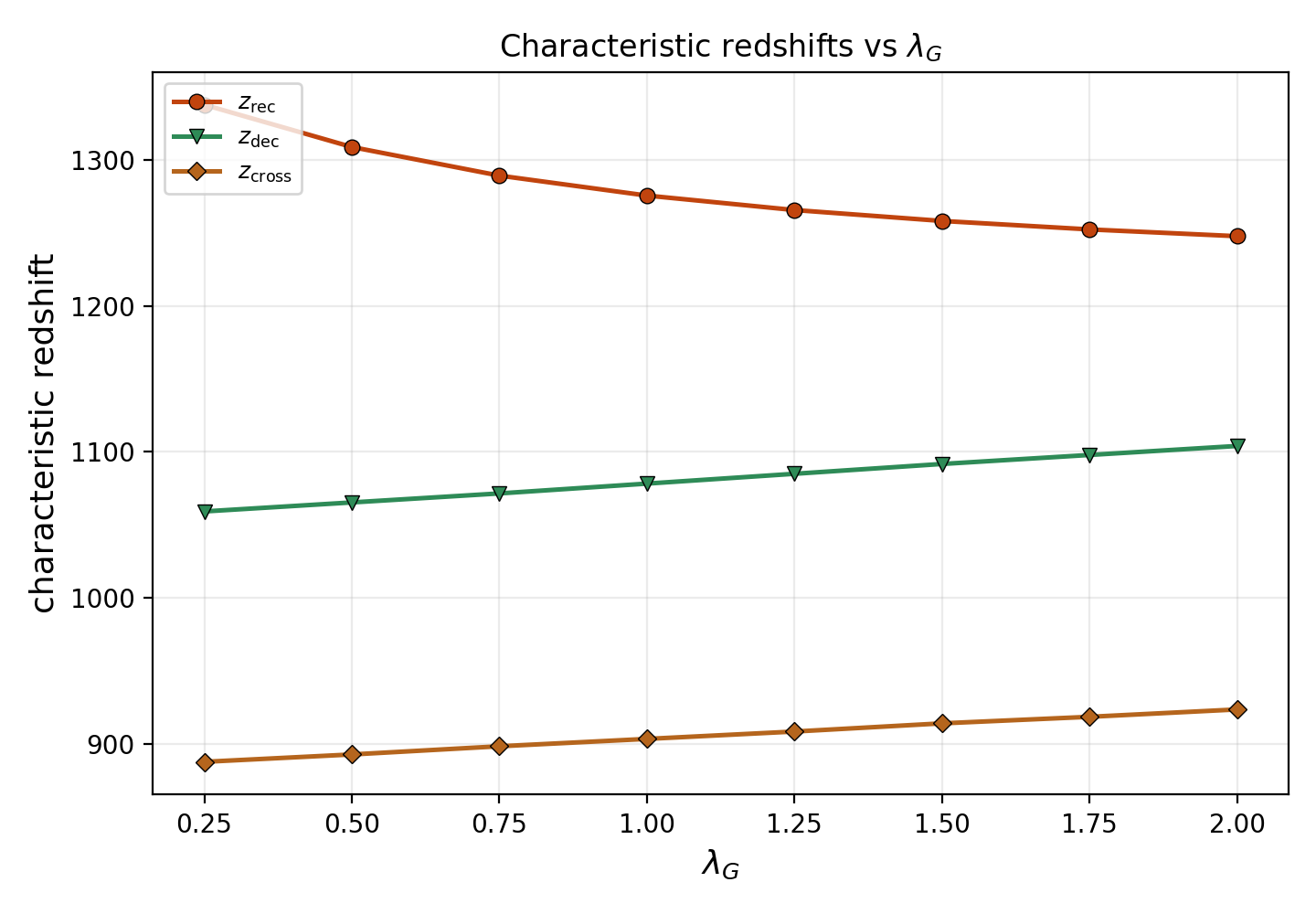}
\caption{Characteristic redshifts of Table~\ref{tab:redshifts} as
  functions of $\lambda_G$. The recombination redshift $z_{\rm rec}$
  decreases with $\lambda_G$ while $z_{\rm dec}$ and $z_{\rm cross}$
  increase, while $z_{\rm eq}$ is invariant. }
\label{fig:redshifts}
\end{figure}

The visibility peak $z_{\rm dec}$ and the recombination midpoint
$z_{\rm rec}$ do not coincide, and, interestingly, they move in
\emph{opposite} directions with $G$. A larger $G$ shifts the
recombination transition to \emph{lower} redshift ($z_{\rm rec}$
decreases, Fig.~\ref{fig:xe} and Fig.~\ref{fig:redshifts}) because the
faster expansion reduces $\Gamma_{\rm rec}/H$, so a given ionization
level such as $\xe = 1/2$ is reached later. The visibility peak, by
contrast, moves to \emph{higher} redshift with $G$, because a larger
$\lG$ lowers $\Gamma_{\rm T}/H$ at any given redshift, so the
accumulated optical depth reaches unity higher up. The two redshifts
respond to distinct rate ratios and move oppositely, a reminder that
no single ratio captures the whole sequence of events.

One check is called for before going on, because at the upper end of
our scan helium is not a trace species. At $\lG = 2$ the
self-consistent BBN value is $Y_{\rm He} = 0.370$, so that
$n_{\rm He}/n_{\rm H} = 0.147$ and a fully ionized plasma has
$\xe = 1.29$ rather than $1.16$. One must therefore ask whether helium
recombination, which the hydrogen-centric algebra of
Sec.~\ref{sec:recombination} does not describe, ever intrudes into the
last-scattering shell, which at large $\lG$ is both wider and at
higher redshift. It does not. The midpoint of
He\,\textsc{iii}\,$\to$\,He\,\textsc{ii} sits at $z \simeq 5880$ to
$5915$ across the family, essentially fixed, and the midpoint of
He\,\textsc{ii}\,$\to$\,He\,\textsc{i} moves from $z = 2320$ at
$\lG = 0.25$ to $z = 1886$ at $\lG = 2$. The two trends do close the
gap, since $z_{\rm dec}$ rises and the shell widens with $G$, but even
in the most extreme model the second helium transition is completed
$661$ redshift units above the top of the shell, more than five times
its half-width. Helium is thus fully neutral well before last
scattering in every model and supplies no free electrons there: the
whole of its influence on the scalings reported below passes through
the counting factor $1-Y_{\rm He}$ in $n_{\rm H}$, which is precisely
the channel that Secs.~\ref{sec:thermal} and \ref{sec:obsconseq} will
find to be dominant. The margin is not large enough, however, for
the helium lines to be neglected, and in a model with substantially
larger $G$ than we consider they would have to be treated inside the
shell.

Having established how the shell forms and how wide it is, we close
this section by collecting the various characteristic redshifts and
showing how the whole sequence responds to $G$.

Table~\ref{tab:redshifts} collects the characteristic redshifts that
punctuate the recombination--decoupling sequence, with their
definitions and their values for $\lG = 1$. They are markedly
different from one another, which is the whole point: no single one of
them captures the process, and each answers a distinct physical
question. We also list $z'_{\rm cross}$, the redshift at which the
\emph{recombination} rate drops to the expansion rate,
$\Gamma_{\rm rec} = H$ (for $\lG = 1$); it lies well below all the
others, at $z'_{\rm cross} \simeq 405$, marking the point where
recombination itself freezes out, long after Thomson scattering has
ceased to keep pace at $z_{\rm cross}$.

That last redshift deserves a word on how it is obtained, because it is the
one entry of the table that could depend on the choice of case-B coefficient.
It does not, because we do not use one. Below $z \simeq 700$ the
photoionization term of Eq.~\eqref{eq:peebles} is exponentially dead and
$C_{\rm P} \to 1$, so that $\dd\xe/\dd t = -\Gamma_{\rm rec}\xe$ exactly and
\begin{equation}
\frac{\Gamma_{\rm rec}}{H} \;=\; (1+z)\left|\frac{\dd\ln\xe}{\dd z}\right|\,,
\label{eq:gammarec_free}
\end{equation}
which is read directly off the computed $\xe(z)$ with no rate coefficient
entering at all. This is how $z'_{\rm cross}$ and its sensitivity are obtained
here. For comparison, reconstructing $\Gamma_{\rm rec}$ explicitly with the
Seaton coefficient of Sec.~\ref{sec:rate} instead returns $420$, and with the
\citet{Pequignot1991} fit used by \texttt{RECFAST} it returns $432$; the
spread of these three determinations, about $6\%$, is a fair measure of how
well a case-B coefficient represents a multilevel calculation, and it is why
we prefer Eq.~\eqref{eq:gammarec_free}.

\begin{table*}
\centering
\caption{Characteristic redshifts of the recombination--decoupling
  sequence, ordered from early to late, with values for $\lG = 1$ and
  the local logarithmic sensitivities $\kappa_X$ of
  Eq.~\eqref{eq:kappa_def} applied to $1+z$. Only $z_{\rm eq}$ is
  strictly invariant under a rescaling of $G$, the others shift with
  $\lG$, and they do not all shift in the same direction.}
\label{tab:redshifts}
\begin{tabular}{llrr}
  \hline\hline
  Symbol & Definition & Value ($\lG=1$) & $\kappa$ \\
  \hline
  $z_{\rm eq}$    & matter--radiation equality, $\rho_m = \rho_r$      & $3405.8$ & $+0.000$ \\
  $z_{\rm rec}$   & recombination midpoint, $\xe = 1/2$ & $1275$ & $-0.036$ \\
  $z_{\rm dec}$   & peak of the visibility function $g(z)$             & $1078$ & $+0.025$ \\
  $z_\ast$        & unit optical depth, $\tau = 1$                     & $1085$ & $+0.039$ \\
  $z_{\rm cross}$ & Thomson rate equals expansion, $\Gamma_{\rm T} = H$ & $904$ & $+0.023$ \\
  $z_{\rm end}$   & recombination complete, $\xe = 3.5\times10^{-3}$   & $798$ & $-0.122$ \\
  $z'_{\rm cross}$ & recombination rate equals expansion, $\Gamma_{\rm rec} = H$ & $405$ & $+0.040$ \\
  \hline
\end{tabular}
\end{table*}

\begin{figure}[t]
\centering
\includegraphics[width=\columnwidth]{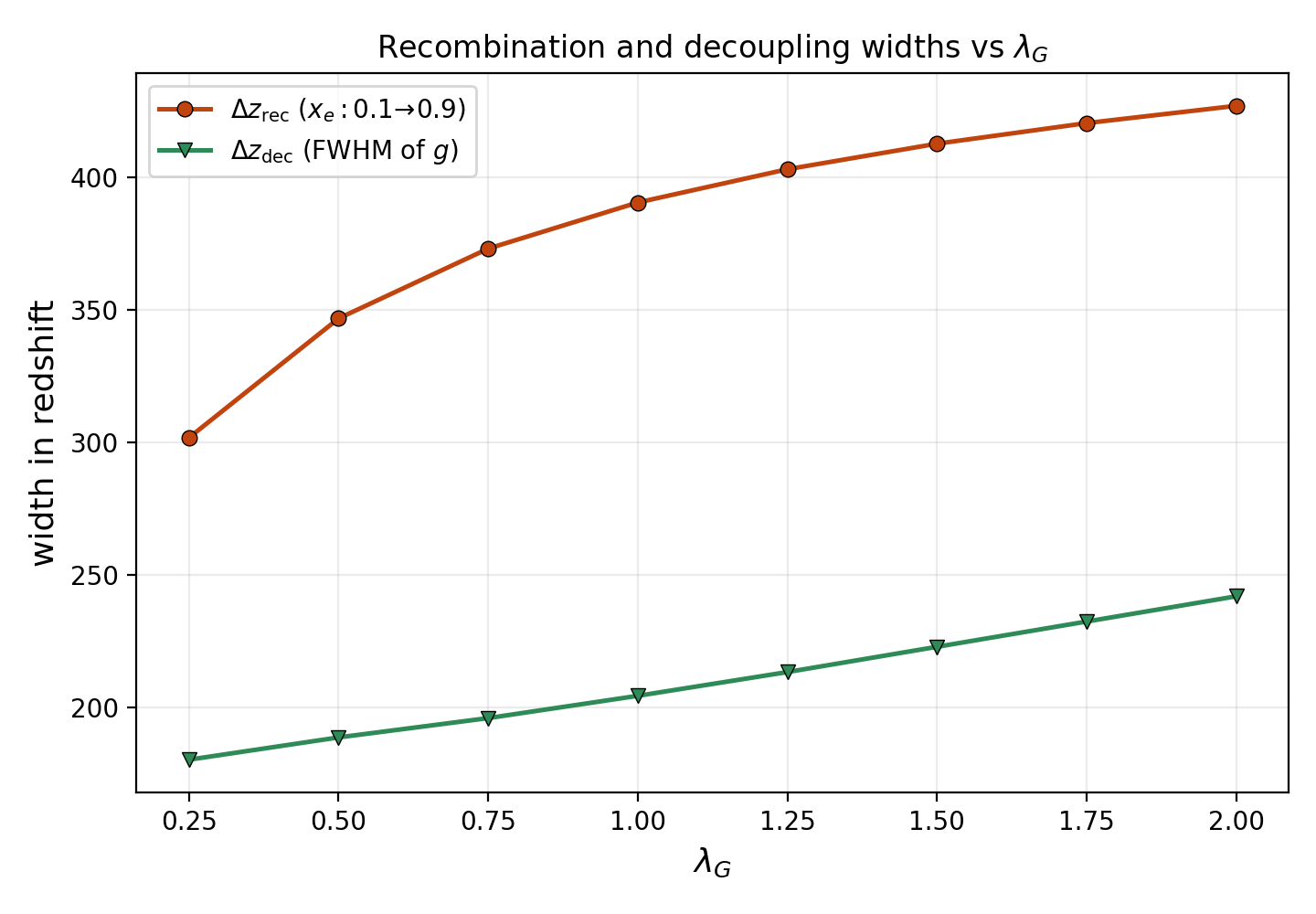}
\caption{Recombination width $\Delta z_{\rm rec}$ (from $\xe = 0.1$ to
  $0.9$) and decoupling width $\Delta z_{\rm dec}$ (FWHM of the
  visibility) as functions of $\lambda_G$. Both widen with
  $\lambda_G$: a larger $G$ makes recombination more gradual and the
  last-scattering surface thicker. Their logarithmic sensitivities at
  $\lG = 1$ are $+0.151$ and $+0.168$ respectively
  (Table~\ref{tab:sensitivities}).}
\label{fig:widths}
\end{figure}

Figure~\ref{fig:redshifts} shows how these redshifts evolve with
$\lG$. The pattern is instructive. $z_{\rm eq}$ is flat, as it must
be, while the four others split into two groups moving in opposite
directions. The recombination midpoint $z_{\rm rec}$ \emph{decreases}
with $\lG$, whereas $z_{\rm dec}$ and $z_{\rm cross}$ both
\emph{increase}. The ordering
$z_{\rm rec} > z_{\rm dec} > z_{\rm cross}$ is preserved throughout,
but the spacings widen. Figure~\ref{fig:widths} shows the companion
trend for the two characteristic widths. Both $\Delta z_{\rm rec}$ and
$\Delta z_{\rm dec}$ grow monotonically with $\lG$, the
last-scattering surface thickening as gravity strengthens. Together
the two figures make explicit that a rescaling of $G$ does not simply
translate the recombination--decoupling sequence but stretches and
reorders it.

\subsection{Logarithmic sensitivities}
\label{sec:sensitivities}

A compact way to summarise these trends, and to compare them with one
another, is to quote logarithmic sensitivities around the standard
model, in the spirit of the sensitivity coefficients used for BBN
abundances \citep{DSW2007}. Here they are a way of \emph{reporting}
results obtained by full computation over the whole scan, not a
substitute for it. Every quantity below is computed exactly for each
model, and the coefficient merely records the local slope at
$\lG = 1$. For any quantity $X$ we define
\begin{equation}
\kappa_X \;\equiv\;
\left.\frac{\partial\ln X}{\partial\ln\lG}\right|_{\lG=1}\,,
\qquad\text{so that}\qquad
\frac{\delta X}{X} \simeq \kappa_X\,\frac{\delta\lG}{\lG}
\label{eq:kappa_def}
\end{equation}
for small departures from $\lG = 1$. We obtain this derivative from the three
closely spaced models $\lG = 0.9, 1, 1.1$ alone, as the linear coefficient of a
quadratic in $\ln\lG$ through them, so that neither an extrapolation nor any
model of the large-$\lG$ behaviour enters. Two independent estimators confirm
the values: a finite difference between $0.9$ and $1.1$ corrected for the
slight asymmetry of that pair in $\ln\lG$, and a quartic fit over the whole
family. They agree with the quoted numbers to $0.0014$ or better on every
entry of Table~\ref{tab:sensitivities}, and to $0.0001$ on $\ldamp$ itself. Table~\ref{tab:sensitivities}
collects these coefficients for the redshifts, widths and lengths
introduced so far, together with the three angular scales, the last of
which is only derived in Sec.~\ref{sec:damping}.

\begin{table}[t]
\centering
\caption{Logarithmic sensitivities
  $\kappa_X = \partial\ln X/\partial\ln\lG$ evaluated at $\lG = 1$,
  Eq.~\eqref{eq:kappa_def}. A value of $-1$ signals a pure $c/H$
  background length while values close to zero signal quantities that
  a rescaling of $G$ leaves essentially untouched. These coefficients
  are \emph{local} slopes at $\lG = 1$, and they are not to be
  confused with the power-law exponents of Eqs.~\eqref{eq:rd} and
  \eqref{eq:decomposition}, which are averaged over the whole range
  $\lG \in [0.25,2]$ and are correspondingly smaller in magnitude. The
  coefficient for $Y_{\rm He}$ is obtained from the modified
  \texttt{PArthENoPE} runs described in Sec.~\ref{sec:framework} and
  agrees with the linear sensitivity of \citet{DSW2007} once their
  $\partial\ln Y/\partial\ln G$ is converted to our quadratic
  convention $\lG = \sqrt{G/G_N}$. The line $1-Y_{\rm He}$ is listed
  separately because it is this combination, and not $Y_{\rm He}$
  itself, that enters $n_{\rm H}$ and hence every recombination rate
  in Secs.~\ref{sec:damping} and \ref{sec:residual}. The entry for
  $\tau(z_{\rm dec})$ is the full optical depth of Eq.~\eqref{eq:tau},
  integrated from the observer and therefore including the
  reionization contribution $\tau_{\rm reio} = 0.054$; excluding the
  latter gives $0.87$ with $\kappa = -0.105$. That value of
  $\tau_{\rm reio}$ is the $\lG = 1$ one: the input is rescaled per
  model through Eq.~\eqref{eq:taureio}, and the integrated $\tau$
  below $z = 50$ runs from $0.275$ at $\lG = 0.25$ to $0.023$ at
  $\lG = 2$.}
\label{tab:sensitivities}
\begin{tabular}{llr}
\hline\hline
Quantity & Value at $\lG=1$ & $\kappa_X$ \\
\hline
$Y_{\rm He}$           & $0.247$             & $+0.715$ \\
$1-Y_{\rm He}$         & $0.753$             & $-0.234$ \\
\hline
$z_{\rm rec}$          & $1275$              & $-0.036$ \\
$z_{\rm dec}$          & $1078$              & $+0.025$ \\
$\Delta z_{\rm rec}$   & $390$               & $+0.149$ \\
$\Delta z_{\rm dec}$   & $204$               & $+0.172$ \\
$\tau(z_{\rm dec})$    & $0.93$              & $-0.174$ \\
\hline
$D_c(z_{\rm dec})$     & $13871\,$Mpc         & $-1.000$ \\
$r_s(z_{\rm dec})$     & $145\,$Mpc           & $-1.016$ \\
$\rdamp$               & $47\,$Mpc            & $-0.696$ \\
\hline
$\ell_{\rm eq}$        & $204$               & $+0.000$ \\
$\ell_A$               & $299.6$             & $+0.017$ \\
$\ldamp$               & $1859$              & $-0.304$ \\
\hline
\end{tabular}
\end{table}

The table makes the structure of the problem visible at a glance. The
two background lengths $D_c$ and $r_s$ both sit at $\kappa \simeq -1$,
the signature of a pure $c/H$ integral. Their near-cancellation leaves
$\kappa_{\ell_A} = +0.017$, the small residual discussed in
Sec.~\ref{sec:acoustic}. The characteristic redshifts barely move,
with $|\kappa|\lesssim0.04$, while the widths respond about five times
more strongly.  The damping length breaks the pattern:
$\kappa_{\rdamp} = -0.696$ is far from $-1$, and the resulting
$\kappa_{\ldamp} = -0.304$ is what makes the damping tail the one
place where a rescaling of $G$ leaves a clear imprint.

\section{The diffusion damping scale}
\label{sec:damping}

The finite thickness of the last-scattering surface established in
Sec.~\ref{sec:decoupling} has a second consequence, which is the
subject of this section. During the time the shell takes to form,
photons do not stream freely: they random-walk through the
baryon-photon plasma, diffusing out of the smallest overdensities
before they decouple. Temperature fluctuations on scales below the
diffusion length are therefore washed out, and the CMB power spectrum
is exponentially suppressed at high multipole. We build this effect up
in the same order as the previous sections. Sec.~\ref{sec:mfp} sets
out the photon mean free path and how it responds to
$G$. Sec.~\ref{sec:silkintegral} then defines the diffusion length
exactly, as an integral over that mean free path, and displays the two
microphysical ingredients that enter it.  Sec.~\ref{sec:silkvis}
builds the diffusion analogue of the visibility function, the Silk
visibility $g_{\rm Silk}(z;k)$, and uses it to identify where, in
redshift and in scale, the damping is actually acquired.
Sec.~\ref{sec:silkscaling} relates the damping multipole to the
diffusion length and derives its scaling with $G$, comparing the
prediction to the \CLASS{} outputs. Finally, Sec.~\ref{sec:tt} turns
to the damping tail of the TT spectra themselves, and determines the
shape of the damping envelope that reproduces this damping scale.

\subsection{The mean free path}
\label{sec:mfp}

The physical picture is a random walk. Between two Thomson scatterings
a photon travels one mean free path,
\begin{equation}
  \lmfp = \frac{1}{a\,n_e\sigma_T} = \frac{1}{a\,
    \xe(1-Y_{\rm He})n_{b,0}(1+z)^3\,\sigma_T}\,,
\label{eq:lmfp}
\end{equation}
here written as a comoving length, and the direction of travel is
randomised at each scattering. It is instructive to express this
length in fundamental constants. With
$\sigma_T = (8\pi/3)\alpha^2\lambdabar_e^2$, where
$\lambdabar_e = \hbar/m_ec$ is the reduced Compton wavelength of the
electron, and with
$n_{b,0} =
\eta_b\,(2\zeta(3)/\pi^2)\,(\theta_0\alpha^2/\lambdabar_e)^3$, where
$\eta_b$ is the baryon-to-photon ratio of Sec.~\ref{sec:rate} and
$\theta_0 \equiv k_BT_0/(\alpha^2m_ec^2)$ the present temperature in
the atomic units used there, Eq.~\eqref{eq:lmfp} becomes
\begin{equation}
\lmfp = \frac{3\pi}{16\zeta(3)}\;\lambdabar_e\;
\eta_b^{-1}\,\alpha^{-8}\,\theta_0^{-3}\;
\frac{1}{\xe\,(1-Y_{\rm He})\,(1+z)^2}\,.
\label{eq:lmfp_micro}
\end{equation}
The only dimensioned quantity involved is $\lambdabar_e$, and no
gravitational constant appears.

It is worth discussing how this length responds to
$G$. Equation~\eqref{eq:lmfp_micro} shows that the response passes
entirely through the two factors $\xe$ and $Y_{\rm He}$, since the
Thomson cross section is a constant and, in our family of models, the
physical baryon density $n_{b,0}(1+z)^3$ is common to all the models
at a given redshift.

The first and dominant one is the ionization fraction $x_e$. As
already explained in Sec.~\ref{sec:recombination}, a larger $G$
makes the plasma dilute faster at the same redshift, leaving electrons
and protons less time to find each other before the dilution drives
them apart. Hence, as can be seen in Fig.~\ref{fig:xe}, at a given
redshift a larger value of $G$ corresponds to a larger residual
fraction of electrons. For example, at $z = 1100$ the ionization
fraction rises from $\xe \simeq 0.047$ at $\lG = 0.25$ to
$\xe \simeq 0.21$ at $\lG = 2$, a factor of $4.5$. A more ionized
plasma is a more opaque one, and $\lmfp$ shortens accordingly.

The second factor acts in the opposite direction. A faster expansion
at the time of nucleosynthesis freezes the neutron-to-proton ratio
earlier, so more neutrons survive to be locked into helium, and
$Y_{\rm He}$ grows with $\lG$. According to Eq.~(\ref{eq:lmfp}), a
larger $Y_{\rm He}$ increases the mean free path. However, this effect
is smaller than the one on $x_e$ so that the net effect is still a
marked reduction of the mean free path, the comoving $\lmfp$ of
Eq.~\eqref{eq:lmfp} at $z = 1100$ falling from $36$ to $12$~Mpc across
our range of $\lG$. This is the microphysical origin of the whole
effect studied in this paper. A stronger gravity leaves a more
ionized, more opaque plasma, photons travel shorter distances between
scatterings.

\subsection{The diffusion length and the Silk integral}
\label{sec:silkintegral}

What matters for the CMB is not the length of a single step but the
net distance a photon covers after very many of them. We therefore
return to the random walk. The mean square comoving displacement grows
by $2 D_{\rm diff}\,\dd t$ per interval of time, with
$D_{\rm diff} = c\,\lmfp/3$ the photon diffusion coefficient. The
factor $1/3$ comes from the isotropic average
$\langle\cos^2\theta\rangle = 1/3$. The diffusion coefficient evolves
as the mean free path $\lmfp$, and hence increases at low redshift as
can be seen in Fig.~\ref{fig:mfp}. At redshift $z$, a photon has
diffused over a comoving distance whose root-mean-square is what is
called $d_{\rm damp}(z)$,
\begin{equation}
  d_{\rm damp}^2(z) \;=\; \langle x^2\rangle \;=\; \int_z^\infty 2\,D_{\rm diff}\,
  \frac{\dd t}{\dd z'}\,\dd z'\,.
\label{eq:rw}
\end{equation}

\begin{figure}[h]
\centering
\includegraphics[width=\columnwidth]{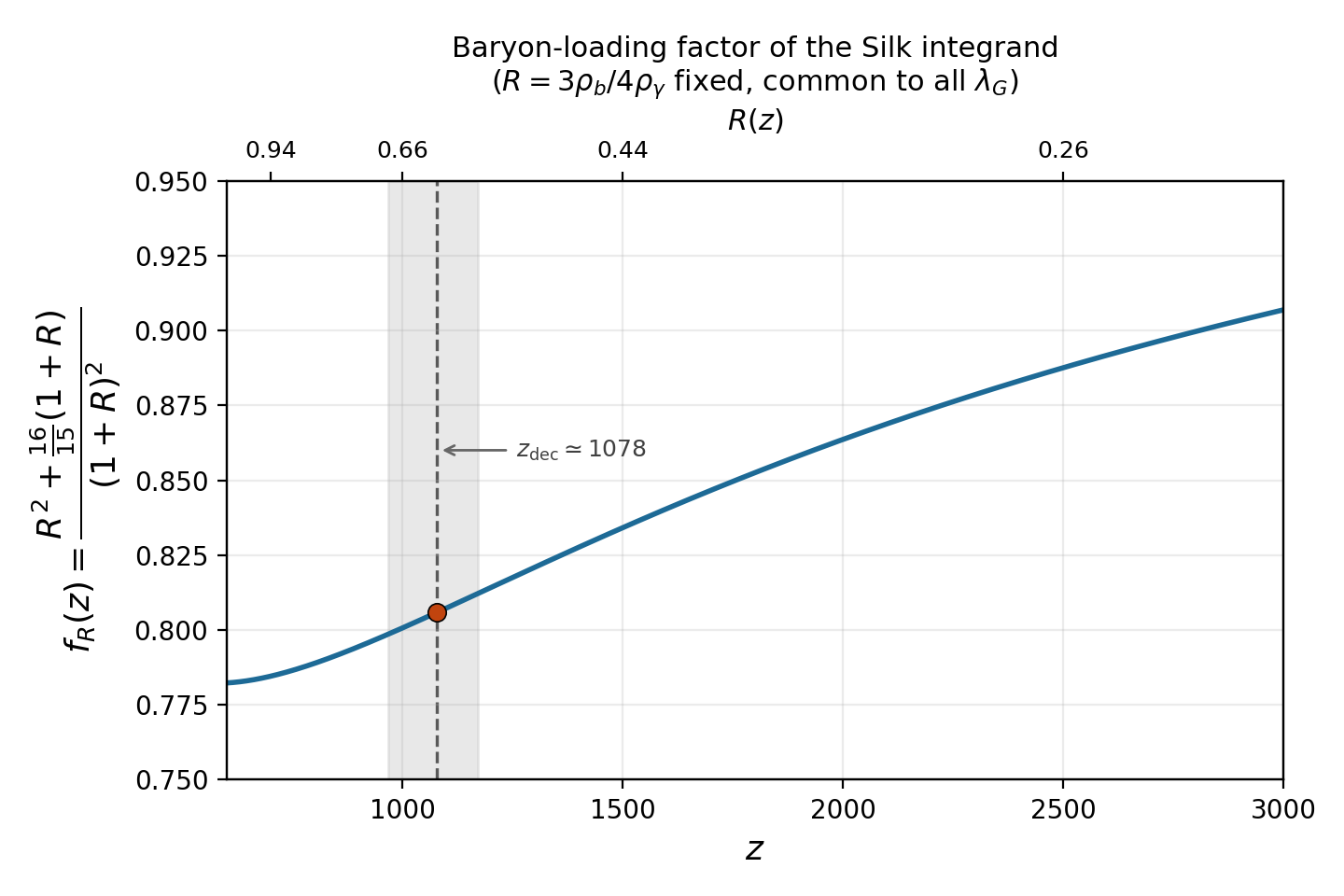}
\caption{Baryon-loading factor
  $f_R(z) = [R^2 + \tfrac{16}{15}(1+R)]/(1+R)^2$ that weights the Silk
  integrand in Eq.~\eqref{eq:silk}. Because the physical densities are
  held fixed, $R(z)$ and hence $f_R(z)$ are independent of $G$. At
  decoupling, $f \simeq 0.81$.}
\label{fig:fR}
\end{figure}

The damping length $d_{\rm damp}(z)$ is thus the diffusion distance,
accumulated from early times down to $z$.  Equation~\eqref{eq:rw} is
in fact a schematic estimate as it treats the photons as a single
species free-streaming between scatterings, with the isotropic
diffusion coefficient $D_{\rm diff} = c\lmfp/3$. In reality, the
photons are not free but tightly coupled to the baryons, and the
finite baryon inertia, measured by $R = 3\rho_b/4\rho_\gamma$, slows
the effective diffusion. Working the tight-coupling expansion of the
photon-baryon system to lowest order in $\lmfp$ \citep{HuSugiyama1995,
  HuWhite1997} replaces the bare coefficient by
$D_{\rm diff} = (c\lmfp/6)\,f_R(z)$, where the baryon-loading factor
\begin{equation}
f_R(z) = \frac{R^2 + \tfrac{16}{15}(1+R)}{(1+R)^2}
\label{eq:fR}
\end{equation}
weights the contribution of each epoch according to how tightly the
baryons are coupled to the photons. It tends to $\tfrac{16}{15}$ in
the radiation-dominated, baryon-free limit ($R\to0$) and decreases
toward unity as $R$ grows, i.e.\ as the baryon-to-photon momentum
ratio $R = 3\rho_b/4\rho_\gamma$ increases toward decoupling and the
baryons contribute a larger share of the fluid inertia. This baryon
loading factor is plotted in Fig.\ref{fig:fR} for the redshift range
of interest, and shows that $f_R(z)$ stays of order unity. The
comoving damping length is then written as
\begin{equation}
d_{\rm damp}^2(z) \;=\; (2\pi)^2\!\!\int_{z}^{\infty}
\lmfp(z')\,R_H(z')\,\frac{f_R(z')}{6}\,\frac{\dd z'}{1+z'}\,,
\label{eq:silk}
\end{equation}
with $\lmfp$ and $R_H$ both comoving, as defined above. The factor
$R_H\,\dd z/(1+z) = c\,\dd z/H$ is simply the comoving line element,
so that $d_{\rm damp}$ comes out as a comoving
length. Equation~\eqref{eq:silk} is thus Eq.~\eqref{eq:rw} with the
free-diffusion coefficient replaced by its tight-coupling form, and
this is the expression we integrate in what follows.

The overall $(2\pi)^2$ deserves to be stated explicitly, since it is a
convention and not a physical factor. The mode suppression of
Sec.~\ref{sec:silkvis} is written $\exp[-k^2/k_D^2(z)]$, and we define
the damping wavenumber as the inverse \emph{wavelength}
\begin{equation}
k_D(z) \;\equiv\; \frac{2\pi}{d_{\rm damp}(z)}\,,
\;\text{so that}\;
\frac{1}{k_D^2(z)} = \int_{z}^{\infty}\!\lmfp R_H\,\frac{f_R}{6}\,\frac{\dd z'}{1+z'}\,,
\label{eq:kD}
\end{equation}
the right-hand side being Eq.~\eqref{eq:silk} stripped of its
$(2\pi)^2$. This is the convention \CLASS{} uses internally for the
quantity it tabulates as $r_d$, and it is what reconciles the
$46\,$Mpc of Eq.~\eqref{eq:rd} with the $63\,$Mpc one would obtain
from a bare inverse wavenumber.

It is worth being explicit that this choice cannot affect any
exponent, nor even the multipole $\ldamp$ itself. The projection of
Sec.~\ref{sec:silkscaling} is $\ldamp = k_D(z_{\rm dec})\,D_c$, that
is $\ldamp = 2\pi D_c/\rdamp$, so the $2\pi$ introduced in
Eq.~\eqref{eq:silk} and the $2\pi$ of the projection cancel
identically,
\begin{equation}
\ldamp \;=\; \frac{2\pi D_c}{\rdamp}
\;=\; \frac{D_c}{\left[\int_{z_{\rm dec}}^{\infty}\!\lmfp R_H\,\frac{f_R}{6}\,
\frac{\dd z'}{1+z'}\right]^{1/2}}\,,
\label{eq:nocancel}
\end{equation}
which contains no $2\pi$ at all and gives $\ldamp = 1859$ for
$\lG = 1$. The convention therefore fixes only the number of Mpc one
quotes for $\rdamp$.

This also obliges us to be careful about what the comparison with
\CLASS{} does and does not establish. Integrating Eq.~\eqref{eq:silk}
independently and comparing with the damping scale tabulated by
\CLASS{}, its \texttt{r\_d} column, agrees with it to better than $2\times10^{-3}$ across the
eight models, and typically to $10^{-4}$, over a factor of five in
$\rdamp$. That is an exact check that we have correctly identified and
implemented \CLASS{}'s own definition, and nothing more: it is not an
independent confirmation of the scaling, since the two calculations
are the same formula. The genuinely independent check of $\rdamp$ is the
analytic normalisation of Eq.~\eqref{eq:rdamp}. The damping tail of the
\CLASS{} $TT$ spectra, analysed in Sec.~\ref{sec:tt}, adds a consistency test
of a different kind, since it shows that a single slowly varying envelope
shape reproduces this $\rdamp$ across the whole family, but it does not
provide a second determination of it.

\begin{figure}[h]
\centering
\includegraphics[width=\columnwidth]{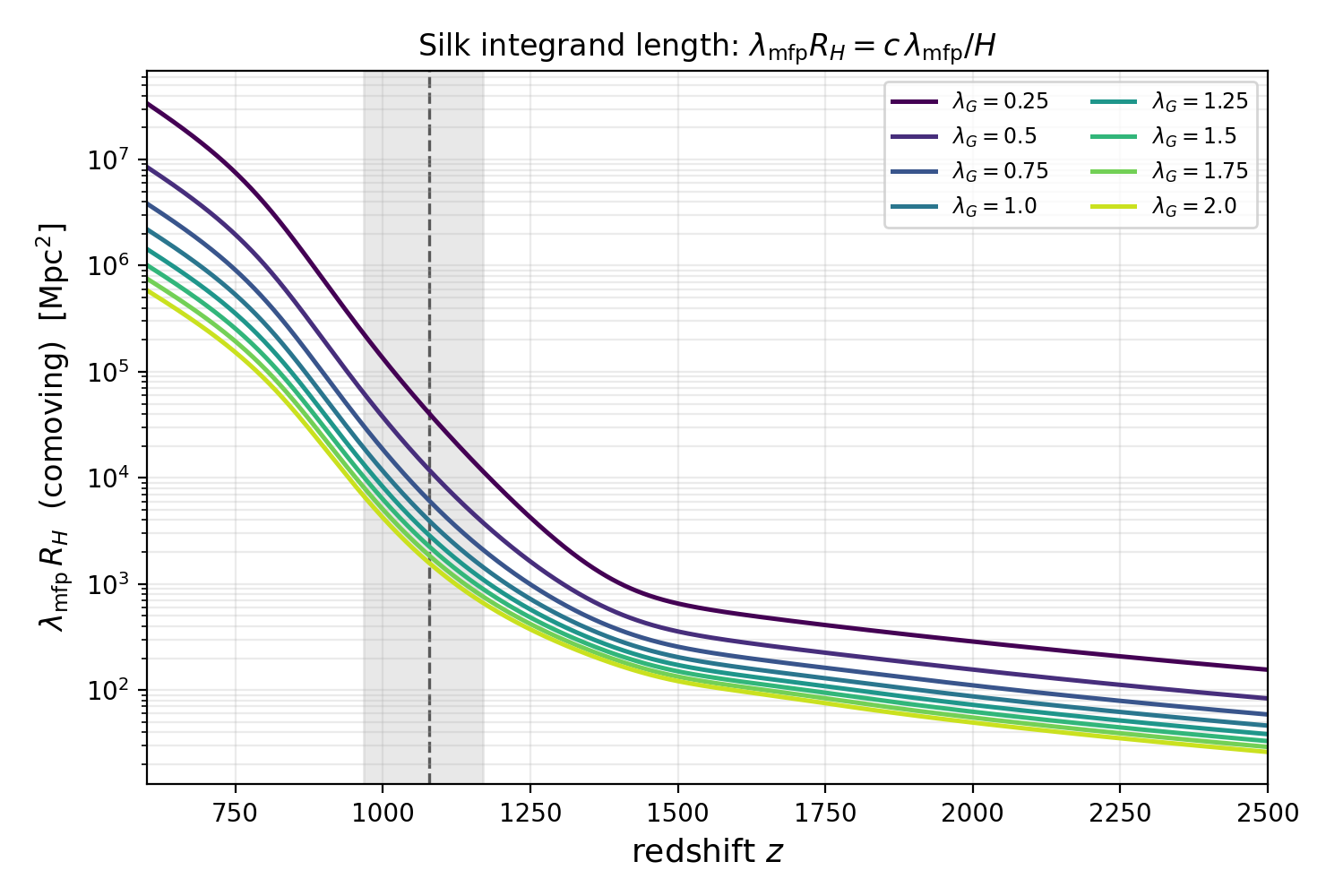}
\caption{The squared comoving length $\lmfp R_H$ (in Mpc$^2$), with
  both lengths comoving, that sets the Silk integrand of
  Eq.~\eqref{eq:silk}, for the eight models. It rises steeply through
  recombination as $\lmfp$ lengthens. At fixed redshift within the
  shell (shaded band) a larger $G$ leaves a shorter mean free path and
  hence a smaller integrand, the microphysical imprint that drives the
  $G$-dependence of $r_{\rm damp}$.}
\label{fig:lmfpRH}
\end{figure}

\begin{figure}[h]
\centering
\includegraphics[width=\columnwidth]{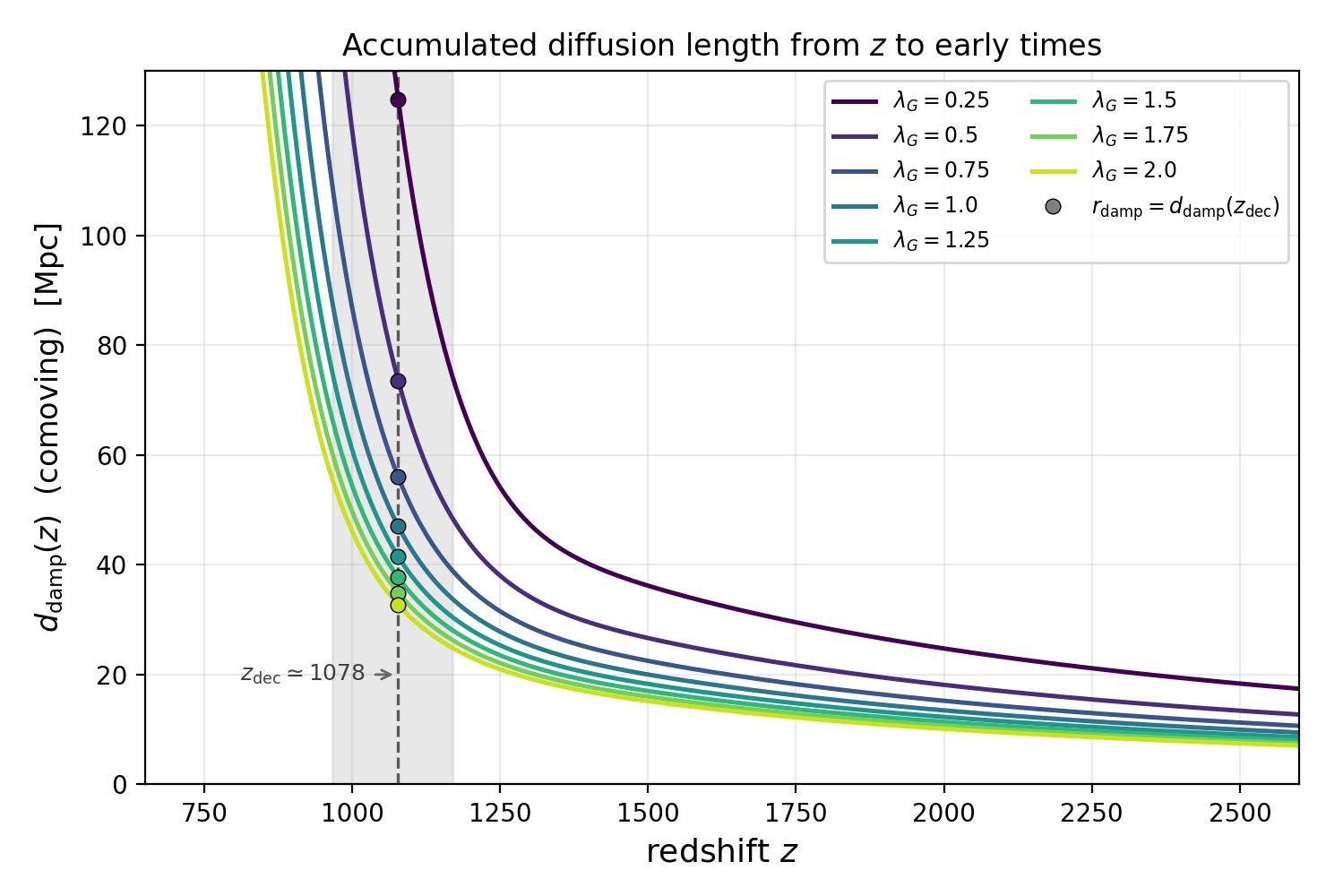}
\caption{Accumulated diffusion length $d_{\rm damp}(z)$, the partial
  Silk integral \eqref{eq:silk} built up from early times down to $z$,
  for the eight models. The circle on each curve marks
  $r_{\rm damp} = d_{\rm damp}(z_{\rm dec})$, the physical damping
  length at decoupling ($\simeq 46$~Mpc for $\lG = 1$, decreasing with
  $G$). The divergence below the shaded band is the unphysical
  post-decoupling growth discussed in the text, cut off at
  $z_{\rm dec}$.}
\label{fig:ddamp}
\end{figure}

The integrand is proportional to $\lmfp R_H$, so $d_{\rm damp}$ is an
integrated geometric mean of $\lmfp$ and $R_H$, weighted by the
baryon-loading factor $f_R(z)$. The damping scale is hence set by an
accumulation of $\lmfp R_H$, not a single value at a given
redshift. The evolution of $\lmfp R_H$ is represented in
Fig.~\ref{fig:lmfpRH}, while $d_{\rm damp}(z)$ is plotted in
Fig.~\ref{fig:ddamp}.

The upper limit of the integral~\ref{eq:silk} is the early-time
plasma, where $\lmfp$ is negligibly small. The physically important
limit is the lower one. Formally $d_{\rm damp}(z)$ keeps growing as
$z\to0$ because $\lmfp$ diverges after recombination, but this growth
is unphysical. Indeed, once the photons have decoupled they
free-stream rather than diffuse, no further damping is imprinted on
the spectrum, and Eq.~(\ref{eq:silk}) coming from the tight-coupling
limit is no longer valid. The relevant damping length is therefore the
value accumulated up to the moment scattering ceases, i.e.\ the
constant $r_{\rm damp} = d_{\rm damp}(z_{\rm dec})$ evaluated at the
visibility peak, and it is this convention we adopt when projecting
onto a multipole in Sec.~\ref{sec:silkscaling}. It is worth being
explicit about what this cut-off means. Stopping the accumulation at
$z_{\rm dec}$ amounts to replacing the visibility by a Dirac spike at
that redshift, so the finite width of the shell does not enter
$\ldamp$ at all. It enters only through the shape of the damping
envelope, as shown in Sec.~\ref{sec:tt}.

We now examine how sensitive the damping length is to this choice of
cut-off in the diffusion integral. Since the cut-off is a convention
rather than a derived quantity, that sensitivity has to be quantified
and not merely asserted. The natural alternative is the redshift
$z_\ast$ defined by $\tau(z_\ast) = 1$. For $\lG = 1$ the two differ
by seven units only, $z_{\rm dec} = 1078$ against $z_\ast = 1085$, and
$\rdamp$ changes by $-2.9\%$, from $46.9$ to $45.5\,$Mpc. That is
reassuring, but it is not the relevant number, because the two
redshifts do not respond to $G$ in the same way. We find
$\kappa_{z_{\rm dec}} = +0.025$ but $\kappa_{z_\ast} = +0.039$, so the
ordering of the two reverses inside the sampled range,
$z_\ast < z_{\rm dec}$ below $\lG \simeq 0.7$ and
$z_\ast > z_{\rm dec}$ above it. The ratio
$d_{\rm damp}(z_\ast)/d_{\rm damp}(z_{\rm dec})$ consequently sweeps
from $1.15$ at $\lG = 0.25$ to $0.94$ at $\lG = 2$, and the exponent
moves with it,
\begin{equation}
\kappa_{\rdamp} =
\begin{cases}
-0.696 & \text{at } z_{\rm dec}\,,\\
-0.753 & \text{at } z_\ast\,,
\end{cases}
\qquad
\kappa_{\ldamp} =
\begin{cases}
-0.304 & \text{at } z_{\rm dec}\,,\\
-0.246 & \text{at } z_\ast\,,
\end{cases}
\label{eq:cutoff}
\end{equation}
and the ratio of Sec.~\ref{sec:finaldiscussion} follows, $\kappa_{\ldamp/\ell_A}$
going from $-0.320$ to $-0.273$. This is the dominant dependence of the
present analysis, and we say so plainly. The shift of $+0.058$ is some sixty times
the $0.001$ to which the close triplet determines the derivative itself, so
what limits the exponent is a definition and not a measurement. For that reason we quote
both values rather than combine them into an error bar, which would suggest an
imprecision that is not there. We retain $z_{\rm dec}$ because the visibility
peak is what the damping envelope of Sec.~\ref{sec:tt} actually selects, and
because it is the convention under which Eq.~\eqref{eq:nocancel} reproduces
the $\ldamp$ that the \CLASS{} spectra deliver.

\subsection{The Silk visibility}
\label{sec:silkvis}

Diffusion does not erase all scales equally. A comoving mode of
wavenumber $k$ corresponds to a fluctuation of comoving wavelength
$2\pi/k$. It is washed out once photons have random-walked a
\emph{net} distance comparable to that wavelength, i.e.\ once the
accumulated diffusion length $d_{\rm damp}$ becomes of order
$2\pi/k$. This relevant net diffusion distance $d_{\rm damp}$ is built
up from many scattering steps, not the single-step mean free path
$\lmfp$, which is far shorter. Since $d_{\rm damp}$ grows as the
Universe expands (see Fig.~\ref{fig:ddamp}), this condition is met
earlier, i.e.\ at higher redshift, for small-scale modes (large $k$),
whose short wavelength is reached while diffusion is still
modest. Meanwhile, large-scale modes (small $k$) are erased only
later, once the diffusion length has grown much larger. Each mode thus
has its own erasure redshift, and it is worth examining in more detail
how the suppression of a given $k$ builds up over cosmic time.

The mean squared diffusion distance accumulated down to redshift $z$
is the running integral Eq.~\ref{eq:silk}. It is convenient to write
it as an inverse squared wavenumber, $k_D(z)=2\pi/d_{\rm damp}(z)$, so
that $k_D(z)$ is the scale that has just been reached by diffusion at
redshift $z$. It decreases monotonically as $z$ falls and the
diffusion length grows. Its final value, at the end of last
scattering, fixes the observed damping scale,
$r_{\rm damp} = 2\pi/k_D(z_{\rm dec})$.

Consider now what this diffusion does to a temperature perturbation. A
photon observed today with a given temperature carries the temperature
of the plasma at the place where it last scattered, not at the place
where it started. Diffusion transports it a random distance before it
starts free streaming. The amplitude of a mode $k$ is then reduced, by
the time the Universe has cooled to redshift $z$, by the factor
$\exp[-k^2/k_D^2(z)]$ \citep{HuWhite1997}. Photons, however, do not
all last scatter at the same redshift. They leave the plasma spread
over the visibility function $g(z) = \dot\tau\,e^{-\tau}$, each
carrying whatever suppression it has accumulated up to its own last
scattering. \citet{HuWhite1997} therefore obtain the net damping by
averaging the suppression factor over that distribution,
\begin{equation}
\mathcal{D}(k) \;=\; \int_0^{\infty} g(z)\;\exp\!\big[-k^2/k_D^2(z)\big]\,\dd z\,.
\label{eq:huwhite}
\end{equation}
The quantity $\mathcal{D}(k)$ is thus the multiplicative factor by
which diffusion reduces the temperature transfer function of mode $k$
relative to what it would be with no diffusion. It is unity for modes
whose wavelength is far larger than the diffusion length and tends to
zero for modes far smaller than it.  Since the temperature power
spectrum is quadratic in the transfer function, the observed $C_\ell$
carries $\mathcal{D}^2(k)$, and it is this factor that produces the
exponential fall-off of power at high multipole, namely the damping
tail we analyse in Sec.~\ref{sec:tt}.

Equation~\eqref{eq:huwhite} gives the total suppression of each mode,
but it does not by itself say \emph{when} that suppression is laid
down. This is the question we need answered, because the redshift at
which a mode is damped is what carries the dependence on $G$. To
isolate it, fix $k$ and read the accumulated suppression
$\exp[-k^2/k_D^2(z)]$ as a function of $z$ alone. It can be
interpreted as a survival factor, equal to $1$ at early times (the
mode is still intact) and falling to $0$ once the mode is fully
erased. Its derivative,
\begin{equation}
g_{\rm Silk}(z;k) \;\equiv\; \frac{\dd}{\dd z}\,e^{-k^2/k_D^2(z)}
\;=\; k^2\,\frac{\lmfp R_H}{6(1+z)}\,f_R\;e^{-k^2/k_D^2(z)}\,,
\label{eq:gsilk}
\end{equation}
is positive and normalised to unity over $z$, and we call it the Silk
visibility. Its meaning is not that of a probability; rather,
$g_{\rm Silk}(z;k)\,\dd z$ is the \emph{fraction of the amplitude of
  mode $k$ that is erased between $z$ and $z+\dd z$}.

As a function of redshift, $g_{\rm Silk}$ is sharply peaked, being the
product of a Silk integrand $\propto f_R\,\lmfp R_H$ that rises as $z$
decreases and of a survival factor that falls once the mode starts to
be erased. The peak marks the narrow redshift window in which that
mode acquires its damping.

\begin{figure}[h]
\centering
\includegraphics[width=\columnwidth]{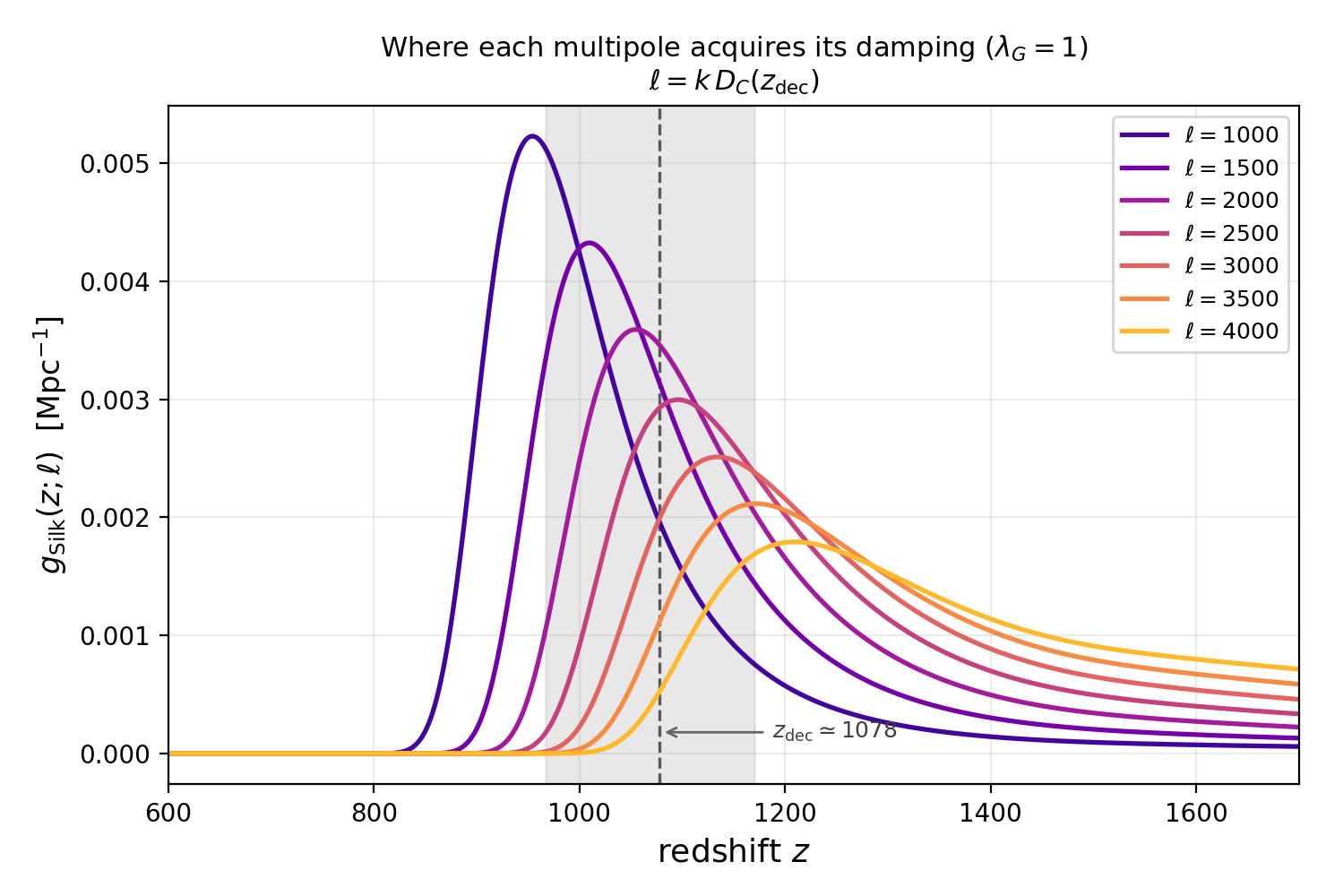}
\caption{Silk visibility $g_{\rm Silk}(z;\ell)$ for the damping-tail
  multipoles $\ell = 1000$--$3000$ at $\lG = 1$, obtained from
  $g_{\rm Silk}(z;k)$ through the projection
  $\ell = k\,D_c(z_{\rm dec})$. Higher multipoles acquire their
  damping earlier (higher $z$), high in the plasma. The dashed line
  and shaded band mark $z_{\rm dec}$. Each curve has unit area, since
  it measures the fraction of the amplitude erased per unit redshift,
  so the earlier-damped, narrower multipoles peak higher. }
\label{fig:gsilkell}
\end{figure}

The cut-off discussed in Sec.~\ref{sec:silkintegral} applies here as
well, and it is worth spelling out what it costs. Only the part of
$g_{\rm Silk}$ lying above $z_{\rm dec}$ is physical, so the total
suppression actually acquired by a mode is
$1-e^{-k^2/k_D^2(z_{\rm dec})}$ rather than unity. For the small-scale
modes that shape the damping tail the distinction is immaterial, since
almost all of their damping is already in place at decoupling, $93\%$
at $\ell = 3000$ and $83\%$ at $\ell = 2500$. It matters for the
lowest multipoles of Fig.~\ref{fig:gsilkell}, where the formal curve
peaks below $z_{\rm dec}$. These modes are only mildly damped, $25\%$
at $\ell = 1000$ and $48\%$ at $\ell = 1500$, the remainder of the
formal curve corresponding to damping that never takes place.

\begin{figure}[h]
\centering
\includegraphics[width=\columnwidth]{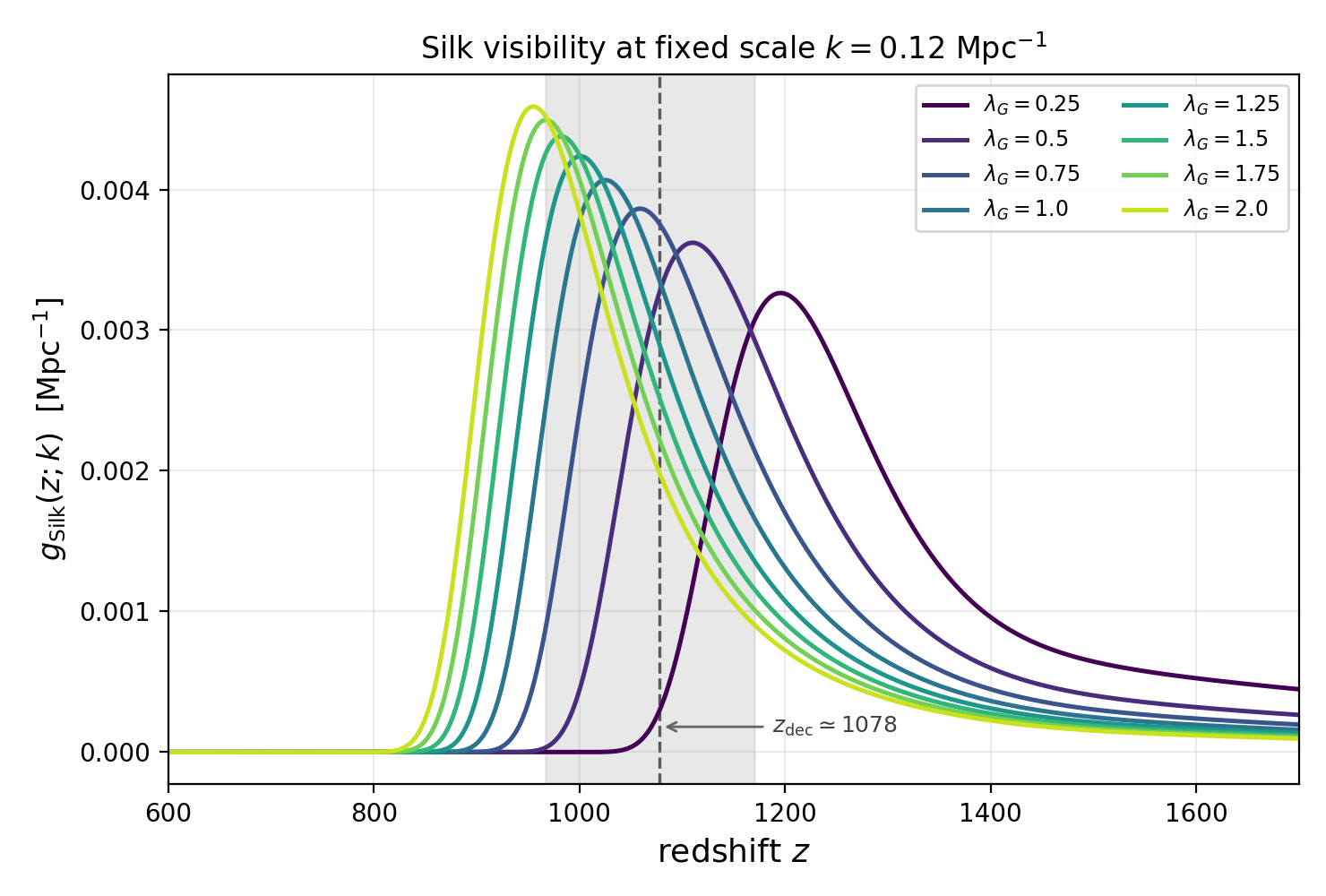}
\caption{Silk visibility $g_{\rm Silk}(z;k)$ at a fixed comoving scale
  $k = 0.12\ {\rm Mpc}^{-1}$ for the eight models. A larger $G$
  (shorter mean free path) moves the damping of this mode to slightly
  higher redshift and narrows its window. $z_{\rm dec}$ for $\lG = 1$
  is marked (dashed line, shaded band). }
\label{fig:gsilklambda}
\end{figure}

The location of the peak is the physically interesting quantity. It is
most useful in multipole space, since it is the angular spectrum
$C_\ell$ that we analyse below. Each mode projects onto
$\ell = k\,D_c(z_{\rm dec})$, giving $g_{\rm Silk}(z;\ell)$
(Fig.~\ref{fig:gsilkell}). Following the argument we began with, a
high-$\ell$ (small-scale) mode is erased early, high in the plasma
where $\lmfp$ is still short, while a low-$\ell$ mode is damped only
near decoupling, or, for the lowest multipoles shown, only partly
damped by then as discussed above. Every multipole therefore has its
own \emph{damping surface}, a characteristic redshift and a finite
width, in the same way that last scattering has its $z_{\rm dec}$ and
its thickness $\Delta z_{\rm dec}$. This ordering, small scales damped
earlier, is what shapes the damping tail whose suppression we fit in
Sec.~\ref{sec:tt}. Increasing $G$ shortens the mean free path, so a
given mode is damped slightly earlier and over a narrower window
(Fig.~\ref{fig:gsilklambda}). This is the physical origin of the
$G$-dependence we quantify next.

It is worth comparing the full result given by Eq.~(\ref{eq:silk})
with the estimate one obtains by counting scatterings over a single
Hubble time. The ratio of Eq.~\eqref{eq:identity},
$\Gamma_{\rm T}/H = R_H/\lmfp$, is precisely the number $N$ of Thomson
scatterings a photon undergoes per $e$-fold of expansion. A random
walk of $N$ steps of length $\lmfp$ produces a net displacement
$\sqrt{N}\,\lmfp$, so that
\begin{equation}
r_{\rm damp} \;\sim\; \sqrt{N}\,\lmfp \;=\; \sqrt{\lmfp R_H}\,,
\label{eq:rich_estimate}
\end{equation}
the geometric mean of the two lengths that cross in
Fig.~\ref{fig:mfp}. This is the classic single-epoch estimate of the
Silk scale \citep[see e.g.][]{HuSugiyama1995,Rich2015}. What is
idealised here is that $\lmfp$ and $R_H$ are read off at the single
epoch $z_{\rm dec}$ and that the diffusion is allowed one Hubble time
at that fixed mean free path.

At $z_{\rm dec}$ and $\lG = 1$ this gives $N \simeq 11$ scatterings
per $e$-fold, with $\lmfp \simeq 19$~Mpc and $R_H \simeq 211$~Mpc in
comoving units, hence $r_{\rm damp} \simeq 63$~Mpc (see
Fig.~\ref{fig:mfp}), against the $46$~Mpc returned by the full
integral of Eq.~\eqref{eq:silk}. The estimate~\ref{eq:rich_estimate}
captures the scale but overshoots it by about a third, because the
true diffusion length is built up over a range of redshifts during
which $\lmfp$ was shorter than its value at decoupling. That
accumulated history is also where a substantial part of the
$G$-dependence resides, as we now show.

\subsection{Scaling of the damping multipole $\ldamp$ with
  $G$}
\label{sec:silkscaling}

In this section, we establish how the diffusion length $d_{\rm damp}$,
and hence $r_{\rm damp}$, depends on $G$. We then deduce the
dependence of the damping multipole $\ell_{\rm damp}$ with $G$. The
two microphysical ingredients of the Silk integrand are shown in
Figs.~\ref{fig:lmfpRH} and \ref{fig:fR}. The combination $\lmfp R_H$
(Fig.~\ref{fig:lmfpRH}) rises steeply through recombination as the
mean free path lengthens. At fixed redshift within the shell a larger
$G$ leaves a shorter mean free path, for the reasons set out after
Eq.~\eqref{eq:lmfp}. The baryon-loading factor $f_R(z)$
(Fig.~\ref{fig:fR}) contains \emph{no} $G$ at all. It is a function of
$R = 3\rho_b/4\rho_\gamma$ alone, and that ratio of two densities at a
given temperature is fixed by the baryon-to-photon ratio,
$R \propto \eta_b/\theta$. The gravitational constant is a fundamental
constant and $\eta_b$ a cosmological parameter, logically independent
of it, so $f_R(z)$ may be treated as a fixed weight.

Before carrying out the integral we write the expansion rate in the
variables of Eq.~\eqref{eq:lmfp_micro}. Above $z_{\rm dec}$ only
matter and radiation contribute, so that, using
$\rho_{m,0} = \mu n_{\gamma,0}$ together with
$n_{\gamma,0} = \eta_b^{-1}\,n_{b,0}$ and
$G\mu = \aG\,c^2\lambdabar_e$,
\begin{equation}
H(z) = \sqrt{\frac{16\zeta(3)}{3\pi}}\;\frac{c}{\lambdabar_e}\;
\alpha^{3}\,\theta_0^{3/2}\,\aG^{1/2}\;(1+z)^{3/2}
\left(1+\frac{1+z}{1+z_{\rm eq}}\right)^{1/2}\,,
\label{eq:Hmicro}
\end{equation}
the bracket being the one already met in Eq.~\eqref{eq:alphaG} and
$z_{\rm eq}$ the invariant of Sec.~\ref{sec:framework}. The prefactor
carries no density parameter, and it reads as the electron Compton
frequency $c/\lambdabar_e \simeq 7.8\times10^{20}$~s$^{-1}$ reduced by
three very small dimensionless factors,
$\alpha^3 \simeq 3.9\times10^{-7}$ for electromagnetism,
$\theta_0^{3/2} \simeq 2.5\times10^{-8}$ for the temperature of the
background and $\aG^{1/2} \simeq 1.1\times10^{-25}\,\lG$ for gravity, the
last carrying the whole $G$-dependence. Their product is
$1.1\times10^{-39}\,\lG$, which is what separates the atomic clock from
the cosmological one.

Inserting this and $n_e = \xe(1-Y_{\rm He})n_{b,0}(1+z)^3$ into the
integral \eqref{eq:silk}, with $\sigma_T n_{b,0}$ read off
Eq.~\eqref{eq:lmfp_micro}, gives the damping length in terms of a
single dimensioned quantity, namely the reduced Compton wavelength of
the electron $\lambdabar_e$,
\begin{equation}
r_{\rm damp} = 1.502\;\lambdabar_e\;
\eta_b^{-1/2}\,\alpha^{-11/2}\,\theta_0^{-9/4}\,\aG^{-1/4}\;\mathcal{K}\,,
\label{eq:rdamp}
\end{equation}
where the dimensionless kinetic factor $\mathcal{K}$ is given by
\begin{equation}
\mathcal{K}^2 = \frac{1}{1-Y_{\rm He}}\int_{z_{\rm dec}}^{\infty}\!
f_R(z)\,
\frac{(1+z)^{-4}\,\dd z}
{\sqrt{(1+z_{\rm eq})^{-1} + (1+z)^{-1}}\;\xe(z)}\,.
\label{eq:Kfactor}
\end{equation}
The prefactor of Eq.~\eqref{eq:rdamp} contains no density parameter
and no Hubble rate. It is built from the fundamental constants
$\lambdabar_e$, $\alpha$ and $\aG$ together with two cosmological
parameters of an elementary kind, the baryon-to-photon ratio $\eta_b$
and the present temperature through $\theta_0$. The only other
cosmological parameter anywhere in Eq.~\eqref{eq:rdamp} is
$z_{\rm eq}$, inside $\mathcal{K}$, and it happens to be an invariant
of this family of models. The
geometric exponent $\aG^{-1/4} \propto \lG^{-1/2}$ therefore appears
directly, and the same $\lambdabar_e$ that sets the mean free path
sets the damping length.

The factor $\mathcal{K}$ also depends on $G$. The ionization history
$\xe(z)$ responds to $G$ through the kinetic effects of
Sec.~\ref{sec:recombination}, and $Y_{\rm He}$ through the BBN
freeze-out of Table~\ref{tab:sensitivities}. The main dependence comes
from $\xe$, and because a larger $\lG$ leaves $\xe$ higher throughout
the decade above $z_{\rm dec}$, $\mathcal{K}$ \emph{decreases} as $G$
increases. Evaluating the logarithmic derivative at $\lG = 1$ in the sense of
Eq.~\eqref{eq:kappa_def}, with the absolute normalisation obtained at
$\lG = 1$ and the normalisation of $\mathcal{K}$ following from
Eq.~\eqref{eq:rdamp},
\begin{equation}
\mathcal{K} \propto \lG^{-0.196}
\quad\text{and}\quad
r_{\rm damp} \simeq 46.9\ {\rm Mpc}\;\lG^{-0.5-0.196}\,.
\label{eq:rd}
\end{equation}
These are local slopes, and the three closely spaced models determine them to
better than $0.001$,
so their precision is not a limiting factor anywhere in what follows. They are
not, however, the exponents of a single power law over the whole family: such a
law reproduces the computed $r_{\rm damp}$ to $4.3\%$ across
$\lG \in [0.25,2]$, but its residuals follow a systematic U-shaped pattern
rather than scattering randomly, and it returns $\mathcal{K} \propto
\lG^{-0.243}$ instead. Section~\ref{sec:wide} takes up that difference and
shows how little it matters within a few percent of $\lG = 1$.

Having the scaling of $r_{\rm damp}$, we project onto a multipole. The
damping length enters the observed spectrum through its angular
projection, and the damping multipole is
\begin{equation}
\ldamp \;=\; 2\pi\,\frac{D_c(z_{\rm dec})}{r_{\rm damp}}\,,
\label{eq:ldamp_def}
\end{equation}
with $D_c(z_{\rm dec})$ the comoving distance to last scattering,
equal to $(1+z_{\rm dec})D_A(z_{\rm dec})$ for any background, and
expressed in the same comoving convention as $r_{\rm damp}$. The
projection brings in one further scaling, that of this distance,
$D_c(z_{\rm dec}) \;\propto\; \lG^{-0.9997 \pm 0.0003}$. The exponent
is $-1$ to four digits, and its uncertainty is two orders of magnitude
smaller than those of Eq.~\eqref{eq:rd} because nothing but geometry
enters here. The residual $3\times10^{-4}$ is not fit noise but the
systematic drift of the upper bound $z_{\rm dec}(\lG)$, which makes
$D_c\,\lG$ creep by $0.07\%$ across the range instead of staying
strictly constant. This is the pure geometric scaling of a $c/H$
integral already met in Sec.~\ref{sec:equality}.

Combining it with Eqs.~\eqref{eq:rd} through
Eq.~\eqref{eq:ldamp_def},
\begin{align}
  \ldamp \;&\propto\;
             \underbrace{
             \underbrace{\lG^{-1.000}}_{D_c}\;\times\;
             \underbrace{\lG^{+0.500}}_{\text{frozen }\rdamp}
             }_{\text{geometric factor}}\;\times\;
             \underbrace{\lG^{+0.196}}_{\substack{\text{kinetic factor }\mathcal{K}\\[2pt]
  \text{(microphysics)}}}\nonumber\\[12pt]
           &\simeq\; 1859\;\lG^{-0.304}\,.
\label{eq:decomposition}
\end{align}
This is the central result of the paper, and the exponent is a logarithmic
derivative at $\lG = 1$, in the sense of Eq.~\eqref{eq:kappa_def}. We quote it
in that form deliberately. Any observational constraint on $G$ from the CMB
lives within a few percent of $\lG = 1$, so the local slope is the number an
observer needs, and it is also the regime in which the comparison of
Sec.~\ref{sec:finaldiscussion} between the recombination-epoch $G$ and the
laboratory one is meaningful at all.

Throughout, we call geometric factor the product of the first two
factors of Eq.~\eqref{eq:decomposition}, the distance and the
$G$-independent part of $\rdamp$, which together give
$\lG^{-1/2}$. The decomposition then separates geometry from
microphysics and makes the central point of the paper quantitative:
the kinetic response of $\mathcal{K}$ through $x_e(z)$ is not a
correction but a substantial part of the effect, taking the exponent
from the $-0.500$ of the geometric factor to $-0.304$.

Two caveats attach to the number itself, and neither is a fit
uncertainty. The derivative is determined to better than $0.001$ by the
three closely spaced models, and the two independent estimators of
Sec.~\ref{sec:sensitivities} agree with it to $0.0001$, so the
precision of the determination is not the limitation. What the
exponent does depend on is the definition of the last-scattering
surface: Eq.~\eqref{eq:decomposition} uses the visibility peak, and
adopting $\tau = 1$ instead gives $-0.246$, as
Sec.~\ref{sec:silkintegral} sets out. We retain the visibility peak
for the reasons given there, and quote the alternative rather than
fold the difference into an error bar, since it is a matter of
convention and not of precision. The second caveat is that a single
exponent describes only the neighbourhood of $\lG = 1$, meaning
$|\ln\lG| \ll 1$. There the linear form is excellent, misstating
$\ldamp$ by $7\times10^{-6}$ at $\lG = 1.01$ and by $0.06\%$ at
$\lG = 1.1$, but over the whole family the curvature is appreciable,
and we return to it in Sec.~\ref{sec:wide}.

It is worth writing $\ldamp$ itself in fundamental
constants. Combining Eq.~\eqref{eq:ldamp_def} with
Eq.~\eqref{eq:rdamp}, and using
$c/H_0 \propto \lambdabar_e\,\alpha^{-3}\,\theta_0^{-3/2}\,\aG^{-1/2}$
up to the density parameters, the Compton wavelength cancels between
the distance and the damping length and
\begin{equation}
\ldamp \;\propto\; \eta_b^{1/2}\,\alpha^{5/2}\,\theta_0^{3/4}\,
\aG^{-1/4}\,\mathcal{K}^{-1}\,,
\label{eq:ldamp_micro}
\end{equation}
the omitted factor being the dimensionless distance
$D_c(z_{\rm dec})H_0/c$, a function of the density parameters and of
$z_{\rm dec}$, together with the numerical coefficients. It equals
$5.1$ for our fiducial cosmology, and since the density parameters are
the same in every model of the family it takes the same value
throughout, which is why it does not enter any of the scalings quoted
here. With the microphysics frozen, that is at fixed
$\mathcal{K}$, the damping multipole therefore scales as
$\aG^{-1/4}$. The same $\alpha_G$ that controls the kinetic lag
through Eq.~\eqref{eq:alphaG}, where it enters as $\alpha_G^{-1/2}$,
thus also controls the geometric factor, and no other gravitational
combination appears. Equation~\eqref{eq:decomposition} is governed
throughout by this single gravitational coupling, its geometric factor
through $\aG^{-1/4}$ and its kinetic factor $\mathcal{K}$ through the
$\aG^{-1/2}$ of Eq.~\eqref{eq:alphaG}.

One further combination does enter, but through a different
interaction and with a more modest weight. It acts on the kinetic
factor $\mathcal{K}$ of Eq.~\eqref{eq:rd}, through the
$(1-Y_{\rm He})$ weighting of the electron density in
Eq.~\eqref{eq:Kfactor}. The helium fraction is set by the freeze-out
of the neutron-to-proton ratio, that is by the competition between the
weak rate $\Gamma_{\rm w} \sim G_F^2T^5$ and the expansion, which
brings in
\begin{equation}
\mathcal{G}_{\rm w} \equiv
\left(\frac{G_F}{(\hbar c)^3}\right)^{\!2}(Qc^2)^3\sqrt{\frac{\hbar c^5}{G}}
\;\simeq\; 3.6\,,
\label{eq:Gw}
\end{equation}
which is dimensionless, $G_F/(\hbar c)^3$ having the dimension of an
inverse energy squared and $\sqrt{\hbar c^5/G}$ being the Planck
energy. Here $Q = m_n - m_p$ is the neutron-proton mass difference,
the only mass scale of the problem. The weak freeze-out, which we label $T_{\rm w}$ to distinguish it from
the much later nucleosynthesis itself, occurs when
$\Gamma_{\rm w} \simeq H$, that is when
$G_F^2T_{\rm w}^5 \simeq T_{\rm w}^2/\sqrt{\hbar c^5/G}$ in natural
units, which gives $T_{\rm w}^3 \propto \sqrt{G}/G_F^2$ and therefore
\begin{equation}
k_BT_{\rm w} = Qc^2\;\mathcal{G}_{\rm w}^{-1/3} \;\propto\; G^{1/6}\,,
\label{eq:Tf}
\end{equation}
the relation being an identity rather than a proportionality once the
freeze-out condition is written in units of $Q$. With
$\mathcal{G}_{\rm w} \simeq 3.6$ this gives
$k_BT_{\rm w} \simeq 0.84$~MeV, the standard value. A larger $G$
raises $Y_{\rm He}$ twice over, through the earlier freeze-out and
through the shorter time left for neutrons to decay before
nucleosynthesis. Adding the two gives
$\partial\ln Y_{\rm He}/\partial\ln G \simeq 0.3$, close to the value
of \citet{DSW2007} quoted in Table~\ref{tab:sensitivities}.

The prediction of Eq.~\eqref{eq:decomposition} can be confronted
directly with the damping scale that \CLASS{} computes
internally. \CLASS{} outputs its own comoving damping length
$r_{\rm damp}$ (from the same Silk integral, evaluated at the
visibility peak). Projecting it through Eq.~\eqref{eq:ldamp_def} with
$D_c$ taken from the background gives an $\ldamp$ that requires no
spectral fit at all. Across the family this reproduces
$\kappa_{\ldamp} = -0.304$ at $\lG = 1$, in agreement with
Eq.~\eqref{eq:decomposition}, and $-0.257$ when forced into a single power law
over the whole family. This route confirms the exponent without
any modelling of the tail, which is its value; it is not however
independent of Eq.~\eqref{eq:silk}, for the reason given in
Sec.~\ref{sec:silkintegral}.

\subsection{The critical value of $G$}
\label{sec:wide}

Equation~\eqref{eq:decomposition} is a statement about the neighbourhood of
$\lG = 1$. The wide family serves a different purpose, and two in fact. It is
what makes the power-law behaviour identifiable in the first place, and what
gives the lever arm needed to test the analytic normalisation of
Eq.~\eqref{eq:ldamp_micro}, whose exponents in $\eta_b$, $\alpha$, $\theta_0$
and $\aG$ hold for any $\lG$ and could not be checked on a narrow scan. The
same is true of the exact invariance of $\ell_{\rm eq}$, of the response of
$Y_{\rm He}$, which runs from $0.048$ to $0.370$ and whose consequences occupy
Sec.~\ref{sec:residual}, and of the approach to Saha equilibrium discussed
below.

It also lets us say how fast the linear description degrades. Forcing a single
power law on the whole family gives $\ldamp \propto \lG^{-0.257}$, with
residuals reaching $4.7\%$ and following a systematic U-shaped pattern rather
than scattering randomly, so the exponent is not constant. Allowing it to
drift, that is writing
\begin{equation}
\ldamp \;\propto\; \lG^{\,n + b\ln\lG}\,,
\label{eq:running}
\end{equation}
absorbs the
curvature almost entirely: fitting this form gives $n = -0.301$, recovering
Eq.~\eqref{eq:decomposition} to better than one percent as it must, together
with $b = -0.069$, and the maximum residual drops from $4.7\%$ to $0.17\%$, a thirtyfold improvement for a
single extra parameter. The effective exponent $n + 2b\ln\lG$ then steepens
from $-0.11$ at $\lG = 0.25$ to $-0.40$ at $\lG = 2$. What matters in practice,
however, is not the exponent but the predicted $\ldamp$, and there the linear
description is far more robust than the drift of the exponent suggests, since
the neglected term is $b(\ln\lG)^2$. Using the single exponent of
Eq.~\eqref{eq:decomposition} rather than the running form misstates $\ldamp$ by
$7\times10^{-6}$ in relative terms at $\lG = 1.01$, by $0.02\%$ at $\lG = 1.05$,
by $0.06\%$ at $\lG = 1.1$, and reaches $1\%$ only at $\lG = 1.5$. The linear
form is therefore amply sufficient for any observationally relevant departure,
whereas the range-averaged $-0.257$ is already off by $15\%$ at $\lG = 1$ and
should not be used there.

This drift also tells us how
specific to our own Universe the result is. A treatment that discards
the kinetics altogether, such as that of \citet{Rich2015}, fares
better for smaller $\aG$ and worse for larger $\aG$. What still needs
to be specified is the reference scale against which $\aG$ is small,
that is, the values of $\aG$ at which the kinetic effects become
comparable to the geometric ones. That
reference value follows from Eq.~\eqref{eq:alphaG}. Equilibrium
requires not $\Gamma_{\rm rec} \gg H$ but
$\Gamma_{\rm rec} \gg (B_1/2k_BT)\,H$, and the reason is that the Saha
fraction is not a fixed target. For $\xe \ll 1$, Eq.~\eqref{eq:saha}
gives $\xe^{\rm eq} \propto e^{-B_1/2k_BT}$, so the equilibrium value
itself falls at the rate
\begin{equation}
\left|\frac{\dd\ln \xe^{\rm eq}}{\dd t}\right| \simeq
\frac{B_1}{2k_BT}\left|\frac{\dd\ln T}{\dd t}\right| =
\frac{B_1}{2k_BT}\,H\,,
\label{eq:target}
\end{equation}
the exponential supplying the large factor $B_1/2k_BT \simeq 22$ at
recombination and the expansion supplying $\dd\ln T/\dd t =
-H$. The same correction arises wherever an equilibrium abundance is
exponentially sensitive to temperature, and its size is the exponent
itself. At the freeze-out of the neutron-to-proton ratio the
equilibrium value is $e^{-Qc^2/k_BT}$ with $Qc^2/k_BT \simeq 1.5$ at
$k_BT_{\rm w} \simeq 0.84\,$MeV, so there the factor is of order unity
and the naive condition $\Gamma_{\rm w} \simeq H$ of
Eq.~\eqref{eq:Tf} is adequate. The deuterium bottleneck is the other
case, with the deuterium binding energy $B_D = 2.22\,$MeV giving
$B_D/k_BT \simeq 30$ at $k_BT \simeq 0.07\,$MeV, which is one
reason nucleosynthesis itself is computed from a reaction network and
not from a threshold condition. Recombination belongs with the second
kind. Tracking equilibrium therefore demands that recombination outrun
this moving target rather than merely outrun the expansion, so the
controlling quantity is
\begin{align}
\mathcal{Q} \equiv \frac{4\theta\,\Gamma_{\rm rec}}{H}
= \;&\frac{64\sqrt{2\zeta(3)}}{3\pi}\,C_{\rm P}\,\xe\,
(1-Y_{\rm He})\,\eta_b \nonumber\\
&\times\alpha^5\,\theta^2\,\phi_B(\theta)
\left(1+\frac{1+z}{1+z_{\rm eq}}\right)^{-1/2}\aG^{-1/2}\,,
\label{eq:Qcrit}
\end{align}
having used $B_1/2k_BT = 1/4\theta$. Evaluating Eq.~\eqref{eq:Qcrit}
at $z_{\rm rec}$ for our own Universe gives $\mathcal{Q} \simeq 0.29$,
so that we sit just on the out-of-equilibrium side of $\mathcal{Q}=1$,
which is why the kinetic term is comparable to, but not larger than,
the geometric one. We evaluate $\mathcal{Q}$ at $z_{\rm rec}$ because
that is where the paper defines the recombination epoch, but the
choice is not a delicate one. Across the full width of the transition,
from $\xe = 0.9$ to $\xe = 0.1$, $\mathcal{Q}$ stays between $0.29$
and $0.42$, so no reasonable alternative marker would move it by as
much as a factor of two.

Locating the boundary itself requires more care, because the explicit
$\aG^{-1/2}$ of Eq.~\eqref{eq:Qcrit} is not the whole of the
$G$-dependence.  The Peebles factor rises with $G$ through
$\Lambda_\alpha \propto H$, as Eqs.~\eqref{eq:CP_exact} to
\eqref{eq:CP_alpha} make explicit, and $z_{\rm rec}$ itself drifts, so
that evaluating $\mathcal{Q}$ at each model's own $z_{\rm rec}$ gives
\begin{equation}
\mathcal{Q} \simeq 0.29\;\lG^{-0.397+0.124\ln\lG}\,,
\label{eq:Qscaling}
\end{equation}
in the running form of Sec.~\ref{sec:wide}, which here reproduces the
whole family to an rms of $0.03\%$ where a single power law is in
error by $5\%$. Its index at $\lG = 1$ is $-0.397$, against the $-0.395$ that the
close triplet gives with no fit at all, the same consistency as for
$\ldamp$. The residual $0.002$ is the truncation of the running form
at second order, and it disappears if one allows more: fitting a
quartic in $\ln\lG$ over the same family returns $-0.3954$, and a
quintic $-0.3952$.

We now give the budget of the $G$-dependence of each
factor of Eq.~\eqref{eq:Qcrit}, evaluated at the model's own
$z_{\rm rec}$, whose drift is
$\dd\ln(1+z_{\rm rec})/\dd\ln\lG = -0.036$,
\begin{equation}
\begin{array}{lr}
\xe = 1/2 \ \text{by definition of } z_{\rm rec} & 0.000\\
\eta_b,\ \alpha^5 \ \text{(invariant)}           & 0.000\\
(1-Y_{\rm He})                                   & -0.234\\
\theta^2                                         & -0.072\\
\phi_B(\theta)                                   & +0.010\\
\big[1+(1+z)/(1+z_{\rm eq})\big]^{-1/2}          & +0.005\\
\aG^{-1/2} = \lG^{-1}                            & -1.000\\
C_{\rm P}                                        & +0.897\\
\hline
\text{total}                                     & -0.395
\end{array}
\label{eq:Qbudget}
\end{equation}
The last entry carries the result. At \emph{fixed} redshift the
Peebles factor contributes $+0.45$ (Sec.~\ref{sec:peeblesfactor}), but
$\mathcal{Q}$ is evaluated at each model's own $z_{\rm rec}$, where
$C_{\rm P}$ also inherits the temperature drift through
$\beta_B \propto e^{-B_2/k_BT}$, with $B_2/k_BT = 1/8\theta \simeq
11.3$ at recombination. A $3.6\%$ drop in $1+z_{\rm rec}$ adds a
further $+0.45$, doubling the contribution. That one effect is what
makes Eq.~\eqref{eq:Qscaling} so much shallower than the explicit
$\lG^{-1}$. The budget closes on the direct evaluation of
Eq.~\eqref{eq:Qcrit} to better than $10^{-4}$.

Nothing in this depends on the case-B expression, since $\mathcal{Q}$
is built from the product $C_{\rm P}\alpha_B$ and
$C_{\rm P} \propto 1/\beta_B \propto 1/\alpha_B$ at
recombination. Replacing Seaton by the \citet{Pequignot1991} fit moves
$\mathcal{Q}(\lG=1)$ by $0.02\%$ and its index in the fourth decimal.

Setting $\mathcal{Q}=1$ in Eq.~\eqref{eq:Qscaling} places the boundary
at $\lG^{\rm crit} \simeq 0.14$, that is
$\aG^{\rm crit} \sim 3\times10^{-52}$, that is
$G^{\rm crit} \simeq 1.4\times10^{-9}$ in cgs units, a factor of
$49$ below the laboratory value. It is worth being explicit about what
this number is made of, since the form of Eq.~\eqref{eq:Qcrit} invites
the reading that it is built from fundamental constants alone. Setting
$\mathcal{Q}=1$ there, with $\xe = 1/2$, gives
\begin{align}
  \left(\aG^{\rm crit}\right)^{1/2} &=
  \frac{32\sqrt{2\zeta(3)}}{3\pi}\;\alpha^5\,\eta_b\;
  C_{\rm P}^{\rm crit}\left(1-Y_{\rm He}^{\rm crit}\right)\nonumber\\
  &\quad\times\left(\theta_{\rm rec}^{\rm crit}\right)^2
  \phi_B\!\left(\theta_{\rm rec}^{\rm crit}\right)
  \left(1+\frac{1+z_{\rm rec}^{\rm crit}}{1+z_{\rm eq}}\right)^{1/2}\,,
  \label{eq:aGcrit}
\end{align}
where a superscript ``crit'' denotes evaluation in the model with
$\lG = \lG^{\rm crit}$, at its own $z_{\rm rec}$. Written this way the
relation is visibly implicit rather than explicit.
$C_{\rm P}^{\rm crit}$ carries $(\aG^{\rm crit})^{1/2}$ itself through
Eq.~\eqref{eq:CP_alpha}, $Y_{\rm He}^{\rm crit}$ is the BBN yield at
that value of $G$, and $\theta_{\rm rec}^{\rm crit}$ is the
recombination temperature of that same model. We therefore locate the
boundary numerically from Eq.~\eqref{eq:Qscaling}, and read
Eq.~\eqref{eq:aGcrit} only for which quantities enter. Of those,
$\alpha$ and $\phi_B$ are atomic, and exactly two cosmological numbers
appear. One is the baryon-to-photon ratio $\eta_b$, which enters
explicitly and also fixes $Y_{\rm He}^{\rm crit}$ and
$\theta_{\rm rec}^{\rm crit}$. The other is the mean mass per photon
$\mu$, which arrives through $(1+z_{\rm rec})/(1+z_{\rm eq})$, a ratio
of temperatures and therefore free of $T_0$. The two cannot be reduced
to one, since $\mu = m_p\eta_b + m_{\rm cdm}\eta_{\rm cdm}$ and the baryons
supply only $16\%$ of it. Of the quantities that $\eta_b$ controls,
the temperature is the mildest, since
it is reached only through a logarithm: the Saha condition of
Sec.~\ref{sec:saha} gives $B_1/k_BT \simeq 42$, half of which is the
logarithm of $\eta_b$ and half purely atomic, so that
$\dd\ln\theta_{\rm rec}/\dd\ln\eta_b \simeq 0.025$ and a factor of two
in $\eta_b$ moves $\theta_{\rm rec}$ by under $2\%$. It is the
explicit factor of $\eta_b$, which is linear, that decides the matter.
Converting $\aG^{\rm crit}$ into $G^{\rm crit}$ brings in $\mu$ as
well. The critical value of the gravitational constant $G^{\rm crit}$
is therefore not a statement about fundamental constants alone. It
depends on the baryon-to-photon ratio and on the mean mass per photon,
and Eq.~\eqref{eq:aGcrit_explicit} below makes that dependence
explicit, so it can be evaluated for any universe and not only for
ours. The dependence is a smooth power law,
$G^{\rm crit} \propto \eta_b^{3/2}\,\theta_{\rm Saha}^{5/4}/\mu$, in
which $\theta_{\rm Saha}$ varies only logarithmically, so a factor of
two in $\eta_b$ moves $G^{\rm crit}$ by nearly a factor of three.

One simplification is worth recording. The Sobolev channel, which is what
makes Eq.~\eqref{eq:aGcrit} implicit through $C_{\rm P}$, becomes negligible
well before the boundary is reached. Since $\Lambda_\alpha/\beta_B$ carries
$\aG^{1/2}$ while $\Lambda_{2s\rightarrow1s}/\beta_B$ carries no $\aG$ at all,
and since $\phi_B$ and the Boltzmann factor cancel between the two, their
ratio falls with $G$, the Sobolev share of the escape going from $36\%$ at
$\lG = 1$ to $9\%$ at $\lG = 0.25$ and some $5\%$ at the boundary. Dropping
it, and using $C_{\rm P} \simeq \mathcal{X}$ since
$\mathcal{X} \sim 10^{-3}$, the explicit $\phi_B$ of Eq.~\eqref{eq:aGcrit}
cancels against the $1/\phi_B$ of Eq.~\eqref{eq:CP_2g}, leaving
\begin{align}
  \left(\aG^{\rm crit}\right)^{1/2} &= 2\sqrt{6\zeta(3)}\;C_{2\gamma}\,
  \alpha^{8}\,\eta_b\left(1-Y_{\rm He}^{\rm crit}\right)
  \theta_{\rm rec}^{\rm crit}\,e^{1/8\theta_{\rm rec}^{\rm crit}}\nonumber\\
  &\quad\times\left(1+\frac{1+z_{\rm rec}^{\rm crit}}{1+z_{\rm eq}}\right)^{1/2}\,,
  \label{eq:aGcrit_closed}
\end{align}
with $2\sqrt{6\zeta(3)} = 5.371$. The case-B coefficient has disappeared
altogether, so the choice between the Seaton and \citet{Pequignot1991} forms
cannot affect $G^{\rm crit}$, and the order in the fine-structure constant is
now visible, $\alpha^8$ rather than the $\alpha^5$ of
Eq.~\eqref{eq:aGcrit}.

Two further approximations close the relation entirely. The first is that
$Y_{\rm He}^{\rm crit} \simeq 0$, since helium is barely produced at that
value of $G$ and the quantity that enters is $1-Y_{\rm He}$, bounded between
$0.753$ and unity. The second is that $\theta_{\rm rec}^{\rm crit}$ may be
taken at its Saha value, and here the argument is self-consistent rather than
approximate. The boundary is the place where the kinetics become marginal, so
recombination there is close to equilibrium by construction. The delay of the
ionization history behind the Saha curve bears this out, falling from
$\Delta z \simeq 120$ at $\lG = 2$ ($95$ at $\lG = 1$) to $\simeq 10$ at the
boundary. Writing the Saha condition at $\xe = 1/2$ with $Y_{\rm He} = 0$,
\begin{equation}
  e^{1/2\theta_{\rm Saha}}\left(\alpha^2\theta_{\rm Saha}\right)^{3/2}
  = \frac{\pi^2}{\zeta(3)\,(2\pi)^{3/2}\,\eta_b}\,,
  \label{eq:saha_crit}
\end{equation}
whose solution is logarithmic in $\eta_b$ and in $\alpha$,
\begin{equation}
  \frac{1}{2\theta_{\rm Saha}} = \ln\frac{1}{\eta_b}
  + 3\ln\frac{1}{\alpha}
  + \frac{3}{2}\ln\frac{1}{\theta_{\rm Saha}}
  + \ln\frac{\pi^2}{\zeta(3)\,(2\pi)^{3/2}}\,,
  \label{eq:saha_log}
\end{equation}
still implicit but converging in a few iterations from any starting
point, the four terms being $21.2$, $14.8$, $6.6$ and $-0.7$. This is
the form behind the statement made above that half of the $42$ is the
logarithm of $\eta_b$ and half is atomic, and it gives
$\dd\ln\theta_{\rm Saha}/\dd\ln\eta_b \simeq 0.025$. The result is
$1/2\theta_{\rm Saha} = 42.0$, that is
$z_{\rm Saha} \simeq 1379$, and noting that $1/8\theta$ is one quarter of
$1/2\theta$, so that the Boltzmann factor of
Eq.~\eqref{eq:aGcrit_closed} is the fourth root of the right-hand side of
Eq.~\eqref{eq:saha_crit}, the exponential cancels and a pure power law
remains,
\begin{align}
  \left(\aG^{\rm crit}\right)^{1/2} &=
  2^{5/8}\sqrt{6}\;\pi^{1/8}\,\zeta(3)^{1/4}\;C_{2\gamma}\;
  \alpha^{29/4}\,\eta_b^{3/4}\,\theta_{\rm Saha}^{5/8}\nonumber\\
  &\quad\times\left(1+\frac{1+z_{\rm Saha}}{1+z_{\rm eq}}\right)^{1/2}\,,
  \label{eq:aGcrit_explicit}
\end{align}
with prefactor $4.564$. It returns $\lG^{\rm crit} \simeq 0.16$, that is
$G^{\rm crit} \simeq G_{\rm N}/40$ and
$\aG^{\rm crit} \simeq 3.1\times10^{-52}$, against the $0.14$,
$G_{\rm N}/49$ and $3\times10^{-52}$ of the numerical determination. The three
approximations behind it are each at the level of a few percent and partly
cancel, so we quote the numerical value and read
Eq.~\eqref{eq:aGcrit_explicit} for its structure.

That structure is now explicit, and the temperature has ceased to be a
difficulty. Where Eq.~\eqref{eq:aGcrit_closed} carried it inside an
exponential, with a one percent error on $\theta_{\rm rec}^{\rm crit}$ moving
$(\aG^{\rm crit})^{1/2}$ by a factor of ten, it survives here only as $\theta_{\rm Saha}^{5/8}$,
and $\theta_{\rm Saha}$ itself depends on $\alpha$ and $\eta_b$ alone, logarithmically.
What is left is $\alpha^{29/4}$ and $\eta_b^{3/4}$, the quarter power of
$\eta_b$ absorbed by the Saha condition, refining the linear dependence read
off Eq.~\eqref{eq:aGcrit}, together with $z_{\rm eq}$, which is where $\mu$
enters.

What Eq.~\eqref{eq:aGcrit_explicit} does say, and this is the point
worth keeping, is that
our own Universe sits close to the boundary, which is why the kinetic
term in Eq.~\eqref{eq:decomposition} is of the same order as the
geometric one rather than negligible beside it. This remains an
extrapolation rather than a measurement, so only its order of
magnitude carries weight, but it is a short one, reaching a quarter of
a decade in $\ln\lG$ beyond the smallest model we compute over a range
that the two-parameter form describes to $0.03\%$. What does carry
weight is the shallowness of
Eq.~\eqref{eq:Qscaling}. Equilibrium is approached very slowly as $G$
decreases, which is the same statement as the slow convergence of the
ionization histories towards Saha noted in Sec.~\ref{sec:numxe}, and
it is why an equilibrium treatment remains inaccurate until $G$ is
some fifty times smaller than its present value. For $\ldamp$ we keep
the single exponent in what follows, since it is what makes the
decomposition into geometric and kinetic factors transparent, but the
running form is the more accurate description should a precise value
be needed at a given $\lG$.

Several closely related exponents have now appeared, and because they
describe different quantities over different ranges it is worth
collecting them in one place, which is the purpose of
Table~\ref{tab:exponents}. Two consistency checks are worth pointing out in
it. The index of the running fit at $\lG = 1$ is $n = -0.301$, against the
$\kappa_{\ldamp} = -0.304$ of Table~\ref{tab:sensitivities}, which is obtained
from the close triplet alone with no fit at all. The two agree to one percent
by genuinely different routes, one local and model-free, the other a curvature
fit over a factor of eight in $G$. And the apparent gap between that value and the
$-0.257$ of a single power law over the whole family is not an inconsistency
but the curvature $b$, as Sec.~\ref{sec:wide} shows. The quoted result is the
local one throughout.

\begin{table}[t]
\centering
\caption{Summary of the exponents quoted in this paper, with their
  definitions. All refer to $\lG$, and $\aG \propto \lG^2$. The first block
  holds the results, which are logarithmic derivatives at $\lG = 1$ in the
  sense of Eq.~\eqref{eq:kappa_def}, obtained from the three models
  $\lG = 0.9, 1, 1.1$ and determined to better than $0.001$. The second shows how they move when the last-scattering
  surface is defined by $\tau = 1$ rather than by the visibility peak, which
  is a matter of convention and is the dominant dependence. The third
  describes the departure from the linear regime, and the fourth gives, for
  reference only, the single power law forced on the whole family, which
  should not be used near $\lG = 1$.}
\label{tab:exponents}
\small
\setlength{\tabcolsep}{4pt}
\begin{tabular}{@{}llr@{}}
  \hline\hline
  Quantity & Definition & Value \\
  \hline
  \multicolumn{3}{@{}l}{Local slopes at $\lG=1$, the quoted results}\\
  $\kappa_{\mathcal{K}}$    & kinetic factor, Eq.~\eqref{eq:rd}  & $-0.196$ \\
  $\kappa_{\rdamp}$         & Eq.~\eqref{eq:rd}                  & $-0.696$ \\
  $\kappa_{\ldamp}$         & Eq.~\eqref{eq:decomposition}       & $-0.304$ \\
  $\kappa_{\ldamp/\ell_A}$  & ratio, Sec.~\ref{sec:finaldiscussion}   & $-0.320$ \\
  \hline
  \multicolumn{3}{@{}l}{Same, with the cut-off at $z_\ast$}\\
  $\kappa_{\rdamp}$         & Eq.~\eqref{eq:cutoff} & $-0.753$ \\
  $\kappa_{\ldamp}$         & Eq.~\eqref{eq:cutoff} & $-0.246$ \\
  $\kappa_{\ldamp/\ell_A}$  & Eq.~\eqref{eq:cutoff} & $-0.273$ \\
  \hline
  \multicolumn{3}{@{}l}{Departure from linearity, $\ldamp \propto \lG^{\,n+b\ln\lG}$}\\
  $n$            & index at $\lG=1$                   & $-0.301$ \\
  $b$            & curvature, Sec.~\ref{sec:wide}     & $-0.069$ \\
  \hline
  \multicolumn{3}{@{}l}{Single power law over $\lG\in[0.25,2]$, for reference}\\
  $\mathcal{K}$  & Sec.~\ref{sec:wide}                & $-0.243$ \\
  $\rdamp$       & Sec.~\ref{sec:wide}                & $-0.743$ \\
  $\ldamp$       & Sec.~\ref{sec:wide}                & $-0.257$ \\
  \hline
\end{tabular}
\end{table}

\subsection{Analysis of the damping tail}
\label{sec:tt}

We now turn to the simulated TT spectra themselves, and ask whether their
damping tails are consistent with the $\ldamp(\lG)$ that the Silk integral
delivers. The question has to be put in that direction, with $\ldamp$ fixed to
the value of Eq.~\eqref{eq:nocancel} and the shape of the envelope left free,
because the tail does not constrain the two independently. The envelope is a
stretched exponential $\exp[-(\ell/\ldamp)^m]$, and over a window of barely
more than a decade a smaller $\ldamp$ with a gentler $m$ is nearly
indistinguishable from a larger one with a steeper $m$. Fitting both at once
is ill-conditioned, as we quantify at the end of the section, whereas fixing
$\ldamp$ and fitting $m$ asks a well-posed question of the spectra.

Describing the tail requires knowing that its shape is not a pure Gaussian,
and the reason is the finite duration of last scattering. The envelope
$\mathcal{D}_\ell$ that multiplies the spectrum is the multipole
counterpart of the mode suppression $\mathcal{D}(k)$ of
Eq.~\eqref{eq:huwhite}, the two being related by the projection
$\ell = k\,D_c(z_{\rm dec})$. Were last scattering instantaneous,
$\mathcal{D}(k)$ would reduce to the single Gaussian
$\exp[-k^2/k_D^2(z_{\rm dec})]$ and the projected envelope would be
exactly $\exp[-(\ell/\ldamp)^2]$ with $\ldamp = 2\pi D_c/\rdamp$. We
have checked this explicitly, and fitting the projected Gaussian returns
an exponent of $2.000$ and $\ldamp = 1857$, which is precisely
$2\pi D_c/\rdamp$ for $\lG = 1$. Averaging over the visibility, as
Eq.~\eqref{eq:huwhite} does, superposes Gaussians of different widths
and flattens the result, so that the envelope is better described by
\begin{equation}
\mathcal{D}_\ell \;=\; \exp\!\left[-(\ell/\ldamp)^m\right]
\label{eq:envelope}
\end{equation}
with an exponent $m$ smaller than $2$. Repeating the fit on
$\mathcal{D}(k)$ itself gives $m = 1.45$. The exponent therefore
measures the departure from instantaneous last scattering, whose
finite duration is quantified by $\Delta z_{\rm dec}$ in
Sec.~\ref{sec:decoupling}, as already noted by \citet{HuWhite1997}.

What we ask of the spectra is then which value of $m$ makes the fitted
envelope reproduce, model by model, the same $\ldamp$ that \CLASS{}
delivers through Sec.~\ref{sec:silkscaling}.

To extract $\ldamp$ from a spectrum we need the whole tail, not the
envelope alone, since the acoustic oscillations are still present
there. We therefore adopt the analytic tail model of \citet{HFZT2001},
also used by \citet{Rich2015}, which collects the ingredients derived
by \citet{HuSugiyama1996} and \citet{HuWhite1997},
\begin{equation}
C_\ell \;\propto\; \ell^{\,n_s-1}\;\mathcal{D}_\ell^2\;P_\ell\left[
\frac{A_\ell^2-1}{1+(\ell_A/2\ell)^6}+2\right]\,.
\label{eq:richmodel}
\end{equation}
The radiation-driving factor
$P_\ell = 1 + A\exp(-1.4\,\ell_{\rm eq}/\ell)$ is obtained from
\citet[Eq.~32]{HuWhite1997}, where the amplitude $A$ follows from the
asymptotic form of \citet[Eq.~33]{HuWhite1997},
\begin{equation}
A = 25\left(1+\tfrac{4}{15}f_\nu\right)^{-2}
\frac{(1+R_*)^{-1/2}+(1+R_*)^{-3/2}}{2}-1\,.
\label{eq:Adrive}
\end{equation}
It gives $A \simeq 12.0$ for our cosmology, with $f_\nu = 0.409$ the
neutrino fraction of the radiation density and
$R_* = R(z_{\rm dec}) \simeq 0.61$. The acoustic modulation $A_\ell$
is carried by
\begin{equation}
  A_\ell = \big[1+R_*\,T(\ell/\ell_{\rm eq})\big]
  \cos\pi\!\left(\ell/\ell_A+\delta\right)
- R_*\,T(\ell/\ell_{\rm eq})\,,
\label{eq:Aell}
\end{equation}
the projected form of the tight-coupling temperature
$[\Delta T_0+R_*\Psi]\cos(ks_*)-R_*\Psi$ of
\citet[Eq.~A1]{HuSugiyama1996}, with $T$ the matter transfer
function. The bracket in Eq.~\eqref{eq:richmodel} deserves a word,
since it is the one ingredient with no derivation behind
it. \citet{HFZT2001} describe it as an offset added to the
oscillations, meant to account crudely for projection smoothing and
for the Doppler contribution, and designed so that the spectrum
returns to $P_\ell$ above the first peak. Its denominator departs from
unity only for $\ell \lesssim \ell_A$, so it plays no role over the
tail. We have checked this by varying the exponent of the
$(\ell_A/2\ell)$ term appearing in that denominator between $2$ and
$8$, which changes $m$ by less than $0.01$.

We set $\ell_A = \ell_A(\lG)$ to the per-model value of
Sec.~\ref{sec:acoustic} (Fig.~\ref{fig:ellA}), evaluated at
$z_{\rm dec}$, rather than to a common constant, so that the acoustic
template tracks the true peak positions of each model. Together with
$\ell_{\rm eq} = 150$ and $R_*$, this fixes every ingredient of
Eq.~\eqref{eq:richmodel} except the shape exponent $m$, an overall
amplitude, $\ldamp$ and the acoustic phase $\delta$. For the latter
\citet{HFZT2001} do provide a prescription, but calibrated on the
first peaks rather than on the tail, so we prefer to let it float. The
value $\ell_{\rm eq} = 150$ is the radiation-driving scale of the
\citet{HFZT2001} template, a fitted convention of that model, and it
should not be confused with the projection $\ell_{\rm eq} = 203.9$ of
Eq.~\eqref{eq:leq}, which is the angular scale of $k_{\rm eq}$ proper.

The extraction then proceeds in two nested steps. At a trial value of
$m$ we fit each spectrum over the fixed window
$\ell \in [1250, 3000]$, letting the amplitude, $\ldamp$ and $\delta$
adjust freely. The exponent $m$ is then varied until the $\ldamp$
returned by that fit coincides with the value \CLASS{} delivers
through Sec.~\ref{sec:silkscaling}.

\begin{figure}[h]
\centering
\includegraphics[width=\columnwidth]{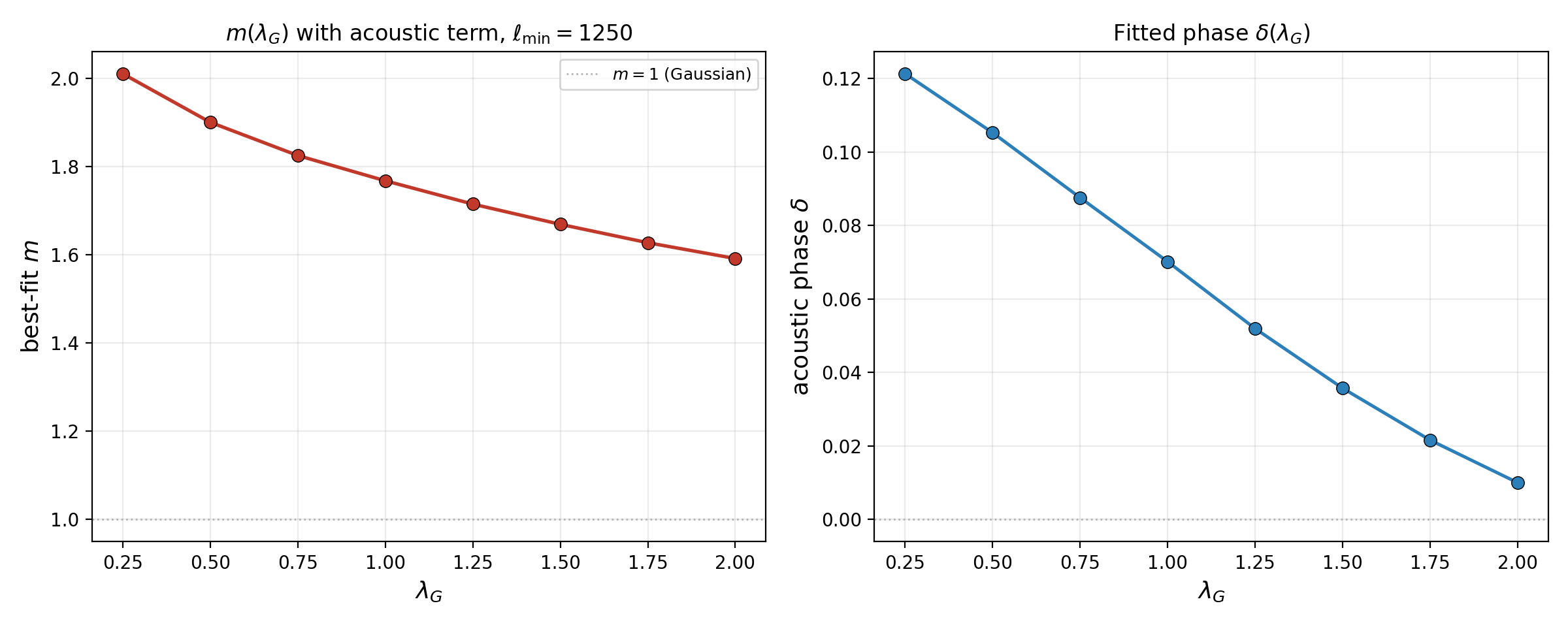}
\caption{\emph{Left}: exponent $m$ that makes the fitted tail envelope
  reproduce the \CLASS{} damping multipole $\ldamp(\lG)$, with the
  acoustic term of Eq.~\eqref{eq:richmodel} included and
  $\ell_{\rm min} = 1250$. \emph{Right}: the fitted acoustic phase
  $\delta$, which drifts smoothly with $G$. Both are extracted model
  by model. The value of $\ldamp$ is matched exactly with \CLASS{}
  value by construction.}
\label{fig:mlambda}
\end{figure}

\begin{figure}[t]
\centering
\includegraphics[width=\columnwidth]{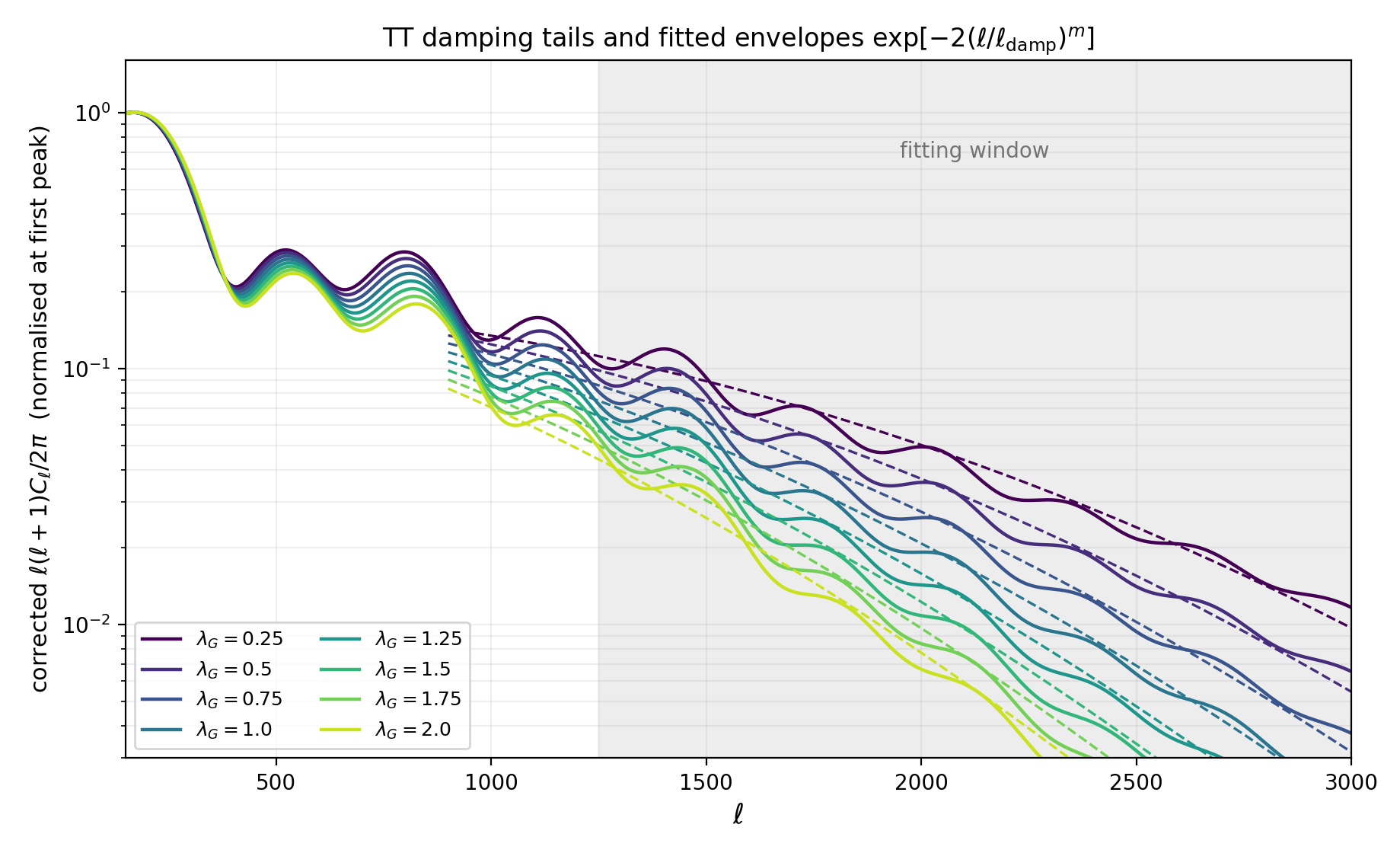}
\caption{Damping tails of the eight simulated TT spectra, divided by
  the primordial tilt, the radiation-driving envelope $P_\ell$ and the
  reionisation factor, and normalised at the first acoustic peak. The
  dashed curves are the fitted envelopes $\exp[-2(\ell/\ldamp)^m]$ of
  Eq.~\eqref{eq:envelope}, drawn with the $(m,\ldamp)$ of
  Table~\ref{tab:mldamp} and multiplied by the acoustic bracket
  averaged over the oscillations. The shaded band is the fitting
  window $\ell \in [1250,3000]$.}
\label{fig:ttdamping}
\end{figure}

The lower bound is fixed at $\ell_{\rm min} = 1250$, and we have
verified that $m$ changes by less than $1\%$ when
$\ell_{\rm min} \in [1200,1400]$. Going higher buys little and starts
to erode the fitting window, which must still contain the
high-$\ldamp$ models ($\ldamp = 2471$ at $\lG = 0.25$,
Table~\ref{tab:mldamp}). The result is shown in Fig.~\ref{fig:mlambda}
and collected in Table~\ref{tab:mldamp}, the fitted envelopes being
displayed against the spectra themselves in
Fig.~\ref{fig:ttdamping}. The exponent decreases monotonically with
$G$, from $m \simeq 2.01$ at $\lG = 0.25$ to $m \simeq 1.59$ at
$\lG = 2$, passing through $m \simeq 1.77$ at $\lG = 1$ (see
Table~\ref{tab:mldamp}). This is larger than the $m \simeq 1.45$
quoted above for $\mathcal{D}(k)$, and the difference comes almost
entirely from the constraint imposed on $\ldamp$, which is the same
degeneracy announced at the outset. Fitting $\mathcal{D}(k)$ with both free
returns $\ldamp \simeq 1630$ together with $m \simeq 1.45$, whereas forcing
$\ldamp$ to the \CLASS{} value of $1860$ raises the exponent to
$m \simeq 1.72$, within a few percent of what the full spectra
give. With the per-model $\ell_A(\lG)$ in the template, the fitted
phase $\delta$ is small and slowly varying, drifting only from $+0.12$
to $+0.01$ across the range. That a single, slowly varying $m$
reproduces the CLASS damping scale across a factor of eight in $G$,
with the acoustic oscillations explicitly modelled rather than
averaged away, is the tail-based counterpart of the semi-analytic
result of Sec.~\ref{sec:silkscaling}.

Three remarks close this section.

The degeneracy announced at the outset can now be given a number. Scanning $m$
from $1.0$ to $2.1$ with $\ldamp$ refitted at each step moves $\ldamp$ from
$949$ to $2073$, a factor of $2.2$, while the rms residual in
$\ln\mathcal{D}_\ell$ changes by only $12\%$, from $0.192$ to $0.216$. Letting
both float lands at $m = 1.18$ and $\ldamp = 1238$, better than the
constrained solution by $6\%$ in rms, which is far smaller than the residual
itself. That residual is dominated by acoustic structure the template of
Eq.~\eqref{eq:richmodel} does not reproduce, not by the envelope, which is
what makes the two parameters inseparable here. The tail therefore confirms
the exponent of $\ldamp(\lG)$ given an external normalisation, and does not
determine $\ldamp$ on its own.

That leaves the question of what a real likelihood does measure, since a
tilt of the primordial spectrum changes the slope of the tail as well, a
degeneracy already noted by \citet{Galli2009}. The
answer lies in the acoustic structure, and it is the very structure that the
template of Eq.~\eqref{eq:richmodel} fails to reproduce. A change of $n_s$ is
a smooth power law in $\ell$ and multiplies peaks and troughs alike. A change
of $\ldamp$ at fixed $\ell_A$ does not. It damps peaks and troughs unequally,
and therefore alters the peak-to-trough contrast, which no tilt can imitate.

We have checked this quantitatively, by comparing the response of the spectra
to $\lG$ with their responses to the standard parameters, each multipole
weighted by the Planck noise and by cosmic variance. The part of the $\lG$
response that no combination of $n_s$, $\omega_b$ and $A_L$ can reproduce
oscillates with the acoustic period, $\Delta\ell \simeq 320$ against
$\ell_A = 299.6$. About a seventh of the response survives marginalisation
over the six $\Lambda$CDM parameters together with $A_L$, which in a Fisher
estimate calibrated so as to return the published Planck 2018 errors on those
six parameters corresponds to $\sigma(G)/G \simeq 2\%$, consistent with the
$1.8\%$ of \citet{Lamine2025}. The lensing amplitude is nearly orthogonal to
the $\lG$ response and costs only a tenth of that error, so what limits the
measurement is the tilt rather than the lensing of the tail.

\label{sec:lensingtest}
The spectra used here are lensed, and lensing redistributes power in the very
region where the envelope is measured. Since $\ldamp$ is fixed externally,
this cannot affect $\ldamp(\lG)$, nor the ratio $\ldamp/\ell_A$ of
Sec.~\ref{sec:finaldiscussion}. It can only shift $m$, and repeating the fit on the
unlensed spectra of the same eight runs shows that it does so weakly, raising
$m$ by $2.1\%$ at $\lG = 1$ and by at most $3.2\%$ across the range, in the
expected direction since lensing transfers power to high $\ell$ and flattens
the tail.

\begin{table}[t]
\centering
\caption{Damping multipole delivered by \CLASS{} through
  Sec.~\ref{sec:silkscaling} and shape exponent $m$ that reproduces it
  when the tail is fitted with Eq.~\eqref{eq:richmodel} over
  $\ell \in [1250,3000]$.}
\label{tab:mldamp}
\begin{tabular}{lrr}
\hline\hline
$\lG$ & $\ldamp$ & $m$ \\
\hline
$0.25$ & $2471$ & $2.01$ \\
$0.50$ & $2213$ & $1.90$ \\
$0.75$ & $2015$ & $1.82$ \\
$1.00$ & $1860$ & $1.77$ \\
$1.25$ & $1731$ & $1.71$ \\
$1.50$ & $1625$ & $1.67$ \\
$1.75$ & $1535$ & $1.63$ \\
$2.00$ & $1460$ & $1.59$ \\
\hline
\end{tabular}
\end{table}

A word on the scope of what precedes. All the determinations of
$\ldamp$ in this section are made on the temperature spectrum, and the
tail model of Eq.~\eqref{eq:richmodel} is specific to it. The
polarization spectra carry a damping envelope driven by the same Silk
diffusion and cut off by the same visibility function, so one would
expect them to respond to $G$ in a closely related way, but
establishing that requires a tail model for $TE$ and $EE$ which we
have not built. We therefore make no claim about the polarization
exponent, and every result below refers to $TT$.

\section{The residual ionization}
\label{sec:residual}

\subsection{Definition and governing equation}

Recombination never goes to completion. The freeze-out of
Eq.~\eqref{eq:peebles} leaves a small residual free-electron fraction
which persists through the dark ages, and which we define
operationally as
\begin{equation}
\xeres \;\equiv\; \min_{z_{\rm reio} \lesssim z \lesssim z_{\rm rec}} \xe(z)\,,
\label{eq:xeres_def}
\end{equation}
the minimum of the ionization history between the end of recombination
and the rise of reionization. Those minima all sit just above the
onset of reionization, at $z \simeq 11$ for $\lG = 1$ and drifting
only between $z \simeq 11$ and $12$ over the whole family
$\lG \in [0.25,2]$ of $\lambda_G$ models. Once $\xe \ll 1$, well past
the recombination peak, the photoionization term of
Eq.~\eqref{eq:peebles} is exponentially suppressed and the evolution
reduces to a Bernoulli equation in $\xe$, which integrates exactly,
\begin{equation}
\frac{1}{\xe(z)} = \frac{1}{\xe(z_f)} + \int_z^{z_f}
\frac{C_{\rm P}\,\alpha_B(T_b)\,n_{\rm H}}{(1+z')\,H(z')}\,\dd z'\,,
\label{eq:bernoulli}
\end{equation}
where $z_f$ is any redshift low enough for that suppression to hold,
so that the free electrons can only be removed by recombination and
are no longer replenished by photoionization. Its precise value is
immaterial, since deep in the ionization tail the integral dominates
the term $1/\xe(z_f)$. This is the property that matters for what
follows, $\xeres$ is an \emph{integrated} quantity rather than a local
one.

\subsection{Thermal feedback and the helium offset}
\label{sec:thermal}

\begin{figure}[t]
\centering
\includegraphics[width=\columnwidth]{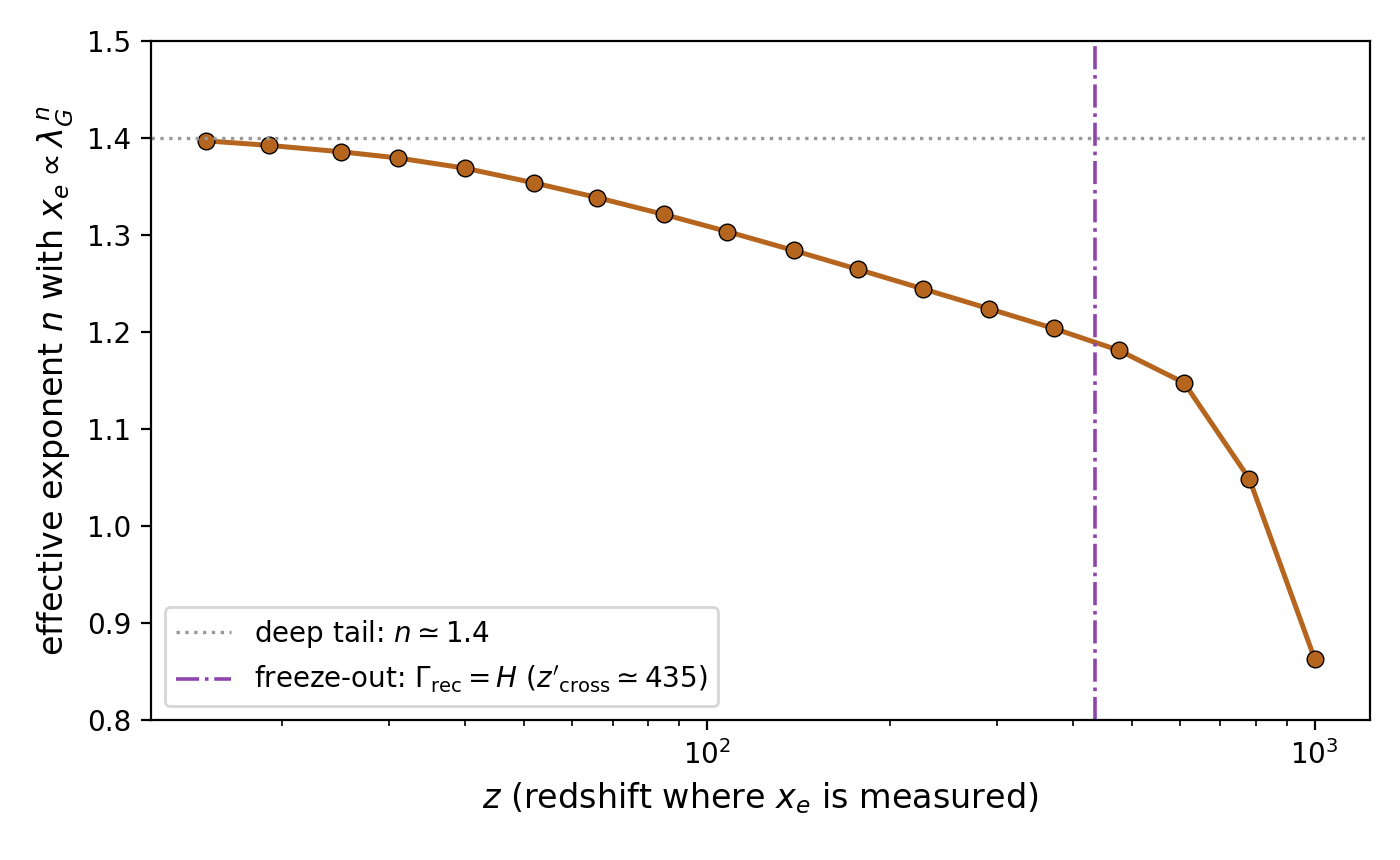}
\caption{Effective exponent $n$ of $\xe \propto \lG^{\,n}$, obtained
  by fitting the eight histories at fixed redshift, as a function of
  that redshift. The curve rises steadily into the tail, reaching
  $\simeq 1.20$ at the freeze-out redshift $z'_{\rm cross} \simeq 405$
  where $\Gamma_{\rm rec} = H$ (dot-dashed line) and $1.40 \pm 0.01$
  over $15 \lesssim z \lesssim 30$. The curve has not reached an
  asymptote at the lower end of that window, and below $z \simeq 15$
  the models enter reionization at different redshifts, so the
  single-exponent description degrades.}
\label{fig:exponent}
\end{figure}

We now look for a scaling of $x_e^{\rm res}$ with $\lambda_G$. In
order to do so, one seeks a power-law description of the integral
Eq.~\eqref{eq:bernoulli}. Fitting $\xe \propto \lG^{\,n}$ at fixed
redshift across the eight models gives an exponent $n$ that is not
constant but grows steadily as one descends into the ionization tail,
as displayed in Fig.~\ref{fig:exponent}. This exponent is
$n \simeq 0.86$ at $z = 1000$ and tends to $\simeq 1.4$ at low
redshift, in the bottom of the ionization tail. The power law is an
excellent description of the eight $\lambda_G$ models throughout the
ionization tail, the fit residuals staying below $1.5\%$ for
$z \lesssim 800$ and below $0.7\%$ around $z = 400$. It degrades only
close to recombination, where the deviation reaches $6\%$ at
$z = 1000$ so that a single exponent is correspondingly less
meaningful. The deep-tail value deserves a quantitative error bar,
since Sec.~\ref{sec:obsconseq} uses it. Over the window
$15 \lesssim z \lesssim 30$, which is below the last decade of the
tail and above the onset of reionization, we find $n = 1.40 \pm
0.01$. The formal uncertainty of the eight-point power-law fit is only
$\pm0.003$, with a maximum residual of $0.8\%$; the quoted $\pm0.01$
is dominated by the fact that $n(z)$ is still drifting upward at the
bottom of the window, from $1.381$ at $z = 30$ to $1.397$ at $z =
15$. Below $z \simeq 15$ the comparison breaks down, because the eight
models enter reionization at slightly different redshifts and the
residual rises past $3\%$. The asymptote is therefore not reached
within the computed range, and $n \simeq 1.4$ should be read as the
value at the bottom of the tail rather than as a limit.

No single exponent $n$ therefore describes the scaling of the residual
ionization tail. Two distinct effects are at work, and it is useful to
separate the \emph{offset} of $n$ above unity from its \emph{drift}
with redshift. The offset is a helium effect and the drift is a
thermal one.

Start from the baseline. In Eq.~\eqref{eq:bernoulli} the only
explicitly $G$-dependent factor is $H \propto \lG$, so freezing
everything else gives $\xe \propto \lG$, that is $n = 1$
exactly. Integrating Eq.~\eqref{eq:bernoulli} numerically with
$\alpha_B$ evaluated at a $G$-independent temperature and with a
frozen $n_{\rm H}$ indeed returns $n = 1.00$, flat across the whole
tail, which is the diagnostic quoted below.

The offset comes from $n_{\rm H} = (1-Y_{\rm He})\,n_b$. The physical
baryon density $n_b$ is invariant in our family by construction, but
$Y_{\rm He}$ is not a free parameter: it is fixed by primordial
nucleosynthesis, hence by $G$ through the freeze-out temperature of
Eq.~\eqref{eq:Tf}, and it runs from $0.048$ at $\lG = 0.25$ to $0.370$
at $\lG = 2$ (Table~\ref{tab:sensitivities}).  Since $1/\xe$
accumulates $\alpha_B n_{\rm H}/H$, one has
$\xe \propto \lG/(1-Y_{\rm He})$ and therefore
\begin{equation}
n \;\simeq\; 1 - \frac{\dd\ln(1-Y_{\rm He})}{\dd\ln\lG}
\;=\; 1 + \frac{Y_{\rm He}}{1-Y_{\rm He}}\,\kappa_{Y_{\rm He}}\,,
\label{eq:nhelium}
\end{equation}
in the limit where the integral of Eq.~\eqref{eq:bernoulli} dominates
over the initial term $1/\xe(z_f)$. A larger $G$ means more helium,
hence fewer hydrogen nuclei at fixed baryon density, hence a smaller
recombination integral and a larger residual $\xe$. This has nothing
to do with temperature; it is a counting effect, and it saturates once
$Y_{\rm He}$ is fixed by nucleosynthesis.

Equation~\eqref{eq:nhelium} is an asymptotic statement and the
comparison with the measured offset requires two corrections, both of
which we can make exactly. The first is that the exponent $n$ is a
power law fitted over the whole sampled range, so the relevant
logarithmic derivative of $1-Y_{\rm He}$ is the range-averaged
$-0.202$ and not the local $-0.234$ of
Table~\ref{tab:sensitivities}. The second is that the initial term
$1/\xe(z_f)$ carries no helium dependence of its own in our
integration, since it is taken from the computed history, so the
helium channel is diluted by the fraction
\begin{equation}
f(z) \;=\; \frac{\xe^{-1}(z)-\xe^{-1}(z_f)}{\xe^{-1}(z)}
\label{eq:dilution}
\end{equation}
of $1/\xe$ actually accumulated by the integral. The prediction is
then $-0.202\,f(z)$, and it reproduces the numerical decomposition at
every redshift:
\begin{center}
\small
\setlength{\tabcolsep}{3.5pt}
\begin{tabular}{lcccccc}
\hline\hline
$z$                 & 700 & 500 & 400 & 300 & 200 & 20 \\
\hline
$f(z)$              & 0.548 & 0.793 & 0.841 & 0.872 & 0.896 & 0.934 \\
$-0.202\,f(z)$      & 0.111 & 0.160 & 0.170 & 0.176 & 0.181 & 0.189 \\
measured            & 0.110 & 0.160 & 0.169 & 0.176 & 0.180 & 0.187 \\
\hline
\end{tabular}
\end{center}
The local form of Eq.~\eqref{eq:nhelium}, which would give $+0.24$,
overpredicts for both reasons at once, and the asymptotic
range-averaged value $+0.202$ is approached only deep in the tail
where $f \to 1$. This also explains why the helium offset still creeps
upward below $z = 200$ even though $Y_{\rm He}$ is long since frozen.

The drift is the thermal feedback. The case-B recombination
coefficient $\alpha_B$ depends on the \emph{matter} temperature $T_b$,
and $T_b$ carries a strong $G$-dependence that develops only in the
tail. The baryon temperature histories show this directly, the ratio
$T_b/T_\gamma$ at $z = 200$ running from $0.49$ at $\lG = 0.25$ to
$0.95$ at $\lG = 2$. A larger $\lG$ means a faster expansion and hence
a larger residual $\xe$. That larger $\xe$ sustains the Compton
coupling to the CMB longer, so the gas decouples thermally later and
stays hotter. A hotter gas in turn recombines less efficiently, since
$\alpha_B \propto T_b^{-0.6}$, which keeps $\xe$ higher
still. Integrated along the tail, this positive feedback lifts the
effective exponent above unity.

Both effects can be isolated numerically, and together they close the
budget.  Integrating Eq.~\eqref{eq:bernoulli} downward from
$z_f = 800$ with \CLASS{}'s own $T_b(z)$, $H(z)$ and $Y_{\rm He}$ as
inputs, and switching the two $G$-dependencies on one at a time, gives
at $z = 200$
\begin{equation}
\underbrace{1.003}_{H \text{ only}}
\;\xrightarrow{\;+0.180\;}\;
\underbrace{1.184}_{+\,(1-Y_{\rm He})}
\;\xrightarrow{\;+0.067\;}\;
\underbrace{1.251}_{+\,T_b(z)}
\quad\text{against}\quad
\underbrace{1.255}_{\CLASS{}}\,,
\label{eq:nbudget}
\end{equation}
so the reconstruction agrees with the full calculation to $0.3\%$. At
the bottom of the tail, $z = 20$, the two contributions have become
equal, $+0.187$ each, for a total of $1.376$ against $1.391$ from
\CLASS{}. The helium term is essentially constant once $Y_{\rm He}$ is
set, while the thermal term grows monotonically into the tail, which
is precisely why the offset and the drift have separate origins.

Three remarks follow. First, the dominant contribution is the helium
one, not the thermal feedback, by a factor of nearly three at
$z = 200$. Second, the neglected photoionization term of
Eq.~\eqref{eq:peebles} cannot contribute at all: relative to the
recombination term, the bracket
$4\beta_B\,e^{-(B_1-B_2)/k_BT}(1-\xe)/[\alpha_B n_{\rm H}\xe^2]$ falls
from $2\times10^{-4}$ at $z = 1000$ to $3\times10^{-8}$ at $z = 800$
and $2\times10^{-17}$ at $z = 600$, and the choice of $T_\gamma$ or
$T_b$ in $\beta_B$ changes it by less than a factor of two where it is
still measurable.  Third, the low-temperature behaviour of $\alpha_B$
matters but only marginally. The effective slope $p$ of
$\alpha_B \propto T_b^{-p}$ in the fit we use is not the constant
$0.6$ of the argument above but drifts from $0.750$ at $z = 1000$ to
$0.654$ at $z = 100$, reaching the quoted $0.617$ only asymptotically;
replacing the full fit by a strict $T_b^{-0.6}$ power law moves $n$ at
$z = 200$ from $1.251$ to $1.231$, that is by $0.020$, or $8\%$ of the
excess. We note for completeness that at the temperatures reached in
the deep tail, a few hundred kelvin and below, the case-B fit is
extrapolated well outside the range over which it was calibrated, so
this last $0.020$ should not be taken as a precise number.

\subsection{Observational consequences}
\label{sec:obsconseq}

The residual electrons contribute an additional optical depth through
the dark ages,
\begin{equation}
\tau_{\rm dark} = \int_{z_{\rm reio}}^{z_{\rm end}}
\frac{\xe(z)\, n_{\rm H}\,\sigma_T c}{(1+z)H}\,\dd z\,.
\label{eq:taudark}
\end{equation}
Its response to $G$ involves three channels and not two, and the third
is the one that matters. Besides $\xe \propto \lG^{\,n}$ and
$1/H \propto \lG^{-1}$, the integrand carries
$n_{\rm H} = (1-Y_{\rm He})n_b$, the same counting factor as
Sec.~\ref{sec:thermal}, so that
\begin{equation}
\frac{\dd\ln\tau_{\rm dark}}{\dd\ln\lG} \;=\;
\langle n\rangle - 1 + \frac{\dd\ln(1-Y_{\rm He})}{\dd\ln\lG}\,,
\label{eq:taudark_scaling}
\end{equation}
where $\langle n \rangle$ is the exponent of Sec.~\ref{sec:thermal}
weighted by the integrand itself. Two points about the bounds deserve
to be explicit. The integrand uses the full $\xe(z)$ and not the
constant residual plateau, which would underestimate the result by a
factor $4.6$. And the upper bound $z_{\rm end}$ is the redshift at
which recombination is \emph{complete}, which we take at
$z_{\rm end} \simeq 800$, where $\xe \simeq 3.5\times10^{-3}$. It is
not $z_{\rm rec} = 1275$, which is the midpoint of the transition and
where $\xe = 1/2$ still contributes overwhelmingly, driving the
integral to order unity and making it a recombination quantity rather
than a dark-ages one.

With these bounds, integrating our own histories gives
$\tau_{\rm dark} = 0.0419$ at $\lG = 1$. Across the eight models it
falls monotonically from $0.0446$ at $\lG = 0.25$ to $0.0384$ at
$\lG = 2$, a total variation of $16\%$ over a factor of eight in $\lG$,
that is
\begin{equation}
\tau_{\rm dark} \;\simeq\; 0.042\;\lG^{-0.074}\,,
\label{eq:taudark_fit}
\end{equation}
with a local slope of $-0.099$ at $\lG = 1$. We have checked this
against the optical depth tabulated by \CLASS{} itself, which it
reproduces to $0.2\%$.

Equation~\eqref{eq:taudark_scaling} accounts for this almost
exactly. The integrand-weighted exponent over the window is
$\langle n \rangle = 1.129$, and the range-averaged helium derivative
is $-0.202$, so that
\begin{equation}
\underbrace{+0.129}_{\langle n\rangle - 1}
\;\underbrace{-\,0.202}_{(1-Y_{\rm He})}
\;=\; -0.073
\qquad\text{against}\qquad
\underbrace{-0.074}_{\text{direct fit}}\,,
\label{eq:taudark_budget}
\end{equation}
a closure at the percent level. It is worth being precise about what
cancels here, because it is not what one might expect. The exponent
$n(z)$ exceeds unity \emph{everywhere} in the integration window, from
$1.03$ at $z_{\rm end}$ to $1.40$ at $z_{\rm reio}$, so there is no
compensation between regions with $n>1$ and $n<1$. Nor is the integral
dominated by the deep tail where $n$ is largest: $64\%$ of
$\tau_{\rm dark}$ accumulates above $z = 600$ and only $4\%$ below
$z = 200$, which is why $\langle n \rangle = 1.13$ sits far closer to
the value just below recombination than to the deep-tail $1.40$. The
near-cancellation is instead between the $\xe/H$ combination, worth
$+0.13$, and the helium counting factor, worth $-0.20$, and the latter
slightly overshoots the former, which is why $\tau_{\rm dark}$
decreases rather than increases with $G$.

The practical conclusion is unchanged, and if anything firmer. A
percent change in $\lG$ moves $\tau_{\rm dark}$ by seven parts in ten
thousand in relative terms, that is by $3\times10^{-5}$ in absolute terms, far
below any foreseeable sensitivity. The residual-ionization sector is
therefore sensitive to $G$ but very weakly, and the dominant CMB
signature of a modified $G$ remains the damping tail
(Eq.~\ref{eq:decomposition}), together with the displacement and
broadening of the visibility function.

\section{Discussion}
\label{sec:finaldiscussion}

The results are listed in Sec.~\ref{sec:conclusion}. What they rest on
is the dropping of three approximations, each harmless on its own and
together responsible for the whole effect. Recombination is not in
equilibrium, the ionization history lagging the Saha curve by
$\Delta z \simeq 95$ at half ionization. Last scattering is not
instantaneous, and the finite thickness of the shell, while it leaves
the position of $\ldamp$ alone, governs the shape of the envelope. And
the photon mean free path is not the Hubble radius at the visibility
peak, where the two differ by a factor of about $11$. That last
identification is legitimate only at the rate crossing
$\Gamma_{\rm T} = H$, and it is the step that suppresses the signal
in \citet{Rich2015}.

Two caveats attach to the exponent itself, and neither is a matter of
precision. Three closely spaced models fix it to better than $0.001$,
so the determination is not the limitation. The first is the
convention that defines the last-scattering surface, the visibility
peak giving $-0.30$ and $\tau = 1$ giving $-0.25$; we quote both
rather than merge them, since the difference is one of definition. The
second is that the damping scale \CLASS{} computes internally agrees
with ours and calls for no spectral fit, but it is the same
calculation rather than an independent one. The genuinely independent
check is the analytic normalisation of Eq.~\eqref{eq:rdamp}. The
damping tail is therefore the place where a rescaling of $G$ shows up,
the acoustic and equality scales being almost untouched.

This last point has a consequence worth stating on its own, and it is
best put as a counting argument. Each of the three multipoles is the
ratio of the comoving distance to last scattering to a comoving length
imprinted \emph{before} last scattering. A late-time modification of
gravity, or of the expansion history, therefore rescales all three
multipoles by one common factor and leaves their ratios untouched. By
the same token $H_0$, which enters only through $D_c$, is absorbed
into $\ell_A$ and drops out of the ratios. What the CMB constrains
without any assumption about the late-time expansion is thus a set of
distance-free numbers, three of which matter here,
\begin{equation}
\frac{\ldamp}{\ell_A} \propto \frac{r_s}{\rdamp}\,,
\qquad
\frac{\ell_{\rm eq}}{\ell_A} \propto k_{\rm eq}\,r_s\,,
\qquad
R_\ast \equiv \left.\frac{3\rho_b}{4\rho_\gamma}\right|_{z_{\rm dec}}\,,
\label{eq:threeratios}
\end{equation}
the last being the baryon loading at decoupling, measured directly by
the relative heights of the acoustic peaks. Against these stand
three unknowns of the pre-decoupling era, $\lG$ and the two physical
densities $\omega_b^{\rm phys}$ and $\omega_{\rm cdm}^{\rm phys}$, provided
that $Y_{\rm He}$ is tied to them by BBN and that the radiation
content is held fixed. Three numbers for three unknowns, and $G$ is
determined. The fourth observable, $\ell_A$ itself, then returns $D_c$
and hence $H_0$ for a given late-time model, which is the usual CMB
inference and plays no part in the determination of $G$.

The two closure conditions are worth a word, because they are where
this paper meets the global fits. Releasing $Y_{\rm He}$ adds a fourth
unknown acting on $\rdamp$ through $n_e$, and with
$\kappa_{Y_{\rm He}} = +0.715$ that channel is far from small, which
is why a BBN helium prior enters the data combination of
\citet{Lamine2025}. Releasing $N_{\rm eff}$ adds another, since it too
changes the expansion rate before decoupling, and with three numbers
for four unknowns the counting no longer closes. The two are not
equivalent, however. At fixed physical densities $\lG$ leaves
$z_{\rm eq}$ strictly invariant, whereas $N_{\rm eff}$ moves it, so
$\ell_{\rm eq}/\ell_A$ responds to one and not to the other, and the
full spectrum, which holds more than these three numbers, retains the
means to separate them.

What makes the system well conditioned rather than merely closed is
that it is almost triangular. From Table~\ref{tab:sensitivities}, and
using $R_\ast \propto \omega_b^{\rm phys}/(1+z_{\rm dec})$ at fixed
physical densities,
\begin{equation}
\kappa_{R_\ast} = -0.025\,,
\quad
\kappa_{\ell_{\rm eq}/\ell_A} = -0.017\,,
\quad
\kappa_{\ldamp/\ell_A} = -0.320\,,
\label{eq:triangular}
\end{equation}
so the first two respond to $G$ some twenty times more weakly than the
third. $R_\ast$ fixes $\omega_b^{\rm phys}$ while being essentially
blind to $\lG$; $\ell_{\rm eq}/\ell_A$ then fixes
$\omega_{\rm cdm}^{\rm phys}$, almost as blind, because it is anchored to
$z_{\rm eq}$; and only at the third step does $\ldamp/\ell_A$ carry
$G$. The determination therefore does not rest on inverting a
near-degenerate system, but on one scale responding where the others
do not. Up to the caveats listed below, the ratio $\ldamp/\ell_A$
provides a direct comparison between $G$ at recombination and the
value measured today in the laboratory, with no assumption about the
expansion history in between.

That statement rests on four assumptions which should be made
explicit. First, every other constant is held fixed. A varying
fine-structure constant moves $\ldamp$ through the $\alpha^{5/2}$ and
$\theta_0^{3/4}$ of Eq.~\eqref{eq:ldamp_micro} and is strongly
degenerate with a rescaling of $G$ in this ratio, so the comparison
tests $G$ only within a theory in which $\alpha$ and $m_e$ are
fixed. Second, spatial flatness is assumed, since it is what lets
$D_c$ be identified with $(1+z)D_A$; we have not verified that the
cancellation survives at the percent level in a curved
background. Third, the modification of gravity must be confined to
$z \ll z_{\rm dec}$. Early dark energy, or any pre-recombination
change of $H(z)$, moves the three scales differently and voids the
argument, which is precisely why the ratio is diagnostic rather than
merely degenerate. Fourth, the gravitational lensing of the damping
tail is a genuine late-time contaminant. It does not touch the ratio quoted
here, since $\ldamp$ is taken from the Silk integral and the background
distances rather than from a spectral fit, and on the tail itself it shifts
only the envelope exponent, by a few percent
(Sec.~\ref{sec:lensingtest}). Turning the argument into a measurement
requires extracting $\ldamp$ from real data, where it is degenerate with the
envelope shape. What survives that degeneracy, as Sec.~\ref{sec:tt} shows, is
the peak-to-trough contrast of the acoustic structure at high $\ell$, and the
resulting sensitivity to $G$ is consistent with the published
determination. A screening mechanism making the locally
measured $G$ differ from the cosmological one would of course also
break the comparison, but at the level of the interpretation rather
than of the observable. This is also the cleanest answer to the
concern that a CMB bound on $G$ might be absorbed into late-time new
physics, since such physics cannot move this ratio.

The residual ionization, by contrast, is a much weaker channel. It
scales as $\xeres \propto \lG^{\,n(z)}$ with $n$ rising from
$\simeq 0.9$ near recombination to $\simeq 1.4$ in the deep ionization
tail, already $\simeq 1.2$ at freeze-out, an excess traced to a
positive thermal feedback between the residual electrons and the
gas. The associated optical depth through the dark ages changes only
at the per mille level for percent-level variations of $\lG$, so this
sector, while predictively clean, carries little observational weight.

\section{Conclusions}
\label{sec:conclusion}

The results of this paper are the following, each with the equation
that states it.

\begin{itemize}

\item Since the Saha equilibrium contains no $G$,
  Eq.~\eqref{eq:saha}, the gravitational constant affects
  recombination only through the rate at which the Universe expands,
  and therefore only out of equilibrium.

\item The gravitational effect on non-equilibrium recombination is
  governed by a single dimensionless combination,
  the gravitational coupling between an electron and the mean mass of
  matter per photon, $\aG = Gm_e\mu/\hbar c$, which
  Eq.~\eqref{eq:alphaG} isolates in the ratio of the recombination
  rate to the expansion rate. The Peebles factor itself reduces to
  dimensionless quantities, Eqs.~\eqref{eq:CP_exact}
  to \eqref{eq:CP_alpha}. The same $\aG$ governs the geometric part of
  the damping scale, so the whole gravitational sensitivity rests on
  it, the helium fraction being the only channel that calls on a
  separate, weak-interaction combination, $\mathcal{G}_{\rm w}$ of
  Eq.~\eqref{eq:Gw}.

\item Equilibrium requires not $\Gamma_{\rm rec} \gg H$ but
  $\Gamma_{\rm rec} \gg (B_1/2k_BT)\,H$, Eq.~\eqref{eq:target},
  because the Saha target is itself moving. The controlling quantity
  is therefore $\mathcal{Q}$ of Eq.~\eqref{eq:Qcrit}, equal to $0.29$ in
  our own Universe, which places us just on the out-of-equilibrium
  side of $\mathcal{Q} = 1$.

\item The diffusion length and the damping multipole follow in
  fundamental constants, Eqs.~\eqref{eq:rdamp}
  and \eqref{eq:ldamp_micro}, both built on the reduced Compton
  wavelength of the electron and carrying a single $\aG^{-1/4}$ of
  gravity.

\item The central result is the decomposition of
  Eq.~\eqref{eq:decomposition}, $\ldamp \simeq 1859\,\lG^{-0.304}$,
  whose exponent is $-0.500$ of geometry and $+0.196$ of recombination
  kinetics. Over the wide family the exponent runs,
  Eq.~\eqref{eq:running} with $n = -0.301$ and $b = -0.069$, the local
  slope steepening from $-0.11$ at $\lG = 0.25$ to $-0.40$ at
  $\lG = 2$.

\item The shape of the damping envelope is a separate observable. Its
  exponent in Eq.~\eqref{eq:envelope} is $m \simeq 1.45$ rather than
  $2$, and it measures the finite thickness of last scattering and not
  the position of the cut-off.

\item There is a critical value of $G$ below which recombination is in
  equilibrium and the entire effect disappears,
  Eq.~\eqref{eq:aGcrit_explicit}. It is a pure power law of $\alpha$
  and $\eta_b$, and the numerical determination gives
  $\lG^{\rm crit} \simeq 0.14$, that is $G^{\rm crit} \simeq
  G_{\rm N}/49$. Its only cosmological inputs are $\eta_b$ and the
  mean mass per photon $\mu$. Our own Universe sits close to that
  boundary, which is why the kinetic and the geometric terms of
  Eq.~\eqref{eq:decomposition} are comparable.

\item What the CMB constrains with no assumption about the late-time
  expansion is a set of distance-free ratios, three of which matter
  here, Eq.~\eqref{eq:threeratios}, against three unknowns of the
  pre-decoupling era. The system is well conditioned because it is
  almost triangular, Eq.~\eqref{eq:triangular}.

\item The residual ionization of the dark ages carries the same
  physics but little observational weight. Its optical depth obeys
  Eq.~\eqref{eq:taudark_fit},
  $\tau_{\rm dark} \simeq 0.042\,\lG^{-0.074}$, so percent-level
  changes of $\lG$ move it only at the per mille level.

\end{itemize}

\begin{acknowledgements}
  This work made use of the \CLASS{} code \citep{Blas2011}. Claude
  (Anthropic) was used for analytical cross-checks, numerical
  verification and manuscript editing. Every result reported here as
  new, in particular the normalisation of Eq.~\eqref{eq:rdamp} and the
  decomposition of Eq.~\eqref{eq:decomposition}, has been
  independently reproduced by the authors.
\end{acknowledgements}

\end{document}